\documentclass[11pt]{article}
\usepackage{authblk}
\usepackage{enumerate}
\usepackage{amssymb}
\usepackage{amsmath}
\usepackage{graphicx}
\usepackage{amsthm}
\usepackage{bm}
\usepackage{multicol}  
\usepackage[all]{xy}
\usepackage{times}
\usepackage{graphicx}
\usepackage{tikz}
\usepackage{url}
\usepackage{mathrsfs}

\usepackage{thmtools}
\usepackage{thm-restate}
\usepackage{cleveref}
\usepackage[shortlabels]{enumitem}
\usepackage{moreenum}

\usetikzlibrary{automata, positioning, arrows}
\usetikzlibrary{patterns}

\newcommand{\proptoinverse}{\mathrel{\mskip-2mu\reflectbox{{\scriptsize$\propto$}}}}

\newcommand{\Alphabet}{\Sigma}

\newcommand{\fa}{\forall}
\newcommand{\Ga}{\Gamma}
\newcommand{\Gas}{\Gamma^{<\omega}}
\newcommand{\Si}{\Alphabet}

\newcommand{\macroforcomplement}{-}

\newcommand{\ite}{\item}

\newcommand{\tla}{\twoheadleftarrow}
\newcommand{\strongtla}{\twoheadleftarrowtail}
\newcommand{\strongtlascripstyle}{\twoheadleftarrowtailscripstyle}
\newcommand*{\twoheadleftarrowtail}{\hbox{${\mathrel{\textcolor{black}{\leftarrowtail}\kern-3.08ex\twoheadleftarrow}}$}}
\newcommand*{\twoheadleftarrowtailscripstyle}{\hbox{${\mathrel{\scriptstyle\leftarrowtail\kern-1.96ex\twoheadleftarrow}}$}}
\newcommand{\Linfinite}{{\scriptstyle\omega}\text{-}\mathscr{L}}
\newcommand{\Lfinite}{{\scriptstyle<\omega}\text{-}\mathscr{L}}
\newcommand{\Lboth}{{\scriptstyle\leq\omega}\text{-}\mathscr{L}}

\newcommand{\skiponesturn}{{\scriptstyle\quq{\text{skip}}}}

\DeclareMathAlphabet{\mathbbold}{U}{bbold}{m}{n}

\newcommand{\borel}{\mathbf{\Delta}^{1}_{1}}

\newcommand{\borone}{\mathbf{\Delta}^{0}_{1}}

\newcommand{\boraone}{\mathbf{\Sigma}^{0}_{1}}

\newcommand{\blank}{\texttt{d}}

\newcommand {\bconeinf}{\ifmmode\leq_{\mathbb{BC}_1}\else$\leq_{\mathbb{BC}_1}$\fi}
\newcommand {\winf}{\ifmmode\leq_{\mathbb{W}}\else$\leq_{\mathbb{W}}$\fi}
\newcommand {\wsup}{\ifmmode\geq_{\mathbb{W}}\else$\geq_{\mathbb{W}}$\fi}

\newcommand {\wequ}{\ifmmode\equiv_{\mathbb{W}}\else$\equiv_{\mathbb{W}}$\fi}

\newcommand {\emptysequence}{\ifmmode\varepsilon\else$\varepsilon$\fi}

\newcommand{\tlaBaire}{\textcolor{blush}{\tla}}

\newcommand {\one}{\ifmmode\hbox{$I$}\else$I$\fi}
\newcommand {\two}{\hbox{$I$\hspace{-.6ex}$I$}}

\newcommand {\Gcont}[1]{\ifmmode\hbox{\bf G}_{\textit{cont.}}\left(#1\right)\else$\hbox{\bf G}_{\textit{cont.}}\left(#1\right)$\fi}
\newcommand {\GWadge}[2]{\mathbb{W}\left(#1,#2\right)}
\newcommand {\GWadgeBCone}[2]{\mathbb{W}_{\mathbb{BC}_1}\left(#1,#2\right)}

\newcommand {\GBaire}[1]{\ifmmode\hbox{\bf G}_{\mathbb{BC}_1}\left(#1\right)\else$\hbox{\bf G}_{\mathbb{BC}_1}\left(#1\right)$\fi}

\newcommand {\set}[1]{\left\{#1\right\}}

\newcommand {\setbig}[1]{\big\{#1\big\}}
\newcommand {\setBig}[1]{\Big\{#1\Big\}}

\newcommand {\parentheses}[1]{\left(#1\right)}
\newcommand {\parenthesesbig}[1]{\big(#1\big)}

\newcommand {\parenthesesBig}[1]{\Big(#1\Big)}

\newcommand {\quq}[1]{\text{``}\,#1\,\text{''}}

\newcommand {\auxiliarymove}{\mathbb{A}}
\definecolor{airforceblue}{rgb}{0.36, 0.54, 0.66}
\definecolor{ao}{rgb}{0.0, 0.0, 1.0}
\definecolor{brandeisblue}{rgb}{0.0, 0.44, 1.0}
\definecolor{darkterracotta}{rgb}{0.8, 0.31, 0.36}
\definecolor{denim}{rgb}{0.08, 0.38, 0.74}
\definecolor{glaucous}{rgb}{0.38, 0.51, 0.71}
\definecolor{darkcandyapplered}{rgb}{0.64, 0.0, 0.0}
\definecolor{blush}{rgb}{0.87, 0.36, 0.51}
\definecolor{olive}{rgb}{0.5, 0.5, 0.0}
\definecolor{officegreen}{rgb}{0.0, 0.5, 0.0}
\definecolor{cambridgeblue}{rgb}{0.64, 0.76, 0.68}
\definecolor{egyptianblue}{rgb}{0.06, 0.2, 0.65}
\definecolor{internationalkleinblue}{rgb}{0.0, 0.18, 0.65}
\definecolor{awesome}{rgb}{1.0, 0.13, 0.32}
\definecolor{internationalkleinblue}{rgb}{0.0, 0.18, 0.65}
\definecolor{hookersgreen}{rgb}{0.0, 0.44, 0.0}
\definecolor{tomato}{rgb}{1.0, 0.39, 0.28}
\definecolor{buff}{rgb}{0.94, 0.86, 0.51}
\definecolor{dogwoodrose}{rgb}{0.84, 0.09, 0.41}

\newcommand {\colorforyes}{hookersgreen}
\newcommand {\colorforno}{tomato}
\newcommand {\Tvar}[1]{{\textcolor{internationalkleinblue}{#1}}}%internationalkleinblue
\newcommand {\tildevar}[1]{{\textcolor{internationalkleinblue}{\ddot{#1}}}}%internationalkleinblue
\newcommand {\tildeT}{\Tvar{\ddot{T}}}
\newcommand {\tildex}{\Tvar{\ddot{x}}}
\newcommand {\tildexprime}{\Tvar{\ddot{x}'}}
\newcommand {\T}{\Tvar{T}}
\newcommand {\x}{\Tvar{x}}
\newcommand {\anovar}{\textcolor{darkterracotta}{\auxiliarymove}}

\newcommand {\xvar}[1]{\textcolor{internationalkleinblue}{x_{#1}}}%internationalkleinblue
\newcommand {\tildexvar}[1]{\Tvar{\ddot{x}_{(#1)}}}
\newcommand {\avar}[1]{\textcolor{darkterracotta}{\auxiliarymove_{#1}}}
\newcommand {\avarprime}[1]{\textcolor{darkterracotta}{\auxiliarymove'_{#1}}}

\newcommand {\code}[1]{\ulcorner#1\urcorner}
\newcommand {\codeinv}[1]{\llcorner#1\lrcorner}
\newcommand {\N}{\mathbb{N}}

\newcommand {\checked}[1]{{#1}^{-}}

\newcommand {\thecomplement}[1]{\checked{#1}}

\newtheorem{theorem} {Theorem}
\newtheorem{definition} [theorem] {Definition}
\newtheorem{corollary} [theorem] {Corollary}
\newtheorem{claim}{Claim}
\newtheorem{notation}[theorem]{Notation}
\newtheorem*{notation*}{Notation}
\newtheorem{example} {Example}
\newtheorem{lemma} [theorem] {Lemma}
\newtheorem{proposition} [theorem] {Proposition}
\newtheorem{remark} {Remark}
\newtheorem{remarks} [remark] {Remarks}
\newcommand {\proofpro}[1]{\noindent{\it Proof of  Proposition~$\ref{#1}$}: }

\newcommand {\proofthm}[1]{\noindent{\it Proof of Theorem~$\ref{#1}$}: }

\newcommand {\prooflem}[1]{\noindent{\it Proof of Lemma~$\ref{#1}$}: }

\newcommand {\proofcor}[1]{\noindent{\it Proof of Corollary~$\ref{#1}$}: }
\newcommand {\proofclaim}[1]{\noindent{\it Proof of  Claim~$\ref{#1}$}: }

\newcommand {\boxpro}[1]{\hfill $\Box~${\footnotesize\hbox{$\ref{#1}$}}\vspace{1.2ex}}
\newcommand {\boxcor}[1]{\hfill $\Box~${\footnotesize\hbox{$\ref{#1}$}}\vspace{1.2ex}}

\newcommand {\boxthm}[1]{\hfill $\Box~${\footnotesize\hbox{$\ref{#1}$}}\vspace{1.2ex}}

\newcommand {\boxlem}[1]{\hfill $\Box~${\footnotesize\hbox{$\ref{#1}$}}\vspace{1.2ex}}

\newcommand {\boxclaim}[1]{\hfill $\Box~${\footnotesize\hbox{$\ref{#1}$}}\vspace{1.2ex}}

\usepackage{tikz-cd}

\date{\today}

\def\ufootnote#1{\let\savedthfn\thefootnote\let\thefootnote\relax
\footnote{#1}\let\thefootnote\savedthfn\addtocounter{footnote}{-1}}

\title{Some complete $\omega$-powers of a one-counter language in the difference hierarchy}
\author[1,2]{Jacques DUPARC}
\author[3]{Olivier FINKEL}
\author[4,5]{Dominique LECOMTE}

\affil[1]{Universit\'e de Lausanne\\

D\'epartement des op\'erations, HEC Lausanne\\

CH-1015, Lausanne, Switzerland\\

Jacques.Duparc@unil.ch\medskip}

\affil[2]{EPFL\\

Mathematics Section, School of Basic Sciences\\

CH-1015, Lausanne, Switzerland\\

Jacques.Duparc@epfl.ch\medskip}

\affil[3]{CNRS, Universit\'e Paris Cit\'e and Sorbonne Universit\'e\\

Institut de Math\'ematiques de Jussieu-Paris Rive Gauche\\

F-75013 Paris, France\\

finkel@math.univ-paris-diderot.fr\medskip}

\affil[4]{Sorbonne Universit\'e and Universit\'e Paris Cit\'e\\

 CNRS, Institut de Math\'ematiques de Jussieu-Paris Rive Gauche\\

F-75005 Paris, France\\

dominique.lecomte@upmc.fr\medskip}

\affil[5]{Universit\'e de Picardie, I.U.T. de l'Oise, site de Creil\\

13, all\'ee de la fa\"\i encerie\\

 60107 Creil, France}

\date{}                     %% if you don't need date to appear
\begin{document}
  \maketitle

\ufootnote{{\it 2020 Mathematics Subject Classification.}~Primary: 03E15, 03D05 Secondary: 54H05, 68Q45, 68R15}

\ufootnote{{\it Keywords and phrases.}~Borel class, complete, context-free, difference hierarchy, $\omega$-power, one-counter automaton}

\abstract{ Fix a finite alphabet $\Alphabet$ having at least two letters, and natural numbers $n,k\geq 1$. We denote by $D_{n}(\mathbf{\Pi}^{0}_{k})$ the class of differences of $n$ subsets of the Cantor space $\Alphabet^{\omega}$ in $\mathbf{\Pi}^{0}_{k}$, and by $\checked{D}_{n}(\mathbf{\Pi}^{0}_{k})$ the class of complements of sets in $D_{n}(\mathbf{\Pi}^{0}_{k})$. We prove, for any natural numbers $n,k\geq 1$,  the existence of a finite alphabet $\Alphabet$ having at least two letters and a finitary language $U_{n,k}$ over $\Alphabet$ accepted by a one-counter automaton such that the $\omega$-power 
$$U_{n,k}^\infty:=\{ w_0w_1\cdots\in\Alphabet^{\omega}\mid\fa i\in\omega  ~~w_i\in U_{n,k}\}$$ 
is $D_{2n-1}(\mathbf{\Pi}^{0}_{k})$-complete. We also prove a conditional result for the dual classes 
$\checked{D}_{2n}(\mathbf{\Pi}_k^0)$:  under the assumption  that there exists a finitary language $V_{n,3}$ accepted by a one-counter automaton,  such that its $\omega$-power $V_{n,3}^\infty$ is 
$\checked{D}_{2n}(\mathbf{\Pi}_3^0)$-complete, there would also exist, for any integer $k>3$,  a finitary language $V_{n,k}$ accepted by a one-counter automaton,  such that its $\omega$-power $V_{n,k}^\infty$ is $\checked{D}_{2n}(\mathbf{\Pi}_k^0)$-complete.
Finally we add another  conditionnal result about Wadge degrees of rational $\mathbf{\Delta}^{0}_{2}$ omega powers, which completes the recent results of \cite{DBLP:journals/tcs/FinkelGD26}.

\section{Introduction}\label{introduction}\indent

 We pursue in this paper the study of  the topological complexity of $\omega$-powers of languages of finite words over a finite alphabet $\Alphabet$. A finitary language over a finite alphabet $\Alphabet$ is a subset $A$ of the set $\Alphabet^{<\omega}$ of finite words with letters in $\Alphabet$. The set of infinite words over the alphabet $\Alphabet$, i.e., of sequences of length $\omega$ of letters of $\Alphabet$, is denoted by $\Alphabet^{\omega}$. The $\omega$-\textbf{power} associated with 
 $L\subseteq\Alphabet^{<\omega}$ is the set $L^\infty$ of all infinite words  that can be constructed from $L$ by concatenation, i.e., 
${L^\infty:=\{ w_{0}w_{1}\cdots\in\Alphabet^{\omega}\mid\forall i\in\omega~\ w_{i}\in L\}}$. Note that we use $L^\infty$ to denote the $\omega$-power associated with $L$ (as in \cite{Lecomte05,Fink-Lec2,Fink-Lec4}), although it is often referred to as $L^{\omega}$ in articles on theoretical computer science. We reserve the notation $L^{\omega}$ to denote the Cartesian product of a countable number of copies of $L$, as it will be used frequently in this article.
\medskip

The $\omega$-powers occur very naturally in the theory of formal languages of infinite words accepted by various kinds of automata.  
For instance, it features in the characterization of the class $\mbox{REG}_{\omega}$ of $\omega$-regular languages as the $\omega$-Kleene closure of the class $\mbox{REG}$ of regular finitary languages. It also plays the same role for the class   $\mbox{CF}_{\omega}$ of context-free $\omega$-languages, as the $\omega$-Kleene closure of the class 
$\mbox{CF}$ of context-free finitary languages \cite{Staiger97}. 
 The $\omega$-\textbf{Kleene closure} of a class $\mathcal{C}$  of languages of finite words over finite alphabets is the class of $\omega$-languages of the form $\bigcup_{1\leq j \leq n}~U_j V_j^{\infty}$, for some finitary languages $U_j$ and $V_j$ in the class $\mathcal{C}$, $1\leq j\leq n$. Once the set $\Alphabet^{\omega}$  equipped with the usual Cantor topology, the question of  the topological  complexity of  $\omega$-powers of finitary  languages naturally arose and was  raised in particular by Niwinski \cite{Niwinski90}, Simonnet \cite{Simonnet92},  and  Staiger \cite{Staiger97}. In addition,  $\omega$-powers have also been studied from the perspective of descriptive set theory  \cite{Fin01a,{Fin03a},Fin04-FI,Lecomte05,Fin-Dup07,Fin-Lec1,Fink-Lec2,Fink-Lec4,Fink-Lec5,DBLP:journals/tcs/FinkelGD26}.\medskip

An  $\omega$-power $L^\infty$ is an analytic set since the  function from $L^{\omega}$ onto $L^\infty$ which maps a given infinite sequence $(w_{i})_{i\in\omega}$ of finite words to the concatenated word $w_{0}w_{1}\cdots$ is continuous. Moreover,  in \cite{Fin03a}, the second author provided a context-free language $L$ such that $L^\infty$ is analytic but not Borel. Surprisingly , this language $L$ is  very simple to describe and is even accepted by a simple one-counter automaton. Louveau has proved independently that analytic-complete 
$\omega$-powers exist, but the existence was proved in a non effective way (this is non-published work). We refer the reader to \cite{ABB96} for basic notions about context-free languages.%
\medskip

With regard to Borel $\omega$-powers, the second author showed in  \cite{Fin01a} that, for each integer $n\geq 1$, there exist some $\omega$-powers of  context-free languages that are $\mathbf{\Pi}_{n}^0$-complete Borel sets. Subsequently, he proved in \cite{Fin04-FI} that there exists a finitary language $L$ such that $L^\infty$ is a Borel set of infinite rank, and,  with the first author,  in \cite{Fin-Dup07} that there exists a context-free language $M$ such that $M^\infty$ is Borel above $\mathbf{\Delta}_{\omega}^{0}$.
\medskip

The second and third authors  showed that $\omega$-powers exhibit great topological complexity. Namely, for each countable ordinal  $\xi\geq 1$, there exist both $\mathbf{\Pi}^{0}_{\xi}$-complete  and $\mathbf{\Sigma}^{0}_{\xi}$-complete  $\omega$-powers \cite{Fin-Lec1, Fink-Lec2}.  These results also have effective aspects: let $\omega_1^{\text{CK}}$ stand for the first non-recursive ordinal, for every recursive ordinal $\xi <\omega_1^{\text{CK}}$ 
there exist recursive finitary languages $P_{\xi}$ and $S_{\xi}$ such that $P_{\xi}^\infty$ is $\mathbf{\Pi}^{0}_{\xi}$-complete and $S_{\xi}^\infty$ is $\mathbf{\Sigma}^{0}_{\xi}$-complete.%
\medskip

Moreover it is proved in \cite{Fink-Lec4} that one can effectively construct some one counter automata accepting some finitary languages $P_n$ and $S_n$, for $n\geq 1$, 
 such that $P_{n}^\infty$ is $\mathbf{\Pi}^{0}_{n}$-complete and $S_{n}^\infty$ is $\mathbf{\Sigma}^{0}_{n}$-complete.

\medskip

Many questions remain unanswered concerning the topological complexity of the $\omega$-powers of languages in a given class, such as the class of context-free languages, one-counter languages, recursive languages, or, more generally,  languages of finite words accepted by certain specific types of automata.

\medskip

Fix a finite alphabet $\Alphabet$ having at least two letters, and natural numbers $n,k\geq 1$. We denote by $D_{n}(\mathbf{\Pi}^{0}_{k})$ the class of differences of $n$ subsets of the Cantor space 
$\Alphabet^{\omega}$ in $\mathbf{\Pi}^{0}_{k}$, and by $\checked{D}_{n}(\mathbf{\Pi}^{0}_{k})$ the class of complements of sets in $D_{n}(\mathbf{\Pi}^{0}_{k})$. In this paper we obtain the following new results about $\omega$-powers of languages accepted by one-counter automata, obtained in \cite[main theorem]{Fink-Lec5} when $k= 1$, and in \cite[Theorem 2]{DBLP:journals/tcs/FinkelGD26} when $k= 2$ (with finitary regular languages in both cases). In this article, our main result is the following.

\begin{restatable*}{theorem}{themaintheorem} \label{thm:themaintheorem} For every natural numbers $n,k\geq 1$, there exists

\begin{enumerate}[(a)]
  \item a finitary language $U_{n,k}$ accepted by a one-counter automaton,  such that its $\omega$-power $U_{n,k}^\infty$ is $D_{2n-1}(\mathbf{\Pi}_k^0)$-complete.

    \item  Moreover, a one-counter automaton accepting this finitary language $U_{n,k}$  can be constructed effectively.

  \end{enumerate}

In addition, under the assumption  that there exists a finitary language $V_{n,3}$ accepted by a one-counter automaton,  such that its $\omega$-power $V_{n,3}^\infty$ is $\checked{D}_{2n}(\mathbf{\Pi}_3^0)$-complete, there would also exist, for $k>3$,  a finitary language $V_{n,k}$ accepted by a one-counter automaton,  such that its $\omega$-power $V_{n,k}^\infty$ is $\checked{D}_{2n}(\mathbf{\Pi}_k^0)$-complete.

\end{restatable*}

{

 This article is organized as follows. Notions of automata and formal language theory are recalled in Section \ref{section-automata}. Basic notions of topology and useful notions from descriptive set theory such as the Borel and the Hausdorff-Kuratowski hierarchies, as well as fundamental  concepts from Wadge's theory, are recalled in Section \ref{section-topology}. The definition and certain properties of the exponentiation operation on sets, which allows to climb up the Borel hierarchy, are given in Section \ref{section-exponentiation-on-sets}. Our main result ---Theorem \ref{thm:themaintheorem} --- is presented in Section \ref{differences-of-complete-powers}. 
\medskip

\section{Automata}\label{section-automata}\indent

 We assume that the reader is familiar with the concepts of formal languages as found, for example, in  \cite{HopcroftMotwaniUllman2001,Thomas90}. First, let us review some definitions and results concerning pushdown automata and context-free  languages that can be found in \cite{ABB96,CohengGold77a,CohengGold77b,Staiger97}.

 Given  $\Alphabet$ any non-empty finite alphabet, a  \textbf{finite word} over $\Alphabet$ is a sequence $w= a_0\cdots a_{l-1}$, where $l\geq 1$ is a natural number, and $a_i\in\Alphabet$ holds for each $i< l$. The integer $l$ denotes the \textbf{length} of $w$, denoted by $|w|$. If $|w|= 0$, then $w$ is the \textbf{empty word}, denoted by $\emptysequence$. When $w$ is a finite word over $\Alphabet$, we write $w= w(0)w(1)\cdots w(l -  1)$, and for each $i\leq l$, we write $w\upharpoonright i$ for its prefix $w(0)w(1)\cdots w(i-1)$ of length $i$. We also use the notation $u\subseteq v$ to indicate that the word $u$ is a prefix of the finite word $v$. We denote by $\Alphabet^{<\omega}$ the set of finite words over $\Alphabet$, and by  $\Alphabet^+$ the set of non-empty finite words over $\Alphabet$. A \textbf{language} over $\Alphabet$ is any subset of $\Alphabet^{<\omega}$. For $L\subseteq\Alphabet^{<\omega}$, its complement $\Alphabet^{<\omega}\smallsetminus L$ is denoted by  $\thecomplement{L}$.

\medskip

\vfill\eject

As usual, the first infinite ordinal\footnote{This ordinal is no other than the set $\N$ of all (positive) integers.} is denoted by $\omega$. An $\omega$-\textbf{word} over $\Alphabet$ is an $\omega$-sequence $(a_i)_{i<\omega}$, where $a_i\in\Alphabet$ hols for each natural number $i$. Given $\sigma$ any $\omega$-word over $\Alphabet$, we denote 
 $\sigma$ by $\sigma(0)\sigma(1)\cdots$, and, for any $i<\omega$,  its prefix  $\sigma(0)\sigma(1)\cdots\sigma(i -  1)$ of length $i$ 
 by $\sigma\upharpoonright i$.
 We also write $u\subseteq\sigma$ to denote that the finite word $u$ is a prefix of the $\omega$-word $\sigma$. The set of $\omega$-words over $\Alphabet$ is denoted by $\Alphabet^{\omega}$. An 
$\omega$-\textbf{language} over $\Alphabet$ is any subset of $\Alphabet^{\omega}$. For $A\subseteq\Alphabet^{\omega}$, its complement $\Alphabet^{\omega}\smallsetminus A$  is denoted by $\thecomplement{A}$.
\medskip

 The usual \textbf{concatenation} product of two finite words $u$ and $v$ is denoted by $u^\smallfrown v$ (more often just $uv$). This product is extended to the product of a finite word $u$ and an $\omega$-word $\sigma$: the infinite word $u^\smallfrown\sigma$ is then the $\omega$-word such that 
 $(u^\smallfrown\sigma )(k)= u(k)$ if $k< |u|$, and $(u^\smallfrown\sigma )(k)=\sigma (k -  |u|)$ if $k\geq |u|$. If $E$ is a set, $l\in\omega$ and $(e_i)_{i<l}\in E^l$, then 
 ${^\smallfrown}_{i<l}\ e_i$ stands for the concatenation $e_0^{\smallfrown}\cdots^{\smallfrown}e_{l-1}$. Similarly, ${^\smallfrown}_{i\in\omega}\ e_i$ stands for  the infinite concatenation $e_0^{\smallfrown}e_1^{\smallfrown}\cdots$.

\begin{definition} 
A \textbf{pushdown automaton} is a 7-tuple $\mathcal{A}= (Q,\Alphabet ,\Ga ,q_0,Z_0,\delta ,F)$, where \begin{itemize}
  \item $Q$ is a non-empty finite set of states,
    \item $\Alphabet$ is a non-empty finite input alphabet,
      \item  $\Gamma$ is a non-empty finite pushdown alphabet,
        \item  $q_0\in Q$ is the initial state,
          \item  $Z_0\in\Ga$ is the starting symbol which always remains at the bottom of the pushdown stack, 
           \item  $\delta$ is a mapping from $Q\times (\Alphabet\cup\{\emptysequence\} )\times\Ga$ into the set of finite subsets of $Q\times\Gas$,
            \item  $F\subseteq Q$ is the set of final states.

\end{itemize} The automaton $\mathcal{A}$ is said to be 
\textbf{real-time} if there is no $\emptysequence$-transition, i.e., if the domain of $\delta$ is 
$Q\times\Alphabet\times\Ga$.\medskip

 If $\gamma\in\Ga^+$ describes the pushdown stack content, then the leftmost symbol will be assumed to be on the ``top" of the stack. A \textbf{configuration} of the pushdown automaton 
$\mathcal{A}$ is a pair $(q,\gamma )$, where $q\in Q$ and $\gamma\in\Gas$. For 
$a\in\Alphabet\cup\{\emptysequence\}$, $\gamma ,\beta\in\Gas$ and $Z\in\Ga$, if $(p,\beta )$ is in 
$\delta (q,a,Z)$, then we write $a: (q,Z\gamma )\mapsto_{\mathcal{A}} (p,\beta\gamma )$.\smallskip
  
 Let $w= a_0\cdots a_{l-1}$ be any finite word over $\Alphabet$, a  sequence of configurations 
$r= (q_i,\gamma_i)_{i<N}$ is called a \textbf{run of $\mathcal{A}$ on $\boldsymbol{w}$ starting in the configuration} $(p,\gamma )$ if
\begin{enumerate}
\ite[(1)] $(q_0,\gamma_0)= (p,\gamma)$,

\ite[(2)]  for each integer $i< N -  1$, there exists $b_i\in\Alphabet\cup\{\emptysequence\}$ satisfying 
$b_i: (q_i,\gamma_i)\mapsto_{\mathcal{A}} (q_{i+1},\gamma_{i+1})$ such that 
$a_0\cdots a_{l-1}= b_0\cdots b_{N-2}$.
\end{enumerate}

 A run $r$ of $\mathcal{A}$ on $w$ starting in configuration $(q_0,Z_0)$ will be simply called a 
\textbf{run of $\boldsymbol{\mathcal{A}}$ on $\boldsymbol{w}$}. This run is \textbf{accepting} if it ends in a final state.\smallskip

 The language $\Lfinite(\mathcal{A})$ accepted by $\mathcal{A}$ is the set of finite words that admit an accepting run.  We denote by $CFL$ (for \textbf{context-free languages}) the class of all (finitary) languages accepted by pushdown automata. \smallskip

 A \textbf{one-counter automaton} is a pushdown automaton with a pushdown alphabet of the form 
${\Ga=\{ Z_0,z\}}$, where $Z_0$ is the starting symbol that always remains at the bottom of the pushdown stack. A \textbf{one-counter language} is a (finitary) language which is accepted by a one-counter automaton. 
\end{definition}

\begin{remarks}\rm
\begin{enumerate}
  \item   The above definition of a pushdown automaton refers to a {\it non-deterministic} automaton.  In the following, we will indicate most often when the automaton considered can be taken as deterministic or, conversely, when non-determinism is essential  to its behavior.

  \item  In the above definition, the acceptance condition is given by the final states. Other acceptance conditions have been considered.  For instance, a run of a  pushdown automaton is accepting if it ends in a final state with an empty stack   \cite{ABB96}. This latter acceptance condition will be addressed in the final few sections of this article.

\end{enumerate}
\end{remarks}

\begin{definition} Let $\Alphabet ,\Gamma$ be finite alphabets,

\begin{itemize}
  \item  a $(\Alphabet ,\Gamma )$-\textbf{substitution} is a mapping 
$f:\Alphabet\rightarrow 2^{\Gamma^{<\omega}}$;

\item  this mapping extends to $\Alphabet^{<\omega}$ by setting 
$f({^\smallfrown}_{i<l}\ a_i):=\{ {^\smallfrown}_{i<l}\ w_i\mid\forall i< l~~w_i\in f(a_i)\}$, where 
$l\in\omega$ and $a_0, \cdots ,a_{l-1}\in\Alphabet$;

\item this mapping extends further to $2^{\Alphabet^{<\omega}}$ by setting 
$f(L):=\bigcup_{w\in L}~f(w)$.

\item For $\mathcal{F}$  any family of languages, a $(\Alphabet ,\Gamma )$-substitution $f$ is called an $\mathcal{F}$-\textbf{substitution} whenever $f(a)\in\mathcal{F}$ holds for each $a\in\Alphabet$.

\item Given any families of (finitary) languages $\mathcal{E}$ and $\mathcal{F}$ , we set 
\[\mathcal{E}~\square ~\mathcal{F}:=\{ f(L)\mid L\in\mathcal{E}\mbox{ and }f\mbox{ is a }\mathcal{F}\mbox{-substitution}\}.
\]

\end{itemize} 
\end{definition}

 The operation of substitution gives rise to an infinite hierarchy of context-free finitary languages defined as follows.

\begin{definition} Let  $\mbox{OCL}(0):=\mbox{REG}$ be the class of regular languages, $\mbox{OCL}(1):=\mbox{OCL}$ be the class of one-counter languages, and 
$\mbox{OCL}(k+ 1):=\mbox{OCL}(k)~\square ~\mbox{OCL}$, for $k\geq 1$.\end{definition}

It is well known that the hierarchy generated by the families of languages $\mbox{OCL}(k)$ is strictly increasing. There is also a characterization of these languages using automata.

\begin{proposition}[\cite{ABB96}]  A language $L$ is in $\mbox{OCL}(k)$ if and only if $L$ is accepted by a pushdown automaton such that, during any computation, the words in the pushdown stack remain in a bounded language of the form $(z_{k-1})^{<\omega}\cdots (z_0)^{<\omega}Z_0$, where 
$\{ Z_0, z_0, \cdots ,z_{k-1}\}$ is the pushdown alphabet. Such an automaton is called a 
$k$-\textbf{iterated counter automaton}. The union 
$\mbox{ICL}:=\bigcup_{k\geq 1}~\mbox{OCL}(k)$ is called the family of 
\textbf{iterated counter languages}, which is the closure under substitution of the $\mbox{OCL}$ family. 
\end{proposition}

 Note that we can consider a  $k$-iterated counter automaton to be a $k$-counter automaton  in the
following way.  If the content of the pushdown stack of a $k$-iterated counter automaton is equal to $(z_{k-1})^{n_{k-1}}\cdots (z_0)^{n_0}Z_0$ for some natural numbers 
$n_0, \cdots ,n_{k-1}$, then the numbers $n_0, \cdots ,n_{k-1}$ are the contents of  counters 1, $\cdots$, $k$ of the $k$-counter automaton. Moreover, it is then clear that the content of counter $i$ can only be modified if the content of all counters $i+1, i+2,\cdots,k-1$ is equal to zero.
 We now recall the formal definition of a $k$-counter automaton. 

\begin{definition} Let $k\geq 1$ be an integer. A $k$-\textbf{counter automaton} is a 5-tuple 
$\mathcal{A}= (Q,\Alphabet ,q_0,\Delta ,F)$  where,

\begin{enumerate}
  \item  $Q$ is a finite set of states, 
    \item $\Alphabet$ is a finite input alphabet, 
      \item $q_0\in Q$ is the initial state,  
        \item $\Delta\subseteq Q\times (\Alphabet\cup\{\emptysequence\} )\times\{ 0,1\}^k\times Q\times
\{0,1,-1\}^k$
is the transition relation,
  \item  $F\subseteq Q$ is the set of final states. 
 \end{enumerate} 
  
  The $k$-counter automaton $\mathcal{A}$ is said to be 
\textbf{real-time} if there is no $\emptysequence$-transition, which means that 
$\Delta\subseteq Q\times\Alphabet\times\{ 0,1\}^k\times Q\times\{0,1,-1\}^k$.\smallskip

 If the machine $\mathcal{A}$ is in the state $q$ and $c_i\in\omega$ is the content of the $i^{\text{th}}$ counter $\mathcal{C}_i$, then the current \textbf{configuration} (or global state) of $\mathcal{A}$ is the $(k + 1)$-tuple $(q,c_0, \cdots,c_{k-1})$.\smallskip

For $a\in\Alphabet\cup\{\emptysequence\}$, states  $q, q'\in Q$, and $(c_0, \cdots ,c_{k-1})\in\omega^k$, we write\medskip
 
\centerline{$a: (q,c_0, \cdots ,c_{k-1})\mapsto_{\mathcal{A}} (q',c_0+ l_0, \cdots ,
c_{k-1}+ l_{k-1})$}\medskip

when $(q,a,i_0, \cdots ,i_{k-1},q', l_0, \cdots ,l_{k-1})\in\Delta$, where $i_j= 0$ if 
$c_j= 0$ and $i_j= 1$ if $c_j > 0$. Thus the transition relation has the property that if 
$(q,a,i_0, \cdots ,i_{k-1},q',l_0, \cdots ,l_{k-1})\in\Delta$ and $i_m= 0$ for some $m < k$, then $l_m= 0$ or $l_m= 1$ (but $l_m$ cannot be equal to $-1$).\smallskip

 Let $w= a_0\cdots a_{l-1}$ be any finite word over $\Alphabet$. A sequence 
$r= (q_i,c_0^i, \cdots ,c_{k-1}^i)_{i<N}$ of configurations, where $N > l$, is called a 
\textbf{run of $\boldsymbol{\mathcal{A}}$ on $\boldsymbol{w}$ starting in the configuration} $(p, c_0, \cdots ,c_{k-1})$ if
\begin{enumerate}
\ite[(1)]  $(q_0,c_0^0, \cdots ,c_{k-1}^0)= (p,c_0, \cdots , c_{k-1})$,
\ite[(2)] for each $i < N -  1$, there exists $b_i\in\Alphabet\cup\{\emptysequence\}$ such that\medskip
 
\centerline{$b_i : (q_i,c_0^i, \cdots ,c_{k-1}^i)\mapsto_{\mathcal{A}} 
(q_{i+1},c_0^{i+1}, \cdots ,c_{k-1}^{i+1})\mbox{,}$}\medskip

and $a_0\cdots a_{l-1}= b_0\cdots b_{N-2}$.
\end{enumerate}

 A run of $\mathcal{A}$ on $w$ starting in the configuration $(q_0,0, \cdots ,0)$ will be simply referred to as  a \textbf{run of $\boldsymbol{\mathcal{A}}$ on} $\boldsymbol{w}$. Such a run is \textbf{accepting} if it ends in a final state. The language $\Lfinite(\mathcal{A})$ accepted by $\mathcal{A}$ is the set of finite words admitting an accepting run by $\mathcal{A}$.\smallskip
  
 Let $\sigma= a_0a_1\cdots$ be an $\omega$-word over $\Alphabet$. An $\omega$-sequence $r= (q_i,c_0^i, \cdots ,c_{k-1}^i)_{i\in\omega}$ of configurations  is called a 
\textbf{run of $\boldsymbol{\mathcal{A}}$ on $\boldsymbol{\sigma}$ starting in the configuration} $(p,c_0, \cdots ,c_{k-1})$ if
\begin{enumerate}
\item[(1)] $(q_0,c_0^0, \cdots ,c_{k-1}^0)= (p,c_0, \cdots ,c_{k-1})$,

\item[(2)] for each $i\in\omega$, there is $b_i\in\Alphabet\cup\{\emptysequence\}$ such that\medskip
 
\centerline{$b_i : (q_i,c_0^i, \cdots ,c_{k-1}^i)\mapsto_{\mathcal{A}} (q_{i+1},c_0^{i+1}, \cdots ,
c_{k-1}^{i+1})\mbox{,}$}\medskip
 
and either $b_0b_1\cdots= a_0a_1\cdots$, or $b_0b_1\cdots$ is a finite prefix of $a_0a_1\cdots$
\end{enumerate}

 The run $r$ is said to be \textbf{complete} when $a_0a_1\cdots= b_0b_1\cdots$ For every such run, $\mbox{In}(r)$ is the set of all states entered infinitely often during the run $r$. A complete run of $\mathcal{A}$ on $\sigma$ starting in the configuration $(q_0,0, \cdots , 0)$ will  simply be referred to as a \textbf{run of $\boldsymbol{\mathcal{A}}$ on $\sigma$}. The $\omega$-language~accepted by $\mathcal{A}$ is then
\begin{center}
$\Linfinite(\mathcal{A}):=\{\sigma\in\Alphabet^{\omega}\mid\text{ there exists a run }r\text{ of }\mathcal{A}\text{ on }\sigma\text{ such that } \text{In}(r)\cap F\neq\emptyset\}$.
\end{center}
\end{definition}

\begin{remark}\rm
Once again, the condition for accepting finite words is determined here by the final states. Other acceptance conditions have been considered. In particular, in the last sections of this article, we will examine the acceptance of finite words by final states \emph{and} counters with a value of zero.
\end{remark}

Finally, we mention the concept of finite automata, which now that we have  both notions of pushdown automata and counter automata at our disposal, can be seen as a simple restriction of these two types of machines. Namely, a (finite state) automaton is a pushdown automaton deprived of its stack, or equivalently a counter automaton deprived of all its counters \cite{HopcroftMotwaniUllman2001}.
\medskip

Finite automata can be easily represented as labeled graphs. In these, the initial state is represented as the only state with an incoming arrow coming from nowhere; transitions are indicated on the edges; accepting states (also called \emph{final} states) are double-circled.

\vfill\eject

 For instance, in Figure \ref{automatonafour}, $q_4$ stands as the initial state and $q_1$, $q_3$  the accepting states of the automaton $\mathcal{A}_4$. Throughout this article, we will refer to certain rudimentary automata such as the automaton $\mathcal{A}_4$. 
For each of these automata, we will distinguish between three different types of language that they recognize:

\begin{itemize}
  \item the finitary language referred to as $\Lfinite$, which is the set of all finite words for which there exists an accepting run (the acceptance condition being that the run terminates in an accepting state).
    \item The $\omega$-language referred to as $\Linfinite$, which is the set of all infinite words for which there exists an accepting run (the acceptance condition being to visit an accepting state infinitely often).

     \item The language formed by the union of the two previous ones, referred to as $\Lboth$. In other words, for a given finite automaton $\mathcal{A}$, a word belongs to $\Lboth(\mathcal{A})$ if either it is finite and there exists some corresponding run ending on a final state, or it is infinite and there exists some infinite run which visits  a final state  infinitely often.
\end{itemize} 

\begin{example}
For instance, let us consider the automaton $\mathcal{A}_4$ in Figure  \ref{automatonafour}, we have:
\begin{itemize}
  \item  $\begin{array}[t]{lll}
     \Lfinite({\mathcal{A}_4}) & = &   \mathbbold{0}^*\mathbbold{1}\mathbbold{0}^{*}\cup 
     \mathbbold{0}^*\mathbbold{1}\mathbbold{0}^{*}\mathbbold{1}\mathbbold{0}^{*}\mathbbold{1}\mathbbold{0}^{*}\\
   & = &  \setbig{x\in\set{\mathbbold{0},\mathbbold{1}}^{<\omega}\mid x \textit{ contains either one or three }\mathbbold{1}\textit{'s}}
\end{array}$
    \item  $\begin{array}[t]{lll}
     \Linfinite(\mathcal{A}_4)& = &  \mathbbold{0}^*\mathbbold{1}\mathbbold{0}^{\omega}\cup \mathbbold{0}^*\mathbbold{1}\mathbbold{0}^*\mathbbold{1}\mathbbold{0}^*\mathbbold{1}\mathbbold{0}^{\omega}\\
   & = &  \setbig{x\in\set{\mathbbold{0},\mathbbold{1}}^{\omega}\mid x \textit{ contains either one or three }\mathbbold{1}\textit{'s}}
\end{array}$
     \item $\begin{array}[t]{lll}
     \Lboth({\mathcal{A}_4}) & = &   \mathbbold{0}^*\mathbbold{1}\mathbbold{0}^{*}\cup 
     \mathbbold{0}^*\mathbbold{1}\mathbbold{0}^{*}\mathbbold{1}\mathbbold{0}^{*}\mathbbold{1}\mathbbold{0}^{*}\cup  \mathbbold{0}^*\mathbbold{1}\mathbbold{0}^{\omega}\cup \mathbbold{0}^*\mathbbold{1}\mathbbold{0}^*\mathbbold{1}\mathbbold{0}^*\mathbbold{1}\mathbbold{0}^{\omega}\\
   & = &  \setbig{x\in\set{\mathbbold{0},\mathbbold{1}}^{\leq\omega}\mid x \textit{ contains either one or three }\mathbbold{1}\textit{'s}}.
\end{array}$
\end{itemize}

\begin{figure}[h]
\begin{center}
\tikzset{
->,  % makes the edges directed
>=stealth', % makes the arrow heads bold
node distance=22mm, % specifies the minimum distance between two nodes. Change if necessary.
every state/.style={thick, fill=gray!10}, % sets the properties for each ?state? node
initial text=$ $, % sets the text that appears on the start arrow
}
\scalebox{0.64}{\begin{tikzpicture}%[shorten >=1pt,node distance=3cm,on grid]
\node[state, initial] (q4) {$q_4$};
\node[left of=q4] (def) {$\mathcal{A}_4:=$};
\node[state, accepting, right of=q4] (q3) {$q_3$};
\node[state, right of=q3] (q2) {$q_2$};
\node[state, accepting, right of=q2] (q1) {$q_1$};
\node[state, right of=q1] (q0) {$q_0$};
\draw  
(q4) edge[loop above] node{$\mathbbold{0}$} (q4)
(q3) edge[loop above] node{$\mathbbold{0}$} (q3)
(q2) edge[loop above] node{$\mathbbold{0}$} (q2)
(q1) edge[loop above] node{$\mathbbold{0}$} (q1)
(q0) edge[loop above] node{$\mathbbold{0},\mathbbold{1}$} (q0)
(q4) edge[above] node{$\mathbbold{1}$} (q3)
(q3) edge[above] node{$\mathbbold{1}$} (q2)
(q2) edge[above] node{$\mathbbold{1}$} (q1)
(q1) edge[above] node{$\mathbbold{1}$} (q0)
;
\end{tikzpicture}}
\caption{\small{The automaton $\mathcal{A}_4$ \label{automatonafour}}}
\end{center}
\end{figure}
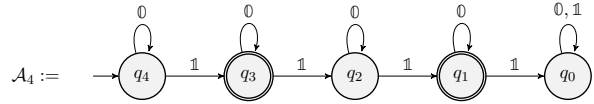
\end{example}

\begin{example}
Let us generalize what we did with automaton $\mathcal{A}_{4}$, to obtain automaton 
$\mathcal{A}_{n}$ as shown in Figure \ref{therealautomatonAnplusonewithnoblanks}. These automata $\mathcal{A}_{n}$ all have trivial loops labeled $\mathbbold{0}$, and arrows labeled $\mathbbold{1}$ that allow you to descend through the sequence of states until you reach a dead end with $q_{0}$. It should also be noted that $q_{n}$ is the initial state, and accepting states are those of the form $q_k$ for odd $k$, so that the right most state ($q_0$) is always rejecting regardless of the parity of $n$.

\begin{figure}[h]
\begin{center}
\tikzset{
%->,  % makes the edges directed
>=stealth', % makes the arrow heads bold
node distance=16mm, % specifies the minimum distance between two nodes. Change if necessary.
every state/.style={thick, fill=gray!10}, % sets the properties for each ?state? node
initial text=$ $, % sets the text that appears on the start arrow
}
\scalebox{0.64}{\begin{tikzpicture}%[shorten >=1pt,node distance=3cm,on grid]
\node[state, initial] (q4) {$q_{2l}$};
\node[left of=q4] (def) {$\mathcal{A}_{2l}:=$};
\node[state, accepting, right of=q4] (q3) {${\scriptstyle q_{2l-1}}$};
\node[rectangle, right of=q3, ] (q3bis) {};
\node[rectangle, right of=q3bis, ] (q3ter) {};
\node[state, right of=q3ter] (q2) {$q_2$};
\node[state, accepting, right of=q2] (q1) {$q_1$};
\node[state, right of=q1] (q0) {$q_0$};
\draw  
(q4) edge[loop above] node{$\mathbbold{0}$} (q4)
(q3) edge[loop above] node{$\mathbbold{0}$} (q3)
(q2) edge[loop above] node{$\mathbbold{0}$} (q2)
(q1) edge[loop above] node{$\mathbbold{0}$} (q1)
(q0) edge[loop above] node{$\mathbbold{0},\mathbbold{1}$} (q0)
(q4) edge[->,above] node{$\mathbbold{1}$} (q3)
(q3) edge[->,above] node{$\mathbbold{1}$} (q3bis)
(q3ter) edge[->,above] node{$\mathbbold{1}$} (q2)
(q2) edge[->,above] node{$\mathbbold{1}$} (q1)
(q1) edge[->,above] node{$\mathbbold{1}$} (q0)
;
\draw[dotted]  (q3bis) -- (q3ter);
\end{tikzpicture}}
\tikzset{
%->,  % makes the edges directed
>=stealth', % makes the arrow heads bold
node distance=16mm, % specifies the minimum distance between two nodes. Change if necessary.
every state/.style={thick, fill=gray!10}, % sets the properties for each ?state? node
initial text=$ $, % sets the text that appears on the start arrow
}
\scalebox{0.64}{\begin{tikzpicture}%[shorten >=1pt,node distance=3cm,on grid]
\node[state, initial,accepting] (q4) {${\scriptstyle q_{2l+1}}$};
\node[left of=q4] (def) {$\mathcal{A}_{2l+1}:=~~~~~$};
\node[state, right of=q4] (q3) {$q_{2l}$};
\node[rectangle, right of=q3, ] (q3bis) {};
\node[rectangle, right of=q3bis, ] (q3ter) {};
\node[state, right of=q3ter] (q2) {$q_2$};
\node[state, accepting, right of=q2] (q1) {$q_1$};
\node[state, right of=q1] (q0) {$q_0$};
\draw  
(q4) edge[loop above] node{$\mathbbold{0}$} (q4)
(q3) edge[loop above] node{$\mathbbold{0}$} (q3)
(q2) edge[loop above] node{$\mathbbold{0}$} (q2)
(q1) edge[loop above] node{$\mathbbold{0}$} (q1)
(q0) edge[loop above] node{$\mathbbold{0},\mathbbold{1}$} (q0)
(q4) edge[->,above] node{$\mathbbold{1}$} (q3)
(q3) edge[->,above] node{$\mathbbold{1}$} (q3bis)
(q3ter) edge[->,above] node{$\mathbbold{1}$} (q2)
(q2) edge[->,above] node{$\mathbbold{1}$} (q1)
(q1) edge[->,above] node{$\mathbbold{1}$} (q0)
;
\draw[dotted]  (q3bis) -- (q3ter);
\end{tikzpicture}}
\end{center}
\caption{\small{The automaton $\mathcal{A}_{n}$\label{therealautomatonAnplusonewithnoblanks}, depending on whether $n$ is even ($n=2l$) or odd ($n=2l+1$)}}
\end{figure}
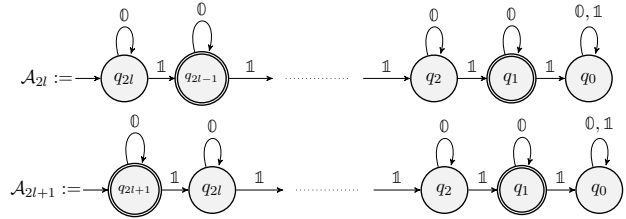

\vfill\eject

 If we set  $\#_\mathbbold{1}(x)$ for how many occurrences of the letter 
$\quq{\mathbbold{1}}$ the either finite or infinite word $x$ contains, then we have:
\begin{itemize}
  \item  $\begin{array}[t]{lll}
     \Lfinite({\mathcal{A}_{n}}) & = &  \setbig{x\in \set{\mathbbold{0},\mathbbold{1}}^{<\omega}\mid \#_\mathbbold{1}(x)< n \textit{ and }\big(n-\#_\mathbbold{1}(x)\big)\textit{ is odd}}\\
\end{array}$
    \item  $\begin{array}[t]{lll}
     \Linfinite({\mathcal{A}_{n}}) & = &  \setbig{x\in \set{\mathbbold{0},\mathbbold{1}}^{\omega}\mid \#_\mathbbold{1}(x)< n \textit{ and }\big(n-\#_\mathbbold{1}(x)\big)\textit{ is odd}}\\
\end{array}$
     \item $\begin{array}[t]{lll}
     \Lboth({\mathcal{A}_{n}}) & = &  \setbig{x\in \set{\mathbbold{0},\mathbbold{1}}^{\leq\omega}\mid \#_\mathbbold{1}(x)< n \textit{ and }\big(n-\#_\mathbbold{1}(x)\big)\textit{ is odd}}.\\
\end{array}$
\end{itemize} 
\end{example}

%
%\noindent\bf Remark.\rm\ The acceptance condition for finite words here is by final states.  Some other acceptance conditions have been considered. In particular, in the last sections of the paper, we will consider acceptance of finite words by final states \emph{and} counters having the value zero.
%

%\begin{definition} A Muller $k$-counter automaton  is a 5-tuple 
%$\mathcal{M}$=$(K,\Alphabet, \Delta, q_0, \mathcal{F})$, 
% where $ \mathcal{M}'$=$(K,\Alphabet, \Delta, q_0)$
%is a $k$-counter machine and $\mathcal{F}$$ \subseteq 2^K$ 
%is the set of accepting sets of  states.
%The \ol~ accepted by $\mathcal{M}$ is 
%\begin{center}
%$L(\mathcal{M})$=$ \{  \sigma\in\Alphabet^{\omega} \mid \mbox{  there exists a  run r
% of } \mathcal{M}$$ \mbox{ on } \sigma \mbox{  such that } 
%\exists F \in \mathcal{F}$$ ~~ \mathrm{In}(r)=F  \}$
%\end{center}
%\end{definition}

\section{Topology}\label{section-topology}\indent

We review the fundamental concepts of topology, assuming that the reader is familiar with the basic notions that can be found in   \cite{moschovakis2009descriptive,kechris1995classical,Staiger97,perrin2004infinite}. The topological spaces we will be working with in this article will be subspaces of $\Alphabet^{\omega}$, where $\Alphabet$  is either a finite set with cardinality at least two (for example, set $2:=\{ \mathbbold{0},\mathbbold{1}\}$), or countably infinite (for example,   $\N$).

\medskip

  The topology on $\Alphabet^{\omega}$ is the product topology of the discrete topology on $\Alphabet$. For $w\in\Alphabet^{<\omega}$, the set $N_w:=\{\alpha\in\Alphabet^{\omega}\mid w\subseteq\alpha\}$ is a basic clopen (i.e., closed and open) set of $\Alphabet^{\omega}$. The open subsets of $\Alphabet^{\omega}$, being   unions of such basic clopen sets, are of the form\medskip
  
\centerline{$W\Alphabet^{\omega}:=\{ w\sigma\mid w\in W\mbox{ and }\sigma\in\Alphabet^{\omega}\}\mbox{,}$}\medskip

\noindent where $W\subseteq\Alphabet^{<\omega}$. When $\Alphabet$ is finite, this topology makes 
$\Alphabet^{\omega}$ a compact space and  is called the \textbf{Cantor topology}. When 
$\Alphabet=\N$, $\Alphabet^{\omega}$ is known as the Baire space, which is homeomorphic to 
$\mathbb{P}_\infty:=
 \{\alpha\in 2^{\omega}\mid\forall i\in\omega\ \exists j\geq i\ \ \alpha (j)= \mathbbold{1}\}$, via the mapping defined on $\N^{\omega}$ by 
$h(\beta ):= \mathbbold{0}^{\beta (0)}\mathbbold{1}\mathbbold{0}^{\beta (1)}\mathbbold{1}\mathbbold{0}^{\beta (2)}\cdots$ The \textbf{prefix metric} is a natural metric on 
$\Alphabet^{\omega}$ that induces this same topology. It is defined as follows: for 
$\sigma\neq\tau\in\Alphabet^{\omega}$, $d(\sigma ,\tau ):= 2^{-l_{\text{pref}(\sigma ,\tau )}}$, where $l_{\text{pref}(\sigma ,\tau )}$ is the least natural number $n$ such that $\sigma (n)\neq\tau (n)$. 
\medskip

There is an elegant characterization of the closed subsets of $\Alphabet^{\omega}$ equipped with the prefix topology: A set $F\subseteq \Alphabet^{\omega}$ is closed if and only if there exists  some tree $T$ on $\Alphabet$ such that $F=[T]$, where a tree on $\Alphabet$ is a subset of 
$\Alphabet^{<\omega}$ closed under prefixes, and $[T]$ stands for the \quq{body} of the 
tree; in other words, the set of all its infinite branches. 

\medskip

We will make use of non-empty closed subspaces of spaces of the form $\Alphabet^{\omega}$, for some countable alphabet $\Alphabet$. Any such non-empty closed subspace  
can easily be thought of as the body ($[T]$)  of some non-empty pruned tree $T$ (a tree which has a root but no leaf), equipped with the prefix topology whose basic open sets are of the form $\set{x\in [T]\mid s\subseteq x}$. These non-empty pruned trees are natural arenas on which to play infinite games, since a player never gets trapped in a dead end, for there is no leaf.

\medskip

 We now define the Borel hierarchy.

\begin{definition}
Let $X$ be any topological space, and $k\geq 1$ be any natural number. The classes 
$\mathbf{\Sigma}_k^0(X)$ and $\mathbf{\Pi}_k^0(X) $ of the \textbf{Borel hierarchy} are inductively defined as follows:

\begin{itemize}
  \item $\mathbf{\Sigma}^{0}_{1}(X) $ is the class of open subsets of $X$

\item $\mathbf{\Sigma}^{0}_{1}(X) $ is the class of closed subsets of $X$

\item $\mathbf{\Sigma}^0_{k+1}(X)$ is the class of countable unions of $\mathbf{\Pi}^{0}_{k}$-subsets of  $X$

\item $\mathbf{\Pi}^0_{k+1}(X)$ is the class of countable intersections of $\mathbf{\Sigma}^{0}_{k}$-subsets of $X$.
\end{itemize}
The Borel hierarchy is also defined for the transfinite levels. For any countable ordinal  $\xi> 1$,
\begin{itemize}
  \item $\mathbf{\Sigma}^{0}_{\xi} (X)$ is the class of countable unions of subsets of $X$ in $\bigcup_{\gamma <\xi}~\mathbf{\Pi}^0_\gamma$

\item $\mathbf{\Pi}^{0}_{\xi} (X)$ is the class of countable intersections of subsets of $X$ in $\bigcup_{\gamma <\xi}~\mathbf{\Sigma}^0_\gamma$.
\end{itemize}
For any countable ordinal  $\xi\geq 1$, we also define the ambiguous class
\begin{itemize}

\item $\mathbf{\Delta}^{0}_{\xi}(X):=\mathbf{\Sigma}^{0}_{\xi}(X)\cap\mathbf{\Pi}^{0}_{\xi}(X)$.
\end{itemize}\end{definition}
\medskip

Given $Y$ a topological space, and $X\subseteq Y$ equipped with the induced topology,  for any countable ordinal $\xi\geq 1$, we have $\mathbf{\Sigma}^{0}_{\xi} (X)=\{ A\cap X\mid A\in\mathbf{\Sigma}^{0}_{\xi} (Y)\}$, and similarly for $\mathbf{\Pi}^{0}_{\xi}$, see \cite[Section 22.A]{kechris1995classical}.\medskip

We write $\checked{\mathbf{\Gamma}}(X):=\set{X\smallsetminus A\mid A\in\mathbf{\Gamma}}$ for the dual class of ${\mathbf{\Gamma}}(X)$. 
 Notice  that we defined the  classes 
$\mathbf{\Sigma}^{0}_{\xi}(X)$, $\mathbf{\Pi}^{0}_{\xi} (X)$, and $\mathbf{\Delta}^{0}_{\xi}(X)$ by mentioning the topological space $X$. 
However, when the context is clear, we will drop the underlying  space, and will write, in general terms, $\mathbf{\Gamma}$ instead of $\mathbf{\Gamma}(X)$.
 For instance, we say a class is \textbf{self-dual} if it satisfies 
 $\mathbf{\Gamma}=\checked{\mathbf{\Gamma}}$, and \textbf{non-self-dual} otherwise.  All classes  $\mathbf{\Delta}^{0}_{\xi}$ are self-dual since we have $\mathbf{\Pi}^{0}_{\xi}=\checked{\mathbf{\Sigma}^{0}_{\xi}}$ (and therefore also 
$ \mathbf{\Sigma}^{0}_{\xi}=\checked{\mathbf{\Pi}^{0}_{\xi}}$). To the contrary, all classes $\mathbf{\Sigma}^{0}_{\xi}$ and $\mathbf{\Pi}^{0}_{\xi}$ are non-self-dual, see 
\cite[Theorem 22.4]{kechris1995classical}. 
\medskip
%

% A class $\mathbf{\Gamma}(X)$ is a \textbf{topological class} if it closed under inverse image by continuous functions on $X$. %Namely, $\mathbf{\Gamma}(X)=\setbig{f^{-1}[A]\mid A\in \mathbf{\Gamma}(X)\text{ and }f:X\xrightarrow{cont.}X}$.
% % }.
   All Borel classes are topological classes, meaning that they are closed under inverse image by continuous functions.  Moreover, $\mathbf{\Sigma}^{0}_{\xi}$ is closed under finite intersections and countable unions, and consequently   $\mathbf{\Pi}^{0}_{\xi}$ is closed under finite unions and countable intersections.
\medskip

 Finally, the class of \textbf{Borel sets} is\medskip
 
\centerline{$\borel:=\bigcup_{1\leq\xi <\omega_1}\mathbf{\Sigma}^{0}_{\xi}=\bigcup_{1\leq\xi <\omega_1}\mathbf{\Pi}^{0}_{\xi}=\bigcup_{1\leq\xi <\omega_1}\mathbf{\Delta}^{0}_{\xi}\text{,}$}\medskip
 
\noindent where $\omega_1$ is the least uncountable ordinal. In all spaces $\Alphabet^{\omega}$, the \textbf{Borel hierarchy} relies on strict inclusions\footnote{Remember that the alphabet $\Alphabet$ contains at least two letters, otherwise these inclusions become equalities.} as depicted  in Figure \ref{figborelhierarchy}.

\medskip

\begin{figure}[h]
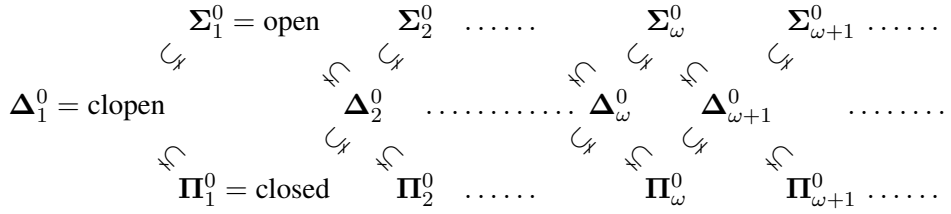


\[
 \arraycolsep=-1.8pt\def\arraystretch{1}
\begin{array}{ccccccccccccccccc }
 & & \mathbf{\Sigma}^0_{1}=\text{open}    &  & &  & \mathbf{\Sigma}^0_{2} &   \ldots\ldots& & &  & \mathbf{\Sigma}^0_{\omega}  &  &  &  & \mathbf{\Sigma}^0_{\omega+1} &   \ldots\ldots\\
 
 & \rotatebox{45}{$\subsetneq$} &   & \rotatebox{-45}{$\subsetneq$}   &  & \rotatebox{45}{$\subsetneq$}   & & &\rotatebox{-45}{$\subsetneq$}  && \rotatebox{45}{$\subsetneq$}&& \rotatebox{-45}{$\subsetneq$}&& \rotatebox{45}{$\subsetneq$}\\
 
  \mathbf{\Delta}^0_{1}=\text{clopen} &   &\     &   &    \mathbf{\Delta}^0_{2}&   &&\ldots \ldots \ldots \ldots&&
 \mathbf{\Delta}^0_{\omega}    &   &   &   &    \mathbf{\Delta}^0_{\omega+1}&   &&\ldots \ldots \ldots \\
 
  & \rotatebox{-45}{$\subsetneq$} &   & \rotatebox{45}{$\subsetneq$} & & \rotatebox{-45}{$\subsetneq$}    && &  \rotatebox{45}{$\subsetneq$} && \rotatebox{-45}{$\subsetneq$}&& \rotatebox{45}{$\subsetneq$}&& \rotatebox{-45}{$\subsetneq$}\\

  &   & \mathbf{\Pi}^0_{1}=\text{closed}  &     & &   & \mathbf{\Pi}^0_{2} &   \ldots\ldots & & & & \mathbf{\Pi}^0_{\omega} &  & &  &  \mathbf{\Pi}^0_{\omega+1} &   \ldots\ldots
\end{array}
\]
\caption{\small{The Borel hierarchy}}
\label{figborelhierarchy}
\end{figure}  

%$$\begin{array}{ll}  
%& \ \ \ \ \ \ \ \ \ \ \ \ \ \ \ \ \ \ \ \ \ \ \ \ \ \boraone=\hbox{\rm open}\ \ \ \ \ \ \ \ \ \ \ \ \ 
%\boratwo \ \ \ \ \ \ \ \  \ \ \ 
%\cdots\ \ \ \ \ \ \ \ \ \ \ \ \boraom\ \ \ \ \ \cdots\cr  
%& \borone=\hbox{\rm clopen}\ \ \ \ \ \ \ \ \ \ \ \ \ \ \ \ \ \ \ \ \ \ \ \ \ \ \ 
%\bortwo\ \ \ \ \ \ \ \ \ \ \ \ \ \ \ \ \ \ \ \ \ \ \ \ \ \ \ \ \ \ \ \borom\ \ \ \ \ \ \ \ \ \ \ \ \ \ \ \ \ \ \ \ \ \ \borel\cr
%& \ \ \ \ \ \ \ \ \ \ \ \ \ \ \ \ \ \ \ \ \ \ \ \ \ \bormone=\hbox{\rm closed}\ \ \ \ \ \ \ \ \ \ \bormtwo \ \ \ \  \ \ \ \ \ \ \ \ \cdots
%\ \ \ \ \ \ \ \ \ \ \ \ \bormom\ \ \ \ \ \cdots
%\end{array}$$

A subset of $\Alphabet^{\omega}$ is a Borel set of \textbf{rank} $\xi$ if it is in $\mathbf{\Sigma}^{0}_{\xi}\cup \mathbf{\Pi}^{0}_{\xi}$ but not in 
$\bigcup_{1\leq\gamma <\xi}~(\mathbf{\Sigma}^0_\gamma\cup \mathbf{\Pi}^0_\gamma)$. We can go further by considering the difference hierarchy on a Cantor space $\Alphabet^{\omega}$, which is the hierarchy of differences of $\mathbf{\Pi}^{0}_{k}$-subsets of $\Alphabet^{\omega}$. We now give the definition for the finite ranks of this hierarchy. 

\begin{definition}
Let $\Alphabet^{\omega}$  be any Cantor space, and $n,k\geq 1$ be natural numbers.  A subset $L$ of 
$\Alphabet^{\omega}$ is in the class $D_{n}(\mathbf{\Pi}^{0}_{k})$ if and only if there is a  sequence 
${G_0}\supseteq {G_1}\supseteq \ldots\supseteq {G_{n-1}}\supseteq {G_{n}}={\emptyset}$  of  $\mathbf{\Pi}^{0}_{k}$-subsets of $\Alphabet^{\omega}$ such that 
\[
L=\bigcup_{j<n,~j\text{ even}}~G_j\smallsetminus G_{j+1}.
\]

The class $\checked{D}_{n}(\mathbf{\Pi}^{0}_{k})$ is the class of complements of sets in $D_{n}(\mathbf{\Pi}^{0}_{k})$. 
\end{definition}

 The difference hierarchy is also defined for transfinite levels, but these will not be needed in this article. 
  In \cite{LouveauSaintRaymond88} Louveau and Saint-Raymond proved the following result 
  which we will use several times for various proofs. 
%  \begin{proposition}[Louveau-Saint Raymond]\label{closureunderpartitionedunions}
%Let $n,k\geq 1$ be natural numbers. The class $D_{n}(\mathbf{\Pi}^{0}_{k})$ is closed under   partitioned  unions with $\mathbf{\Delta}^{0}_{k}$-sets; namely  
% $D_{n}(\mathbf{\Pi}^{0}_{k})=\mathbf{\Delta}^{0}_{k}\mbox{-PU}\big(D_{n}(\mathbf{\Pi}^{0}_{k})\big)$ where 
%\[
%\mathbf{\Delta}^{0}_{k}\mbox{-PU}\big(D_{n}(\mathbf{\Pi}^{0}_{k})\big):=\setbig{\bigcup_{n\in\omega}(P_{n}\cap A_{n})\mid 
%(P_{n})_{n\in\omega}\in (\mathbf{\Delta}^{0}_{k})^{\omega}\mbox{-partition }\wedge 
%(A_{n})_{n\in\omega}\in D_{n}(\mathbf{\Pi}^{0}_{k})^{\omega}}.
%\]
%   \end{proposition}

  \begin{proposition}[Louveau-Saint Raymond]\label{closureunderpartitionedunions}
For any  natural numbers $n,k\geq 1$, the class $D_{n}(\mathbf{\Pi}^{0}_{k})$ is closed under   partitioned  unions with $\mathbf{\Delta}^{0}_{k}$-sets; namely  
$\mathbf{\Gamma} = \mathbf{\Delta}^{0}_{k}\mbox{-PU}(\mathbf{\Gamma})$ where $\mathbf{\Gamma}:= D_{n}(\mathbf{\Pi}^{0}_{k})$ and 
$
\mathbf{\Delta}^{0}_{k}\mbox{-PU}(\mathbf{\Gamma}):=\setBig{\bigcup_{i\in\omega}~(P_{i}\cap A_{i})\mid 
(P_{i})_{i\in\omega}\in (\mathbf{\Delta}^{0}_{k})^{\omega}\mbox{-partition }\wedge 
(A_{i})_{i\in\omega}\in \mathbf{\Gamma}^{\omega}}
$.

   \end{proposition}

The difference hierarchy refines the Borel hierarchy, but one can go even further in refinement by using the notion of reduction by continuous functions.

\begin{definition}\label{definitionofWadgeclass}
Let $X,Y$ be any Polish 0-dimensional topological spaces (equivalently, closed subspaces of the Baire space up to homeomorphism). \begin{itemize} 
 
    \item We say $A\subseteq X$ (continuously) \textbf{reduces} to $B\subseteq Y$, written $A\winf B$, if there exists some continuous mapping $f:X\xrightarrow{cont.}Y$ such that 
    $\forall x\in X~(x\in A \iff f(x)\in B)$, or equivalently $A=f^{-1}[B]$.
    
  \item $B\subseteq X$ is $\mathbf{\Gamma}$-\textbf{hard} if  $A\winf B$ holds for all sets $A\in\mathbf{\Gamma}$.

  \item $B\subseteq X$ is $\mathbf{\Gamma}$-\textbf{complete}  if  $B\in\mathbf{\Gamma}$ and  $B$ is $\mathbf{\Gamma}$-hard.
     
    \item $\mathbf{\Gamma}$ is a \textbf{Wadge class} if it is  a topological class that admits a $\mathbf{\Gamma}$-complete set.

\end{itemize}

\end{definition}

% We now define completeness with respect to reducibility by continuous functions. Let 
%$\mathbf{\Gamma}$ be any class of sets of the form $\mathbf{\Sigma}^{0}_{\xi}$ or $\mathbf{\Pi}^{0}_{\xi}$. A subset $C$ of $\Alphabet^{\omega}$ is said to be $\mathbf{\Gamma}$-\textbf{complete} if $C$ is in $\mathbf{\Gamma}(\Alphabet^{\omega} )$ and, for any finite alphabet $Y$ and any $A\subseteq Y^{\omega}$, $A\in \mathbf{\Gamma}$ if and only if  there exists a continuous function $f: Y^{\omega}\to\Alphabet^{\omega}$ such that $A= f^{-1}(C)$. 

Any non-trivial clopen set is $\mathbf{\Delta}^{0}_{1}$-complete, therefore $\mathbf{\Delta}^{0}_{1}$ is a Wadge class. This result does not extend to higher levels since for $\xi>1$, none of the classes $\mathbf{\Delta}^{0}_{\xi}$  is a Wadge class. To the contrary, for all countable ordinals $\xi\geq 1$, all  classes $\mathbf{\Sigma}^{0}_{\xi}$ and $\mathbf{\Pi}^{0}_{\xi}$ are Wadge classes. 
The $\mathbf{\Sigma}^0_{n}$-complete sets and the $\mathbf{\Pi}^0_{n}$-complete sets were thoroughly characterized in \cite{Staiger86a}. Recall that a subset of $\Alphabet^{\omega}$ is $\mathbf{\Sigma}^{0}_{\xi}$ (respectively $\mathbf{\Pi}^{0}_{\xi}$)-complete if it is in $\mathbf{\Sigma}^{0}_{\xi}$ but not in $\mathbf{\Pi}^{0}_{\xi}$ (respectively in $\mathbf{\Pi}^{0}_{\xi}$ but not in $\mathbf{\Sigma}^{0}_{\xi}$), see \cite[Section 22.B]{kechris1995classical}. For instance, every singleton of $2^{\omega}$ is $\mathbf{\Pi}^{0}_{1}$-complete. The set $\mathbb{P}_\infty$ defined at the beginning of the present section is a well known example of a $\mathbf{\Pi}^{0}_{2}$-complete set. 
\medskip

Beyond the Borel hierarchy, there is another natural one called the projective hierarchy. At the first level of this hierarchy are classes $\mathbf{\Sigma}^{1}_{1}$  of \textbf{analytic} sets  and $\mathbf{\Pi}^{1}_{1}:= \checked{\mathbf{\Sigma}^{1}_{1}}$ of \textbf{co-analytic} sets. A subset $A$ of $\Alphabet^{\omega}$ is \textbf{analytic} if one can find a finite alphabet $Y$ and a Borel subset $B$ of 
$(\Alphabet\times Y)^{\omega}$ such that $x\in A\Leftrightarrow\exists y\in Y^{\omega} ~(x, y)\in B$, where 
$(x, y)\in (\Alphabet\times Y)^{\omega}$ means that $(x, y)(i)=\big( x(i),y(i)\big)$ holds for every natural number $i$. To say it differently, a subset of $\Alphabet^{\omega}$ is analytic if it is  either empty, or the image of the Baire space by a continuous mapping. The class $\mathbf{\Sigma}^{1}_{1}$ of analytic sets contains the class of Borel sets in any of the spaces $\Alphabet^{\omega}$. Note that Borel sets are precisely those that are both analytic and co-analytic:  
$\mathbf{\Delta}_1^1= \mathbf{\Sigma}^{1}_{1}\cap \mathbf{\Pi}^{1}_{1}$.

\medskip
%

%
% There are some subsets of the topological space $\Alphabet^{\omega}$ which are not Borel sets. In particular, there is another hierarchy beyond the Borel hierarchy,  called the projective hierarchy. The first class of the projective hierarchy is the class $\mathbf{\Sigma}^{1}_{1}$ of analytic sets. A subset $A$ of $\Alphabet^{\omega}$ is \textbf{analytic} if we can find a finite alphabet $Y$ and a Borel subset $B$ of 
%$(\Alphabet\times Y)^{\omega}$ such that $x\in A\Leftrightarrow\exists y\in Y^{\omega} ~(x, y)\in B$, where 
%$(x, y)\in (\Alphabet\times Y)^{\omega}$ means that $(x, y)(i)=\big( x(i),y(i)\big)$ for each natural number $i$. A subset of $\Alphabet^{\omega}$ is analytic if it is empty, or the image of the Baire space by a continuous map. The class $\mathbf{\Sigma}^{1}_{1}$ of analytic sets contains the class of Borel sets in any of the spaces $\Alphabet^{\omega}$. Note that 
%$\mathbf{\Delta}_1^1= \mathbf{\Sigma}^{1}_{1}\cap \mathbf{\Pi}^{1}_{1}$, where $\mathbf{\Pi}^{1}_{1}:= \checked{\mathbf{\Sigma}^{1}_{1}}$ is the class of \textbf{co-analytic} sets, i.e., of complements of analytic sets.\medskip
 
 The $\omega$-power of a finitary non-empty language $L$ is always an analytic set. Indeed, if $L$ is finite and has $n$ elements, then $L^{\infty}$  is the continuous image of the compact set 
$\{ 0,1, \cdots ,n - 1\}^{\omega}$. Whereas if $L$ is infinite, then there is a bijection between $L$ and 
$\N$, and $L^{\infty	}$ is the continuous image of the Baire space $\N^{\omega}$ \cite{Simonnet92}.
\medskip

As we can see, the concept of continuous function is central to this article. In the context of the topological spaces we are considering, this concept of continuity is expressed very naturally in the form of a strategy in a well-defined game.

\begin{definition} \label{definitioncontinuousgame} Let $\Alphabet_A,\Alphabet_B$ be non-empty finite alphabets, and $f:\Alphabet_A^{\omega}\to\Alphabet_B^{\omega}$ any function. The game $\Gcont{f}$ is the two-player infinite alternating game represented below. In this game,  Player $\one$ constructs an infinite sequence $x\in \Alphabet_A^{\omega}$, and Player $\two$ answers with a finite or infinite sequence $y\in \Alphabet_B^{\leq\omega}$:
   Player $\one$ starts, and
   Player $\one$'s moves are elements in $\Alphabet_A$, while Player $\two$'s moves are either elements in $\Alphabet_B$ or skips (this means she passes her turn). 
   Player $\two$ wins a play if $f(x)=y$.

%%%%%%%%%%%%%%%%%%%%%%%%%%%%%%%%%%%%%%%%%%%%%%%%%%%%%%%%%%%%%%%%%%%%%%%%%%%%%%%%%%%%%%%%%%%%%%%%%%%%%%%%%%%%%%%%%%%%%%%%%%%%%%%%%%%%%%%%%%%%%%%%%%%%%%%%%%%%%%%%%%%%%%%%%%%%%%%%%%%%%%%%%%%%%%%%%%%%%%%%%%%%%%%%%%%%%%%%%%%%%%%%%%%%%%%%%%%%%%%%%%%%%%%%%%%%%%%%%%%%%%%%%%%%%%%%%%%%%%%%%%%%%%%%%%%%%%%%%%%%%%%%%%%%%%%%%%%%%%%%%%%%%%%%%%%%%%%%%%%%%%%%%%%%%%%%%%%%%%%%%%%%%%%%%%%%%%%%%%%%%%%%%%%%%%%%%%%%%%%%%%%%%%%%%%%%%%%%%%%%%%%%%%%%%%%%%%%%%%%%%%%%%%%%%%%%%%%%%%%%%%%%%%%%%%%%%%%%%%%%%%%%%%%%%%%%%%%%%%%%%%%%%%%%%%%%%%%%%%%%%%%%%%%%%%%%%%%%%%%%%%%%%%%%%%%%%%%%%%%%%%%%%%%%%%%%%%%

  %%%%%%%%%%%%%%%%%%%%%%%%%%%%%%%%%%%%%%%%%%%%%%%%%%%%%%%%%%%%%%%%%%%%%%%%%%%%%%%%%%%%%%%%%%%%%%%%%%%%%%%%%%%%%%%%%%%%%%%%%%%%%%%%%%%%%%%%%%%%%%%%%%%%%%%%%%%%%%%%%%%%%%%%%%%%%%%%%%%%%%%%%%%%%%%%%%%%%%%%%%%%%%%%%%%%%%%%%%%%%%%%%%%%%%%%%%%%%%%%%%%%%%%%%%%%%%%%%%%%%%%%%%%%%%%%%%%%%%%%%%%%%%%%%%%%%%%%%%%%%%%%%%%%%%%%%%%%%%%%%%%%%%%%%%%%%%%%%%%%%%%%%%%%%%%%%%%%%%%%%%%%%%%%%%%%%%%%%%%%%%%%%%%%%%%%%%%%%%%%%%%%%%%%%%%%%%%%%%%%%%%%%%%%%%%%%%%%%%%%%%%%%%%%%%%%%%%%%%%%%%%%%%%%%%%%%%%%%%%%%%%%%%%%%%%%%%%%%%%%%%%%%%%%%%%%%%%%%%%%%%%%%%%%%%%%%%%%%%%%%%%%%%%%%%%%%%%%%%%%%%%%%%%%%%%%%%%

\begin{figure}[h]
\begin{center}
\scalebox{0.85}{\begin{picture}(0,0)%
\includegraphics{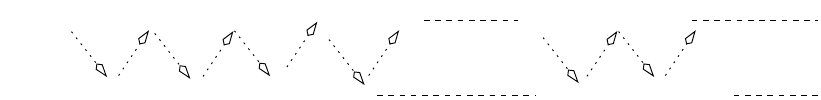}%
\end{picture}}%
\setlength{\unitlength}{0.00083300in}%
\begingroup\makeatletter\ifx\SetFigFont\undefined
% extract first six characters in \fmtname
\def\x#1#2#3#4#5#6#7\relax{\def\x{#1#2#3#4#5#6}}%
\expandafter\x\fmtname xxxxxx\relax \def\y{splain}%
\ifx\x\y   % LaTeX or SliTeX?
\gdef\SetFigFont#1#2#3{%
  \ifnum #1<17\tiny\else \ifnum #1<20\small\else
  \ifnum #1<24\normalsize\else \ifnum #1<29\large\else
  \ifnum #1<34\Large\else \ifnum #1<41\LARGE\else
     \huge\fi\fi\fi\fi\fi\fi
  \csname #3\endcsname}%
\else
\gdef\SetFigFont#1#2#3{\begingroup
  \count@#1\relax \ifnum 25<\count@\count@25\fi
  \def\x{\endgroup\@setsize\SetFigFont{#2pt}}%
  \expandafter\x
    \csname \romannumeral\the\count@ pt\expandafter\endcsname
    \csname @\romannumeral\the\count@ pt\endcsname
  \csname #3\endcsname}%
\fi
\fi\endgroup
\scalebox{.85}{\begin{picture}(6537,781)(2701,-3092)
\put(2701,-2461){\makebox(0,0)[lb]{\smash{\SetFigFont{14}{16.8}{rm}$\one$}}}
\put(2701,-3061){\makebox(0,0)[lb]{\smash{\SetFigFont{14}{16.8}{rm}$\two$}}}
\put(2953,-2461){\makebox(0,0)[lb]{\smash{\SetFigFont{11}{8.4}{rm}:}}}
\put(2953,-3061){\makebox(0,0)[lb]{\smash{\SetFigFont{11}{8.4}{rm}:}}}
\put(3205,-2461){\makebox(0,0)[lb]{\smash{\SetFigFont{11}{8.4}{rm}$x_0$}}}
\put(3877,-2461){\makebox(0,0)[lb]{\smash{\SetFigFont{11}{8.4}{rm}$x_1$}}}
\put(4212,-3061){\makebox(0,0)[lb]{\smash{\SetFigFont{11}{8.4}{rm}$y_1$}}}
\put(6917,-2461){\makebox(0,0)[lb]{\smash{\SetFigFont{11}{8.4}{rm}$x_{n}$}}}
\put(7211,-3061){\makebox(0,0)[lb]{\smash{\SetFigFont{11}{8.4}{rm}$y_{m}$}}}
\put(7547,-2461){\makebox(0,0)[lb]{\smash{\SetFigFont{11}{8.4}{rm}$x_{n+1}$}}}
\put(7883,-3061){\makebox(0,0)[lb]{\smash{\SetFigFont{11}{8.4}{rm}$y_{m+1}$}}}
\put(3541,-3061){\makebox(0,0)[lb]{\smash{\SetFigFont{11}{8.4}{rm}$y_0$}}}
\put(5251,-2461){\makebox(0,0)[lb]{\smash{\SetFigFont{11}{8.4}{rm}$x_3$}}}
\put(5551,-3061){\makebox(0,0)[lb]{\smash{\SetFigFont{11}{8.4}{rm}$y_2$}}}
\put(5851,-2461){\makebox(0,0)[lb]{\smash{\SetFigFont{11}{8.4}{rm}$x_4$}}}
\put(4548,-2461){\makebox(0,0)[lb]{\smash{\SetFigFont{11}{8.4}{rm}$x_2$}}}
\end{picture}}
\end{center}
\caption{\small{The game $\Gcont{f}$}}
\end{figure}
\end{definition}

%%%%%%%%%%%%%%%%%%%%%%%%%%%%%%%%%%%%%%%%%%%%%%%%%%%%%%%%%%%%%%%%%%%%%%%%%%%%%%%%%%%%%%%%%%%%%%%%%%%%%%%%%%%%%%%%%%%%%%%%%%%%%%%%%%%%%%%%%%%%%%%%%%%%%%%%%%%%%%%%%%%%%%%%%%%%%%%%%%%%%%%%%%%%%%%%%%%%%%%%%%%%%%%%%%%%%%%%%%%%%%%%%%%%%%%%%%%%%%%%%%%%%%%%%%%%%%%%%%%%%%%%%%%%%%%%%%%%%%%%%%%%%%%%%%%%%%%%%%%%%%%%%%%%%%%%%%%%%%%%%%%%%%%%%%%%%%%%%%%%%%%%%%%%%%%%%%%%%%%%%%%%%%%%%%%%%%%%%%%%%%%%%%%%%%%%%%%%%%%%%%%%%%%%%%%%%%%%%%%%%%%%%%%%%%%%%%%%%

As shown in the following lemma, the rules of this game are designed to characterize continuous functions.

\begin{lemma}[Wadge] \label{lemgamecontinuity}Let $\Alphabet_A,\Alphabet_B$ be non-empty finite alphabets, and $f:\Alphabet_A^{\omega}\to\Alphabet_B^{\omega}$ any function. Player $\two$ has a winning strategy in the game $\Gcont{f}$ if and only if $f$ is continuous.
\end{lemma}

\prooflem{lemgamecontinuity}  ($\Rightarrow$) Given $\sigma$ any winning strategy  for $\two$, for every finite word $v\in \Alphabet_B^{<\omega}$ we have $f^{-1}[v\Alphabet_B^{\omega}]=\bigcup_{u\in U}u\Alphabet_A^{\omega}$, where $U$ stands for all positions $u$ for Player $\one$ which yield $v$ as Player $\two$'s answer to $u$ according to $\sigma$. ($\Leftarrow$) A winning strategy for Player $\two$ is designed as follows: to determine her next move while being in position $v\in \Alphabet_B^{<\omega}$, Player $\two$ simply considers the partition $\setbig{f^{-1}[vb\Alphabet_B^{\omega}]\mid b\in \Alphabet_B}$  of $f^{-1}[v\Alphabet_B^{\omega}]$ in open sets,  and waits until Player $\one$ reaches a position $u$ such that, for some $b\in \Alphabet_B$,  
$u\Alphabet_A^{\omega}\subseteq f^{-1}[vb\Alphabet_B^{\omega}]$ holds. Then, Player $\two$ plays $b$.\boxlem{lemgamecontinuity}

\medskip

This characterization of continuous functions as strategies in a game allowed Bill Wadge to define the reduction relation by continuous functions in terms of the existence of a winning strategy in a game on which we will rely heavily: the Wadge game.

 %%%%%%%%%%%%%%%%%%%%%%%%%%%%%%%%%%%%%%%%%%%%%%%%%%%%%%%%%%%%%%%%%%%%%%%%%%%%%%%%%%%%%%%%%%%%%%%%%%%%%%%%%%%%%%%%%%%%%%%%%%%%%%%%%%%%%%%%%%%%%%%%%%%%%%%%%%%%%%%%%%%%%%%%%%%%%%%%%%%%%%%%%%%%%%%%%%%%%%%%%%%%%%%%%%%%%%%%%%%%%%%%%%%%%%%%%%%%%%%%%%%%%%%%%%%%%%%%%%%%%%%%%%%%%%%%%%%%%%%%%%%%%%%%%%%%%%%%%%%%%%%%%%%%%%%%%%%%%%%%%%%%%%%%%%%%%%%%%%%%%%%%%%%%%%%%%%%%%%%%%%%%%%%%%%%%%%%%%%%%%%%%%%%%%%%%%%%%%%%%%%%%%%%%%%%%%%%%%%%%%%%%%%%%%%%%%%%%%%%%%%%%%%%%%%%%%%%%%%%%%%%%%%%%%%%%%%%%%%%%%%%%%%%%%%%%%%%%%%%%%%%%%%%%%%%%%%%%%%%%%%%%%%%%%%%%%%%%%%%%%%%%%%%%%%%%%%%%%%%%%%%%%%%%%%%%%%%
  
\begin{definition} \label{definitionWadgegame} Let $\Alphabet_A,\Alphabet_B$ be non-empty finite alphabets, and $A\subseteq\Alphabet_A^{\omega}, B\subseteq\Alphabet_B^{\omega}$. The  Wadge game $\GWadge{A}{B}$ is  identical to the game $\Gcont{f}$  specified in Definition \ref{definitioncontinuousgame}, except for the winning condition which is: Player $\two$ wins a play if and only if $(x\in A \iff y\in B)$.

%%%%%%%%%%%%%%%%%%%%%%%%%%%%%%%%%%%%%%%%%%%%%%%%%%%%%%%%%%%%%%%%%%%%%%%%%%%%%%%%%%%%%%%%%%%%%%%%%%%%%%%%%%%%%%%%%%%%%%%%%%%%%%%%%%%%%%%%%%%%%%%%%%%%%%%%%%%%%%%%%%%%%%%%%%%%%%%%%%%%%%%%%%%%%%%%%%%%%%%%%%%%%%%%%%%%%%%%%%%%%%%%%%%%%%%%%%%%%%%%%%%%%%%%%%%%%%%%%%%%%%%%%%%%%%%%%%%%%%%%%%%%%%%%%%%%%%%%%%%%%%%%%%%%%%%%%%%%%%%%%%%%%%%%%%%%%%%%%%%%%%%%%%%%%%%%%%%%%%%%%%%%%%%%%%%%%%%%%%%%%%%%%%%%%%%%%%%%%%%%%%%%%%%%%%%%%%%%%%%%%%%%%%%%%%%%%%%%%%%%%%%%%%%%%%%%%%%%%%%%%%%%%%%%%%%%%%%%%%%%%%%%%%%%%%%%%%%%%%%%%%%%%%%%%%%%%%%%%%%%%%%%%%%%%%%%%%%%%%%%%%%%%%%%%%%%%%%%%%%%%%%%%%%%%%%%%%%
   
\begin{figure}[h]
\begin{center}
\scalebox{0.85}{\begin{picture}(0,0)%
\includegraphics{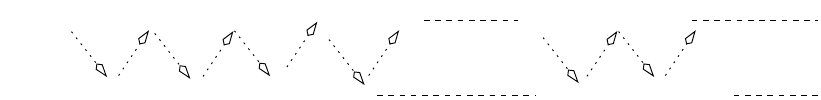}%
\end{picture}}%
\setlength{\unitlength}{0.00083300in}%
\begingroup\makeatletter\ifx\SetFigFont\undefined
% extract first six characters in \fmtname
\def\x#1#2#3#4#5#6#7\relax{\def\x{#1#2#3#4#5#6}}%
\expandafter\x\fmtname xxxxxx\relax \def\y{splain}%
\ifx\x\y   % LaTeX or SliTeX?
\gdef\SetFigFont#1#2#3{%
  \ifnum #1<17\tiny\else \ifnum #1<20\small\else
  \ifnum #1<24\normalsize\else \ifnum #1<29\large\else
  \ifnum #1<34\Large\else \ifnum #1<41\LARGE\else
     \huge\fi\fi\fi\fi\fi\fi
  \csname #3\endcsname}%
\else
\gdef\SetFigFont#1#2#3{\begingroup
  \count@#1\relax \ifnum 25<\count@\count@25\fi
  \def\x{\endgroup\@setsize\SetFigFont{#2pt}}%
  \expandafter\x
    \csname \romannumeral\the\count@ pt\expandafter\endcsname
    \csname @\romannumeral\the\count@ pt\endcsname
  \csname #3\endcsname}%
\fi
\fi\endgroup
\scalebox{.85}{\begin{picture}(6537,781)(2701,-3092)
\put(2701,-2461){\makebox(0,0)[lb]{\smash{\SetFigFont{14}{16.8}{rm}\textit{I}}}}
\put(2701,-3061){\makebox(0,0)[lb]{\smash{\SetFigFont{14}{16.8}{rm}$\two$}}}
\put(2953,-2461){\makebox(0,0)[lb]{\smash{\SetFigFont{11}{8.4}{rm}:}}}
\put(2953,-3061){\makebox(0,0)[lb]{\smash{\SetFigFont{11}{8.4}{rm}:}}}
\put(3205,-2461){\makebox(0,0)[lb]{\smash{\SetFigFont{11}{8.4}{rm}$x_0$}}}
\put(3877,-2461){\makebox(0,0)[lb]{\smash{\SetFigFont{11}{8.4}{rm}$x_1$}}}
\put(4212,-3061){\makebox(0,0)[lb]{\smash{\SetFigFont{11}{8.4}{rm}$y_1$}}}
\put(6917,-2461){\makebox(0,0)[lb]{\smash{\SetFigFont{11}{8.4}{rm}$x_{n}$}}}
\put(7211,-3061){\makebox(0,0)[lb]{\smash{\SetFigFont{11}{8.4}{rm}$y_{m}$}}}
\put(7547,-2461){\makebox(0,0)[lb]{\smash{\SetFigFont{11}{8.4}{rm}$x_{n+1}$}}}
\put(7883,-3061){\makebox(0,0)[lb]{\smash{\SetFigFont{11}{8.4}{rm}$y_{m+1}$}}}
\put(3541,-3061){\makebox(0,0)[lb]{\smash{\SetFigFont{11}{8.4}{rm}$y_0$}}}
\put(5251,-2461){\makebox(0,0)[lb]{\smash{\SetFigFont{11}{8.4}{rm}$x_3$}}}
\put(5551,-3061){\makebox(0,0)[lb]{\smash{\SetFigFont{11}{8.4}{rm}$y_2$}}}
\put(5851,-2461){\makebox(0,0)[lb]{\smash{\SetFigFont{11}{8.4}{rm}$x_4$}}}
\put(4548,-2461){\makebox(0,0)[lb]{\smash{\SetFigFont{11}{8.4}{rm}$x_2$}}}
\end{picture}}
\end{center}
\caption{\small{The Wadge game $\GWadge{A}{B}$}}
\end{figure}
\end{definition}

 This game was designed by Wadge so that Lemma  \ref{lemgamecontinuity} yields the equivalence between Player $\two$ having a winning strategy in $\GWadge{A}{B}$  and  the existence of a continuous mapping $f:\Alphabet_A^{\omega}\to\Alphabet_B^{\omega}$ such that $A=f^{-1}[B]$. In other words, Player $\two$ has a winning strategy in $\GWadge{A}{B}$ if and only if $A\wequ B$. As usual, we write $A\wequ B$ if both $A\winf B$ and $B\winf A$  hold.

\medskip

 In addition to continuous functions, we will also use Baire class 1 functions. 
A function $f:X\to Y$ is Baire class 1 if $f^{-1}[\mathcal{O}]\in \mathbf{\Sigma}^{0}_{2}$ holds for 
 every $\mathcal{O}\in \mathbf{\Sigma}^{0}_{1}$. i.e., the inverse image of every open subset of $Y$ is a countable union of closed sets of $X$. It is well-known that when $X$ and $Y$ are both separable and metrizable, and $X$ is zero-dimensional\footnote{All topological spaces in this article are separable, metrizable and zero-dimensional.}, a function $f:X\to Y$ is Baire class 1  if and only if it is a pointwise limit of continuous functions (see  \cite[Theorem 24.10]{kechris1995classical}). i.e.,   there exists a family $(f_{n})_{n<\omega}$ of continuous functions such that for  all $x\in X$, ${\displaystyle f(x)=\lim_{n<\omega}f_{n}(x)}$ holds. 
\medskip

 This representation of Baire class 1 functions as limits of continuous functions led to the characterization of  the game $\GBaire{f}$, which was introduced in \cite{Duparc01,Semmes09}.
 
\vfill\eject 
 
 It is similar to the game  $\Gcont{f}$
 except that in addition of being able to skip, Player $\two$ can also use an eraser, which is to say 
 that Player $\two$ may come back to any of her previous positions by simply deleting her last moves --- as with one's computer's backspace key ! 
Formally, this comes down to allowing  the second player in the  game $\Gcont{f}$,  to play an extra letter \quq{$\tlaBaire$}
   so that after  $\omega$-many moves  Players $\one$ and $\two$  have respectively produced 
 $x\in\Alphabet_A^{\omega}$ and  $y\in(\Alphabet_{B}\cup\{\tlaBaire\})^{\leq\omega}$. 
  From $y$, the deleting process  yields the sequence $y^{\tlaBaire}$ as defined below.

\begin{definition}\label{definitionoferasingprocess}
Let  $\Alphabet$ be any finite alphabet, $\tlaBaire\notin\Alphabet$ be any letter. Given any  $x\in(\Alphabet\cup\set{\tlaBaire})^{\leq\omega}$, we define $x^{\tlaBaire}$ inductively  as follows.
\begin{itemize}
  \item  If $x=\emptysequence$, then $x^{\tlaBaire}:=\emptysequence$.

     \item  If $x=ua$  for  some $u\in(\Alphabet\cup\set{\tlaBaire})^{<\omega}$ and $a\in\Alphabet\cup\set{\tlaBaire}$, then

$\left\{
\begin{array}{lrll}
\text{ if }a=\tlaBaire,\text{ and }|u^{\tlaBaire}|= 0, &\text{ then } &x^{\tlaBaire}:=&{\emptysequence}\\
\text{ if }a=\tlaBaire,\text{ and }|u^{\tlaBaire}|> 0, &\text{ then } &x^{\tlaBaire}:=&u^{\tlaBaire}\upharpoonright_{(|u^{\tlaBaire}|-1)}\\
\text{ if }a\neq\tlaBaire, &\text{ then }&x^{\tlaBaire}:=& u^{\tlaBaire} a.\\
\end{array}
\right.$

  \item   If $x\in(\Alphabet\cup\set{\tlaBaire})^{\omega}$, then $x^{\tlaBaire}:=\lim_{n\in\omega}~(x\upharpoonright n)^{\tlaBaire}$, where given any word $w\in\Alphabet^{<\omega}$, we have 
  \[
w\subseteq\mbox{lim}_{n\in\omega}~(x\upharpoonright n)^{\tlaBaire}\Longleftrightarrow\exists p\in\omega\,
\forall n> p\,~~w\subseteq(x\upharpoonright n)^{\tlaBaire}.
\]

\end{itemize}

\end{definition}

%
%
%
%
%
%  
%  
%\begin{definition}\label{definitionoferasingprocess}
%Let  $\Alphabet_A$ be any finite alphabet, $\tlaBaire\notin\Alphabet_A$ be any letter, $\Alphabet:=\Alphabet_A\cup\{\tlaBaire\}$. Given any  $x\in\Alphabet^{\leq\omega}$, we define $x^{\tlaBaire}$ inductively  as follows.
%\begin{itemize}
%  \item  If $x=\emptysequence$, then $x^{\tlaBaire}:=\emptysequence$.
%
%     \item  If $x=ua$  for  some $u\in\Alphabet^{<\omega}$ and $a\in\Alphabet$, then 
%
%
%$\left\{
%\begin{array}{lrll}
%\text{ if }a=\tlaBaire,\text{ and }|u^{\tlaBaire}|= 0, &\text{ then } &x^{\tlaBaire}:=&{\emptysequence}\\
%\text{ if }a=\tlaBaire,\text{ and }|u^{\tlaBaire}|> 0, &\text{ then } &x^{\tlaBaire}:=&u^{\tlaBaire}\upharpoonright_{(|u^{\tlaBaire}|-1)}\\
%\text{ if }a\neq\tlaBaire, &\text{ then }&x^{\tlaBaire}:=& u^{\tlaBaire} a.\\
%\end{array}
%\right.$
%
%
%
%
%
%
%
%  \item   If $x\in\Alphabet^{\omega}$, then $x^{\tlaBaire}:=\lim_{n\in\omega}~(x\upharpoonright n)^{\tlaBaire}$, where given any word $w\in\Alphabet_A^{<\omega}$, we have 
%  \[
%w\subseteq\mbox{lim}_{n\in\omega}~(x\upharpoonright n)^{\tlaBaire}\Longleftrightarrow\exists p\in\omega\,
%\forall n> p\,~~w\subseteq(x\upharpoonright n)^{\tlaBaire}.
%\]
%
%\end{itemize}
%
%\end{definition}

\begin{remark}\rm\ For $x\in(\Alphabet\cup\set{\tlaBaire})^{\leq\omega}$, $x^{\tlaBaire}$ denotes the string $x$, once every \quq{$\tlaBaire$} occuring in $x$ has been \quq{evaluated} as the \textit{back space operation} (the one familiar to your computer!), proceeding from left to right inside $x$. In other words, $x^{\tlaBaire}= x$ from which every interval of the form \quq{$a\tlaBaire$} ($a\neq\tlaBaire$) is removed. For instance, no matter whether 
 $x= (a\tlaBaire)^n$ for some $n> 1$, or  $x= (a\tlaBaire)^{\omega}$ or 
$x= (a\tlaBaire\tlaBaire)^{\omega}$, then in all cases we will have $x^{\tlaBaire}=\emptysequence$. If either  $x= (ab\tlaBaire)^{\omega}$ or $x= (\tlaBaire ab)^{\omega}$, we will end up with 
 $x^{\tlaBaire}= a^{\omega}$. Lastly, if  $x= bb(\tlaBaire a)^{\omega}$, then $x^{\tlaBaire}= b$. \end{remark}

\begin{definition} \label{definitionbaireclassonegame} Let $\Alphabet_A,\Alphabet_B$ be non-empty finite alphabets, $\tlaBaire\notin \Alphabet_B$, and $f:\Alphabet_A^{\omega}\to\Alphabet_B^{\omega}$ any function. The game $\GBaire{f}$ is the two-player infinite alternating game represented below, where Player $\one$ constructs an infinite sequence $x\in \Alphabet_A^{\omega}$, and Player $\two$ answers with a finite or infinite sequence $y\in (\Alphabet_B\cup\set{\tlaBaire} )^{\leq\omega}$:
   Player $\one$ starts,
   Player $\one$'s moves are elements in $\Alphabet_A$, while Player $\two$'s moves are either elements in $\Alphabet_B$, skips, or the eraser letter $\tlaBaire$. 
   Player $\two$ wins a play if $f(x)=y^{\tlaBaire}$.

%\begin{itemize}
%  \item Player $\one$ starts,
%  \item Player $\one$'s moves are elements in $\Alphabet_A$, and Player $\two$'s moves are either elements in $\Alphabet_B$ or skips (this means she passes her turn), 
%  \item Player $\two$ wins a play if $f(x)=y$.
%\end{itemize}

%%%%%%%%%%%%%%%%%%%%%%%%%%%%%%%%%%%%%%%%%%%%%%%%%%%%%%%%%%%%%%%%%%%%%%%%%%%%%%%%%%%%%%%%%%%%%%%%%%%%%%%%%%%%%%%%%%%%%%%%%%%%%%%%%%%%%%%%%%%%%%%%%%%%%%%%%%%%%%%%%%%%%%%%%%%%%%%%%%%%%%%%%%%%%%%%%%%%%%%%%%%%%%%%%%%%%%%%%%%%%%%%%%%%%%%%%%%%%%%%%%%%%%%%%%%%%%%%%%%%%%%%%%%%%%%%%%%%%%%%%%%%%%%%%%%%%%%%%%%%%%%%%%%%%%%%%%%%%%%%%%%%%%%%%%%%%%%%%%%%%%%%%%%%%%%%%%%%%%%%%%%%%%%%%%%%%%%%%%%%%%%%%%%%%%%%%%%%%%%%%%%%%%%%%%%%%%%%%%%%%%%%%%%%%%%%%%%%%%%%%%%%%%%%%%%%%%%%%%%%%%%%%%%%%%%%%%%%%%%%%%%%%%%%%%%%%%%%%%%%%%%%%%%%%%%%%%%%%%%%%%%%%%%%%%%%%%%%%%%%%%%%%%%%%%%%%%%%%%%%%%%%%%%%%%%%%%%

  %%%%%%%%%%%%%%%%%%%%%%%%%%%%%%%%%%%%%%%%%%%%%%%%%%%%%%%%%%%%%%%%%%%%%%%%%%%%%%%%%%%%%%%%%%%%%%%%%%%%%%%%%%%%%%%%%%%%%%%%%%%%%%%%%%%%%%%%%%%%%%%%%%%%%%%%%%%%%%%%%%%%%%%%%%%%%%%%%%%%%%%%%%%%%%%%%%%%%%%%%%%%%%%%%%%%%%%%%%%%%%%%%%%%%%%%%%%%%%%%%%%%%%%%%%%%%%%%%%%%%%%%%%%%%%%%%%%%%%%%%%%%%%%%%%%%%%%%%%%%%%%%%%%%%%%%%%%%%%%%%%%%%%%%%%%%%%%%%%%%%%%%%%%%%%%%%%%%%%%%%%%%%%%%%%%%%%%%%%%%%%%%%%%%%%%%%%%%%%%%%%%%%%%%%%%%%%%%%%%%%%%%%%%%%%%%%%%%%%%%%%%%%%%%%%%%%%%%%%%%%%%%%%%%%%%%%%%%%%%%%%%%%%%%%%%%%%%%%%%%%%%%%%%%%%%%%%%%%%%%%%%%%%%%%%%%%%%%%%%%%%%%%%%%%%%%%%%%%%%%%%%%%%%%%%%%%%%
     
\begin{figure}[h]
\begin{center}
\scalebox{0.85}{\begin{picture}(0,0)%
\includegraphics{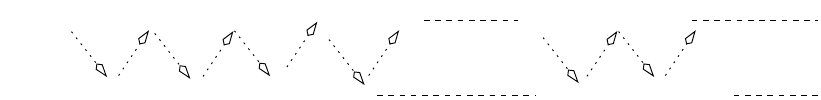}%
\end{picture}}%
\setlength{\unitlength}{0.00083300in}%
\begingroup\makeatletter\ifx\SetFigFont\undefined
% extract first six characters in \fmtname
\def\x#1#2#3#4#5#6#7\relax{\def\x{#1#2#3#4#5#6}}%
\expandafter\x\fmtname xxxxxx\relax \def\y{splain}%
\ifx\x\y   % LaTeX or SliTeX?
\gdef\SetFigFont#1#2#3{%
  \ifnum #1<17\tiny\else \ifnum #1<20\small\else
  \ifnum #1<24\normalsize\else \ifnum #1<29\large\else
  \ifnum #1<34\Large\else \ifnum #1<41\LARGE\else
     \huge\fi\fi\fi\fi\fi\fi
  \csname #3\endcsname}%
\else
\gdef\SetFigFont#1#2#3{\begingroup
  \count@#1\relax \ifnum 25<\count@\count@25\fi
  \def\x{\endgroup\@setsize\SetFigFont{#2pt}}%
  \expandafter\x
    \csname \romannumeral\the\count@ pt\expandafter\endcsname
    \csname @\romannumeral\the\count@ pt\endcsname
  \csname #3\endcsname}%
\fi
\fi\endgroup
\scalebox{.85}{\begin{picture}(6537,781)(2701,-3092)
\put(2701,-2461){\makebox(0,0)[lb]{\smash{\SetFigFont{14}{16.8}{rm}\textit{I}}}}
\put(2701,-3061){\makebox(0,0)[lb]{\smash{\SetFigFont{14}{16.8}{rm}$\two$}}}
\put(2953,-2461){\makebox(0,0)[lb]{\smash{\SetFigFont{11}{8.4}{rm}:}}}
\put(2953,-3061){\makebox(0,0)[lb]{\smash{\SetFigFont{11}{8.4}{rm}:}}}
\put(3205,-2461){\makebox(0,0)[lb]{\smash{\SetFigFont{11}{8.4}{rm}$x_0$}}}
\put(3877,-2461){\makebox(0,0)[lb]{\smash{\SetFigFont{11}{8.4}{rm}$x_1$}}}
\put(4150,-3061){\makebox(0,0)[lb]{\smash{\SetFigFont{11}{8.4}{rm}$\tlaBaire$}}}
\put(6917,-2461){\makebox(0,0)[lb]{\smash{\SetFigFont{11}{8.4}{rm}$x_{n}$}}}
\put(7211,-3061){\makebox(0,0)[lb]{\smash{\SetFigFont{11}{8.4}{rm}$\tlaBaire$}}}
\put(7547,-2461){\makebox(0,0)[lb]{\smash{\SetFigFont{11}{8.4}{rm}$x_{n+1}$}}}
\put(7883,-3061){\makebox(0,0)[lb]{\smash{\SetFigFont{11}{8.4}{rm}$y_{m+1}$}}}
\put(3541,-3061){\makebox(0,0)[lb]{\smash{\SetFigFont{11}{8.4}{rm}$y_0$}}}
\put(5251,-2461){\makebox(0,0)[lb]{\smash{\SetFigFont{11}{8.4}{rm}$x_3$}}}
\put(5551,-3061){\makebox(0,0)[lb]{\smash{\SetFigFont{11}{8.4}{rm}$y_2$}}}
\put(5851,-2461){\makebox(0,0)[lb]{\smash{\SetFigFont{11}{8.4}{rm}$x_4$}}}
\put(4548,-2461){\makebox(0,0)[lb]{\smash{\SetFigFont{11}{8.4}{rm}$x_2$}}}
\end{picture}}
\end{center}
\caption{\small{The game $\GBaire{f}$}}
\end{figure}
\end{definition}

 Note that another equivalent\footnote{These two versions are equivalent in that a player has a winning strategy in one of them if and only if the same\\ player has a winning strategy in the other one.} presentation of this game would be the one where Player $\two$'s moves are finite sequences $y'_{m}\in(\Alphabet_B\cup\set{\tlaBaire})^{<\omega}$
 and the finite or infinite sequence produced by $\two$ is the concatenation $y'=y'_0y'_1\ldots$

\begin{proposition} \label{lemgameBaireClassone}Let $\Alphabet_A,\Alphabet_B$ be non-empty finite alphabets, and $f:\Alphabet_A^{\omega}\to\Alphabet_B^{\omega}$ any function. Player $\two$ has a winning strategy in the game $\GBaire{f}$ if and only if $f$ is Baire class 1.
\end{proposition}

\proofpro{lemgameBaireClassone} 
($\Rightarrow$)  We fix a winning strategy fo Player $\two$  in $\GBaire{f}$, pick any letter $a\in A$, and  let $f_{n}$ be the function defined by\medskip

\centerline{$f_{n}(x)=\langle y_0,y_1,\ldots,y_k\rangle^{\tlaBaire} a^{\omega}.$}\medskip

\noindent where $\langle y_0,y_1,\ldots,y_k\rangle$ is the position that Player $\two$ --- applying her winning strategy --- is in after Player $\one$ has reached the position 
$\langle x_0,x_1,\ldots,x_{n}\rangle$. Thus defined, this function $f_{n}$ is continuous since the following strategy is winning for Player $\two$\  in $\Gcont{f_{n}}$ : she passes her turn $n$ times, then copies $\one$'s first $n$ moves in the game $\GBaire{f}$ and considers the position  $\langle y_0,y_1,\ldots,y_k\rangle$ granted by Player $\two$'s winning strategy in this game, then plays $\langle y_0,y_1,\ldots,y_k\rangle^{\tlaBaire} 0^{\omega}$. By the very definition of Player $\two$'s play $y$ in response to Player $\one$'s play $x$, $f(x)= y^{\tlaBaire}={\displaystyle\lim_{n\in\omega}}\ \  f_{n}(x)$, showing that $f$ is the pointwise limit of a sequence of continuous functions, hence Baire class 1.\medskip
      
  ($\Leftarrow$)    Since $f$ is Baire class 1, there is a sequence of continuous functions 
$(f_{n})_{n<\omega}$ such that, for each $x\in\Alphabet_A^{\omega}$, 
$f(x)=\lim_{n\in\omega} f_{n}(x)$. For each $n$, pick a winning strategy $\sigma_{n}$ for Player $\two$  in the game $\Gcont{f_{n}}$. The following strategy is winning for Player $\two$\  in the version of the game $\GBaire{f}$ where Player $\two$'s moves are finite sequences of the form 
${y'_{m}\in(\Alphabet_B\cup\set{\tlaBaire})^{<\omega}}$: when Player $\one$ reaches the position 
$\langle x_0,x_1,\ldots,x_{n}\rangle$, Player $\two$ (who is actually in the position 
$y'_0y'_1\ldots y'_{n-1}$) replies with $y'_{n}:=$ the shortest finite sequence in 
$(\Alphabet_B\cup\set{\tlaBaire})^{\leq\omega}$ with the property that 
$(y'_0y'_1\ldots y'_{n-1}y'_{n})^{\tlaBaire}=\sigma_{n}*\langle x_0,x_1,\ldots,x_{n}\rangle$, where 
$\sigma_{n}*\langle x_0,x_1,\ldots,x_{n}\rangle$ designates the position that Player $\two$, following her winning strategy $\sigma_{n}$, is in the game $\Gcont{f_{n}}$ after $\one$ has played 
$\langle x_0,x_1,\ldots,x_{n}\rangle$.\boxlem{lemgameBaireClassone}

\medskip

Similar to the Wadge game based on characterizing continuous functions as strategies for the second player in a well-defined game, one defines a game  for reductions by Baire class 1 functions.  Where, given any any topological spaces $X,Y$, $A\subseteq X$  reduces by Baire class 1 functions to $B\subseteq Y$, written $A\bconeinf B$, if there exists some Baire class 1 mapping $f:X\longrightarrow Y$ such that 
    $\forall x\in X~(x\in A \iff f(x)\in B)$, or equivalently $A=f^{-1}[B]$.

%%%%%%%%%%%%

\begin{definition} \label{definitionWadgeBConegame} Let $\Alphabet_A,\Alphabet_B$ be non-empty finite alphabets, and $A\subseteq\Alphabet_A^{\omega}, B\subseteq\Alphabet_B^{\omega}$. The   game $\GWadgeBCone{A}{B}$ is  identical to the game $\GBaire{f}$  specified in Definition \ref{definitioncontinuousgame}, except for the winning condition which is: Player $\two$ wins a play if $(x\in A \iff y\in B)$.

%%%%%%%%%%%%%%%%%%%%%%%%%%%%%%%%%%%%%%%%%%%%%%%%%%%%%%%%%%%%%%%%%%%%%%%%%%%%%%%%%%%%%%%%%%%%%%%%%%%%%%%%%%%%%%%%%%%%%%%%%%%%%%%%%%%%%%%%%%%%%%%%%%%%%%%%%%%%%%%%%%%%%%%%%%%%%%%%%%%%%%%%%%%%%%%%%%%%%%%%%%%%%%%%%%%%%%%%%%%%%%%%%%%%%%%%%%%%%%%%%%%%%%%%%%%%%%%%%%%%%%%%%%%%%%%%%%%%%%%%%%%%%%%%%%%%%%%%%%%%%%%%%%%%%%%%%%%%%%%%%%%%%%%%%%%%%%%%%%%%%%%%%%%%%%%%%%%%%%%%%%%%%%%%%%%%%%%%%%%%%%%%%%%%%%%%%%%%%%%%%%%%%%%%%%%%%%%%%%%%%%%%%%%%%%%%%%%%%%%%%%%%%%%%%%%%%%%%%%%%%%%%%%%%%%%%%%%%%%%%%%%%%%%%%%%%%%%%%%%%%%%%%%%%%%%%%%%%%%%%%%%%%%%%%%%%%%%%%%%%%%%%%%%%%%%%%%%%%%%%%%%%%%%%%%%%%%%

\begin{figure}[h]

\begin{center}
\scalebox{0.85}{\begin{picture}(0,0)%
\includegraphics{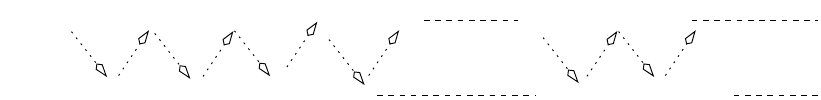}%
\end{picture}}%
\setlength{\unitlength}{0.00083300in}%
\begingroup\makeatletter\ifx\SetFigFont\undefined
% extract first six characters in \fmtname
\def\x#1#2#3#4#5#6#7\relax{\def\x{#1#2#3#4#5#6}}%
\expandafter\x\fmtname xxxxxx\relax \def\y{splain}%
\ifx\x\y   % LaTeX or SliTeX?
\gdef\SetFigFont#1#2#3{%
  \ifnum #1<17\tiny\else \ifnum #1<20\small\else
  \ifnum #1<24\normalsize\else \ifnum #1<29\large\else
  \ifnum #1<34\Large\else \ifnum #1<41\LARGE\else
     \huge\fi\fi\fi\fi\fi\fi
  \csname #3\endcsname}%
\else
\gdef\SetFigFont#1#2#3{\begingroup
  \count@#1\relax \ifnum 25<\count@\count@25\fi
  \def\x{\endgroup\@setsize\SetFigFont{#2pt}}%
  \expandafter\x
    \csname \romannumeral\the\count@ pt\expandafter\endcsname
    \csname @\romannumeral\the\count@ pt\endcsname
  \csname #3\endcsname}%
\fi
\fi\endgroup
\scalebox{.85}{\begin{picture}(6537,781)(2701,-3092)
\put(2701,-2461){\makebox(0,0)[lb]{\smash{\SetFigFont{14}{16.8}{rm}\textit{I}}}}
\put(2701,-3061){\makebox(0,0)[lb]{\smash{\SetFigFont{14}{16.8}{rm}$\two$}}}
\put(2953,-2461){\makebox(0,0)[lb]{\smash{\SetFigFont{11}{8.4}{rm}:}}}
\put(2953,-3061){\makebox(0,0)[lb]{\smash{\SetFigFont{11}{8.4}{rm}:}}}
\put(3205,-2461){\makebox(0,0)[lb]{\smash{\SetFigFont{11}{8.4}{rm}$x_0$}}}
\put(3877,-2461){\makebox(0,0)[lb]{\smash{\SetFigFont{11}{8.4}{rm}$x_1$}}}
\put(4150,-3061){\makebox(0,0)[lb]{\smash{\SetFigFont{11}{8.4}{rm}$\tlaBaire$}}}
\put(6917,-2461){\makebox(0,0)[lb]{\smash{\SetFigFont{11}{8.4}{rm}$x_{n}$}}}
\put(7211,-3061){\makebox(0,0)[lb]{\smash{\SetFigFont{11}{8.4}{rm}$\tlaBaire$}}}
\put(7547,-2461){\makebox(0,0)[lb]{\smash{\SetFigFont{11}{8.4}{rm}$x_{n+1}$}}}
\put(7883,-3061){\makebox(0,0)[lb]{\smash{\SetFigFont{11}{8.4}{rm}$y_{m+1}$}}}
\put(3541,-3061){\makebox(0,0)[lb]{\smash{\SetFigFont{11}{8.4}{rm}$y_0$}}}
\put(5251,-2461){\makebox(0,0)[lb]{\smash{\SetFigFont{11}{8.4}{rm}$x_3$}}}
\put(5551,-3061){\makebox(0,0)[lb]{\smash{\SetFigFont{11}{8.4}{rm}$y_2$}}}
\put(5851,-2461){\makebox(0,0)[lb]{\smash{\SetFigFont{11}{8.4}{rm}$x_4$}}}
\put(4548,-2461){\makebox(0,0)[lb]{\smash{\SetFigFont{11}{8.4}{rm}$x_2$}}}
\end{picture}}
\end{center}
\caption{\small{The Wadge Baire Class 1 game $\GWadgeBCone{A}{B}$}}
\end{figure}
\end{definition}

 Proposition \ref{lemgameBaireClassone} ensures that Player $\two$ has a winning strategy in $\GWadgeBCone{A}{B}$ if and only if there exists some \emph{Baire class 1} mapping $f:\Alphabet_A^{\omega}\to\Alphabet_B^{\omega}$ such that $A=f^{-1}[B]$. To say it differently, $A\bconeinf B$ is equivalent to Player $\two$ having a winning strategy in $\GWadgeBCone{A}{B}$. Notice also that if $A\bconeinf B$ and $B\in D_{n}(\mathbf{\Pi}^{0}_{k})$ (any $n,k\geq 1$), then  $A\in D_{n}(\mathbf{\Pi}^0_{k+1})$.

\section{The operation  ``exponentiation of sets"}\label{section-exponentiation-on-sets}\indent

 The Wadge hierarchy of Borel sets is a profound refinement of both the Borel hierarchy and the Hausdorff-Kuratowski hierarchy of differences. Bill Wadge gave first a description of this hierarchy in \cite{Wadge83}. 
 
\vfill\eject 
 
 The first author obtained in \cite{Duparc01} a new proof of Wadge's results for the case of Borel sets of finite rank. He provided a normal form for finite-rank Borel sets via an inductive construction for each given Wadge degree of a complete set for that degree. These proofs rely on set theoretic operations which are the counterpart of arithmetical operations over ordinals needed to  needed to calculate the level in the Wadge hierarchy at which the set in question is located.
\medskip

 In fact, the first author studied the Wadge hierarchy through the study of the conciliatory hierarchy which orders conciliatory sets, which are sets of  finite {\it or} infinite words over an alphabet $\Alphabet$ (i.e., subsets of $\Alphabet^{<\omega}\cup\Alphabet^{\omega}=\Alphabet^{\leq\omega}$). Among the set theoretic operations  on conciliatory sets introduced in \cite{Duparc01} we will  only need here the exponentiation operation. We now recall the operation $A\mapsto A^\sim$ of exponentiation of conciliatory sets.

\begin{definition}
Let  $\Alphabet_A$ be any finite alphabet, $A\subseteq\Alphabet_A^{\leq\omega}$, and $\tla$ be any letter such that  $\tla\notin\Alphabet_A$. We set 
\[
A^\sim:=\setBig{ x\in\big(\Alphabet_A\cup \set{\tla}\big)^{\leq\omega}\mid x^\tla\in A}.
\]
\end{definition}

 The operation $\sim$ is monotone  with respect to the Wadge order and generates sets of higher topological complexity via the following correspondence. 
 \begin{definition}\label{definitionofblanksoperation}
Given any finite alphabet $\Alphabet$, $A\subseteq\Alphabet^{\leq \omega}$ and $\blank$  a letter such that  $\blank\notin\Alphabet$,  we define
\[
A^{\blank}:=\{x\in (\Alphabet\cup\set{\blank})^{\omega}\mid x_{[\,/\blank]}\in A\},
\]
where $x_{[\,/\blank]}$ represents the sequence obtained from $x$ by removing all occurrences of the letter $\blank$.
\end{definition}

 We recall some results that will be useful later in this article. 
 
\begin{theorem} [Duparc \cite{Duparc01}]\label{thedup} Let $\Alphabet$ be any finite alphabet, and $\blank\notin\Alphabet$.
\begin{enumerate}[label=(\emph{\alph*})]
\item \label{firstitemthedup}
Let $A\subseteq\Alphabet^{\leq \omega}$, and $1\leq k<\omega$. If 
$A^{\blank}\subseteq (\Alphabet\cup\{ d\} )^{\omega}$ is $\mathbf{\Sigma}_k^0$-complete (respectively, $\mathbf{\Pi}_k^0$-complete), then $(A^\sim)^{\blank}$ is $\mathbf{\Sigma}_{k+1}^0$-complete (respectively, $\mathbf{\Pi}_{k+1}^0$-complete).
\item \label{secondtitemthedup}
Let $A\subseteq\Alphabet^{\omega}$,  and $2\leq k<\omega$.  If $A$ is 
$\mathbf{\Pi}_k^0$-complete, then $A^\sim$ is $\mathbf{\Pi}_{k+1}^0$-complete.
\end{enumerate}
\end{theorem}

\begin{remark}\rm In Theorem \ref{thedup}, case {\ref{secondtitemthedup}} follows from  case {\ref{firstitemthedup}}  because

\begin{itemize}
  \item  whenever $A\subseteq\Alphabet^{\omega}$,  $2\leq k<\omega$,  and $A$ is $\mathbf{\Pi}_k^0$-complete, then $A^{\blank}$ is also $\mathbf{\Pi}_k^0$-complete.

\item Whenever $A\subseteq\Alphabet_A^{\omega}$, $3\leq k<\omega$, and 
$A^{\blank}\subseteq (\Alphabet\cup\{ d\} )^{\omega}$ is $\mathbf{\Pi}_k^0$-complete, then $A$ is also 
$\mathbf{\Pi}_k^0$-complete.

\end{itemize}
\end{remark}

 This property proved useful in \cite{Fin01a} for studying  the $\omega$-powers of finitary languages. The second author proved in \cite{Fin01a} that the class $\mbox{CFL}_{\omega}$ of context-free $\omega$-languages, (i.e., those  accepted by  B\"uchi pushdown automata), is closed under the exponentiation operation 
$\sim$. Now we need the following extension of Theorem \ref{thedup}  \textit{\ref{secondtitemthedup}}.

\begin{restatable}{theorem}{thedupthree} \label{thm:thedupthree} Let $\Alphabet$ be any finite alphabet, $A\subseteq\Alphabet^{\omega}$, and $n\geq 1$, $k\geq 2$ be integers. If $A$ is 
$D_{n}(\mathbf{\Pi}_{k}^0)$-complete, then $A^\sim$ is $D_{n}(\mathbf{\Pi}_{k+1}^0)$-complete.\end{restatable}

The proof of this theorem will be a consequence of Lemma \ref{preliminarylemmaforcorollaries} and  Corollary \ref{preliminarycorollary} (\ref{preliminarycorollarya}) to come (see page \pageref{thispageproofthedupthree}).
\medskip

We write $\Lboth({\mathcal{A}_4})$ for the set of \emph{finite} or \emph{infinite} sequences accepted\footnote{We recall that a finite sequence is accepted if its run ends in an accepting state, and an infinite one is accepted\\ if its run visits some accepting state infinitely often.}
 by the automaton  $\mathcal{A}_4$ on Figure \ref{automatonafour}.
 
\vfill\eject 
 
  We also write $\Linfinite(\mathcal{A}_{4^{\blank}})$ for the set of \textit{infinite} sequences accepted
 by the automaton  $\mathcal{A}_{4^{\blank}}$ on Figure \ref{automatonA4d}.
 
 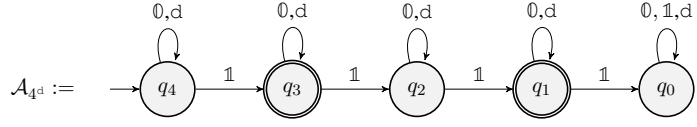
\begin{figure}[h]
\begin{center}
\tikzset{
->,  % makes the edges directed
>=stealth', % makes the arrow heads bold
node distance=22mm, % specifies the minimum distance between two nodes. Change if necessary.
every state/.style={thick, fill=gray!10}, % sets the properties for each ?state? node
initial text=$ $, % sets the text that appears on the start arrow
}
\scalebox{0.75}{\begin{tikzpicture}%[shorten >=1pt,node distance=3cm,on grid]
\node[state, initial] (q4) {$q_4$};
\node[left of=q4] (def) {$\mathcal{A}_{4^{\blank}}:=$};
\node[state, accepting, right of=q4] (q3) {$q_3$};
\node[state, right of=q3] (q2) {$q_2$};
\node[state, accepting, right of=q2] (q1) {$q_1$};
\node[state, right of=q1] (q0) {$q_0$};
\draw  
(q4) edge[loop above] node{$\mathbbold{0}$,\blank} (q4)
(q3) edge[loop above] node{$\mathbbold{0}$,\blank} (q3)
(q2) edge[loop above] node{$\mathbbold{0}$,\blank} (q2)
(q1) edge[loop above] node{$\mathbbold{0}$,\blank} (q1)
(q0) edge[loop above] node{$\mathbbold{0},\mathbbold{1}$,\blank} (q0)
(q4) edge[above] node{$\mathbbold{1}$} (q3)
(q3) edge[above] node{$\mathbbold{1}$} (q2)
(q2) edge[above] node{$\mathbbold{1}$} (q1)
(q1) edge[above] node{$\mathbbold{1}$} (q0)
;
\end{tikzpicture}}
\end{center}
\caption{\small{The automaton $\mathcal{A}_{4^{\blank}}$\label{automatonA4d}}}
\end{figure}

  We have
\begin{itemize}
  \item  $\begin{array}[t]{lll}
     \Lboth({\mathcal{A}_4}) & = &   \mathbbold{0}^*\mathbbold{1}\mathbbold{0}^{*}\cup 
     \mathbbold{0}^*\mathbbold{1}\mathbbold{0}^{*}\mathbbold{1}\mathbbold{0}^{*}\mathbbold{1}\mathbbold{0}^{*}\cup \mathbbold{0}^*\mathbbold{1}\mathbbold{0}^{\omega}\cup \mathbbold{0}^*\mathbbold{1}\mathbbold{0}^*\mathbbold{1}\mathbbold{0}^*\mathbbold{1}\mathbbold{0}^{\omega}\\
   & = &  \setbig{x\in\set{\mathbbold{0},\mathbbold{1}}^{\leq\omega}\mid x \textit{ contains either one or three }\mathbbold{1}\textit{'s}}
\end{array}$
  \item  $\begin{array}[t]{lll}
     \Linfinite(\mathcal{A}_{4^{\blank}})& = &   \set{\mathbbold{0},\blank}^*\mathbbold{1}\set{\mathbbold{0},\blank}^{\omega}\cup \set{\mathbbold{0},\blank}^*\mathbbold{1}\set{\mathbbold{0},\blank}^*\mathbbold{1}\set{\mathbbold{0},\blank}^*\mathbbold{1}\set{\mathbbold{0},\blank}^{\omega}\\
   & = &  \setbig{x\in\set{\mathbbold{0},\mathbbold{1},\blank}^{\omega}\mid x \textit{ contains either one or three }\mathbbold{1}\textit{'s}}.
\end{array}$

\end{itemize} 
 
 Note that we have $\big(\Lboth(\mathcal{A}_4)\big)^{\blank}=\Linfinite(\mathcal{A}_{4^{\blank}})$  \big(see Definition \ref{definitionofblanksoperation} for the definition of $\big(\Lboth(\mathcal{A}_4)\big)^{\blank}$\big).

\begin{lemma} \label{automatonA4dd4closedcomplete} Consider    the automaton $\mathcal{A}_4$ in  Figure \ref{automatonafour}, and $\blank\notin\set{\mathbbold{0},\mathbbold{1}}$. The language 
$$\big(\Lboth(\mathcal{A}_4)\big)^{\blank}\subseteq\set{\mathbbold{0},\mathbbold{1},\blank}^{\omega}$$ 
is $D_4(\mathbf{\Pi}_{1}^0)$-complete.\end{lemma}

\prooflem{automatonA4dd4closedcomplete}   For simplicity, notice that the $\omega$-language $\big(\Lboth(\mathcal{A}_4)\big)^{\blank}$ that we are interested in is rigorously the same as 
the $\omega$-language  $\Linfinite(\mathcal{A}_{4^{\blank}})\subseteq\set{\mathbbold{0},\mathbbold{1},\blank}^{\omega}$ accepted by the  automaton $\mathcal{A}_{4^{\blank}}$ in 
Figure \ref{automatonA4d} above.

\begin{figure}[h]
\begin{center}
\tikzset{
->,  % makes the edges directed
>=stealth', % makes the arrow heads bold
node distance=22mm, % specifies the minimum distance between two nodes. Change if necessary.
every state/.style={thick, fill=gray!10}, % sets the properties for each ?state? node
initial text=$ $, % sets the text that appears on the start arrow
}
\scalebox{0.75}{\begin{tikzpicture}%[shorten >=1pt,node distance=3cm,on grid]
\node[state, initial, accepting] (q4) {$q_4$};
\node[left of=q4] (def) {$\mathcal{G}_{0^{\blank}}:=$};
\node[state, accepting, right of=q4] (q3) {$q_3$};
\node[state, accepting, right of=q3] (q2) {$q_2$};
\node[state, accepting, right of=q2] (q1) {$q_1$};
\node[state, right of=q1] (q0) {$q_0$};
\draw  
(q4) edge[loop above] node{$\mathbbold{0}$,\blank} (q4)
(q3) edge[loop above] node{$\mathbbold{0}$,\blank} (q3)
(q2) edge[loop above] node{$\mathbbold{0}$,\blank} (q2)
(q1) edge[loop above] node{$\mathbbold{0}$,\blank} (q1)
(q0) edge[loop above] node{$\mathbbold{0},\mathbbold{1}$,\blank} (q0)
(q4) edge[above] node{$\mathbbold{1}$} (q3)
(q3) edge[above] node{$\mathbbold{1}$} (q2)
(q2) edge[above] node{$\mathbbold{1}$} (q1)
(q1) edge[above] node{$\mathbbold{1}$} (q0)
;
\end{tikzpicture}}

\scalebox{0.75}{\begin{tikzpicture}%[shorten >=1pt,node distance=3cm,on grid]
\node[state, initial, accepting] (q4) {$q_4$};
\node[left of=q4] (def) {$\mathcal{G}_{1^{\blank}}:=$};
\node[state, accepting, right of=q4] (q3) {$q_3$};
\node[state, accepting, right of=q3] (q2) {$q_2$};
\node[state, right of=q2] (q1) {$q_1$};
\node[state, right of=q1] (q0) {$q_0$};
\draw  
(q4) edge[loop above] node{$\mathbbold{0}$,\blank} (q4)
(q3) edge[loop above] node{$\mathbbold{0}$,\blank} (q3)
(q2) edge[loop above] node{$\mathbbold{0}$,\blank} (q2)
(q1) edge[loop above] node{$\mathbbold{0}$,\blank} (q1)
(q0) edge[loop above] node{$\mathbbold{0},\mathbbold{1}$,\blank} (q0)
(q4) edge[above] node{$\mathbbold{1}$} (q3)
(q3) edge[above] node{$\mathbbold{1}$} (q2)
(q2) edge[above] node{$\mathbbold{1}$} (q1)
(q1) edge[above] node{$\mathbbold{1}$} (q0)
;
\end{tikzpicture}}

\scalebox{0.75}{\begin{tikzpicture}%[shorten >=1pt,node distance=3cm,on grid]
\node[state, initial, accepting] (q4) {$q_4$};
\node[left of=q4] (def) {$\mathcal{G}_{2^{\blank}}:=$};
\node[state, accepting, right of=q4] (q3) {$q_3$};
\node[state, right of=q3] (q2) {$q_2$};
\node[state, right of=q2] (q1) {$q_1$};
\node[state, right of=q1] (q0) {$q_0$};
\draw  
(q4) edge[loop above] node{$\mathbbold{0}$,\blank} (q4)
(q3) edge[loop above] node{$\mathbbold{0}$,\blank} (q3)
(q2) edge[loop above] node{$\mathbbold{0}$,\blank} (q2)
(q1) edge[loop above] node{$\mathbbold{0}$,\blank} (q1)
(q0) edge[loop above] node{$\mathbbold{0},\mathbbold{1}$,\blank} (q0)
(q4) edge[above] node{$\mathbbold{1}$} (q3)
(q3) edge[above] node{$\mathbbold{1}$} (q2)
(q2) edge[above] node{$\mathbbold{1}$} (q1)
(q1) edge[above] node{$\mathbbold{1}$} (q0)
;
\end{tikzpicture}}

\scalebox{0.75}{\begin{tikzpicture}%[shorten >=1pt,node distance=3cm,on grid]
\node[state, initial, accepting] (q4) {$q_4$};
\node[left of=q4] (def) {$\mathcal{G}_{3^{\blank}}:=$};
\node[state, right of=q4] (q3) {$q_3$};
\node[state, right of=q3] (q2) {$q_2$};
\node[state, right of=q2] (q1) {$q_1$};
\node[state, right of=q1] (q0) {$q_0$};
\draw  
(q4) edge[loop above] node{$\mathbbold{0}$,\blank} (q4)
(q3) edge[loop above] node{$\mathbbold{0}$,\blank} (q3)
(q2) edge[loop above] node{$\mathbbold{0}$,\blank} (q2)
(q1) edge[loop above] node{$\mathbbold{0}$,\blank} (q1)
(q0) edge[loop above] node{$\mathbbold{0},\mathbbold{1}$,\blank} (q0)
(q4) edge[above] node{$\mathbbold{1}$} (q3)
(q3) edge[above] node{$\mathbbold{1}$} (q2)
(q2) edge[above] node{$\mathbbold{1}$} (q1)
(q1) edge[above] node{$\mathbbold{1}$} (q0)
;
\end{tikzpicture}}
\end{center}
\caption{\small{The automata $\mathcal{G}_{0^{\blank}}$, $\mathcal{G}_{1^{\blank}}$, $\mathcal{G}_{2^{\blank}}$, and $\mathcal{G}_{3^{\blank}}$\label{automataGs}}}
\end{figure}
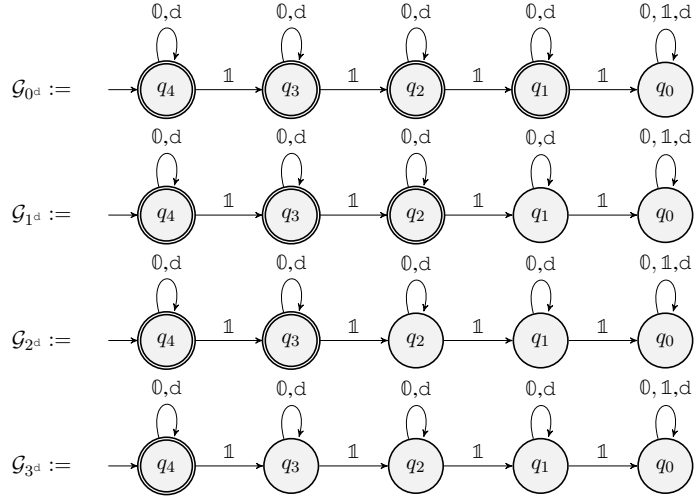

\vfill\eject

 For $0\leq i\leq 3$, let $\mathcal{G}_{i^{\blank}}$  be the automata described in Figure \ref{automataGs}, and let us agree to write $G_i:=\Linfinite(\mathcal{G}_{i^{\blank}})$ (any $0\leq i\leq 3$) to shorten the notation. Note that\begin{itemize}
  \item  each $G_i$ is a closed subset of $\set{\mathbbold{0},\mathbbold{1},\blank}^{\omega}$,
  \item  $G_0\supsetneq G_1\supsetneq G_2\supsetneq G_3$, and
  \item  $\Linfinite(\mathcal{A}_{4^{\blank}})=(G_0\smallsetminus G_1)\cup (G_2\smallsetminus G_3)$,
  \end{itemize}
   which shows that  $\Linfinite(\mathcal{A}_{4^{\blank}})\in D_4(\mathbf{\Pi}_{1}^0)$.

 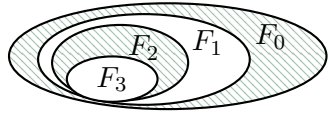
\begin{figure}[h]
\begin{center}
\begin{tikzpicture}
\def\angle{60}%
\pgfmathsetlengthmacro{\xoff}{21mm*cos(\angle)}%
\pgfmathsetlengthmacro{\yoff}{10mm*sin(\angle)}%
\draw [thick,pattern=north west lines, pattern color=cambridgeblue] (0.7*\xoff,0.35*\yoff) circle[x radius=21mm, y radius=7mm] ++(1.3*\xoff,0.3*\yoff) node{$F_0$};
\draw [thick, fill=white!30] (0.4*\xoff,0.3*\yoff) circle[x radius=14mm, y radius=6mm] ++(0.8*\xoff,0.26*\yoff) node{$F_1$};
\draw [thick,pattern=north west lines, pattern color=cambridgeblue] (0.1*\xoff,0.25*\yoff) circle[x radius=9mm, y radius=5mm] ++(0.3*\xoff,0.23*\yoff) node{$F_2$};
\draw [thick, fill=white!30] (0,0) circle[x radius=6mm, y radius=3mm] node{$F_3$};
\end{tikzpicture}
\end{center}
\caption{\small{The set $F=(F_0\smallsetminus F_1)\cup (F_2\smallsetminus F_3)$\label{setasdifferenceoffour}}}
\end{figure}

 To show that $\Linfinite(\mathcal{A}_{4^{\blank}})$ is $D_4(\mathbf{\Pi}_{1}^0)$-hard, we pick any set of the form\medskip
  
\centerline{$F=(F_0\smallsetminus F_1)\cup (F_2\smallsetminus F_3)$}\medskip

\noindent where $F_0\supseteq F_1\supseteq F_2\supseteq F_3$ (as shown in Figure \ref{setasdifferenceoffour}) and every $F_i$ is a closed subset of some given space 
$\Alphabet^{\omega}$. For each $i\leq 3$, we let $T_i$ be any tree on $\Alphabet$ such that $T_3\subseteq T_2\subseteq T_1\subseteq T_0$ and  $[T_i]=F_i$ (we recall that $[T_i]$ stands for the set of infinite branches of $T_i$). A winning strategy for Player $\two$ in 
$\GWadge{F}{\Linfinite(\mathcal{A}_{4^{\blank}})}$ would be to find herself having played exactly as many letters $\mathbbold{1}$ as there are trees $T_i$ that Player $\one$'s position has exited. This means, as long as Player $\one$'s position $u$ remains inside $T_3$, Player $\two$ plays 
$\mathbbold{0}$'s, if $u$ exits $T_3$, then  Player $\two$ plays $\mathbbold{1}$. The same tactics are then used with regards to $T_2$ instead of $T_3$, then $T_1$ instead of $T_2$, and finally $T_0$.\boxlem{automatonA4dd4closedcomplete}

 We now generalize to any non-null integer $n$ what we did with $\parenthesesbig{\Lboth(\mathcal{A}_{4})}^{\blank}$.   

\begin{lemma} \label{automatonAnddnclosedcomplete} Let $n\geq 1$ be any integer. 
 Consider    the automaton $\mathcal{A}_{n}$ in  Figure \ref{therealautomatonAnplusonewithnoblanks}, and $\blank\notin\set{\mathbbold{0},\mathbbold{1}}$. The language $\big(\Lboth(\mathcal{A}_{n})\big)^{\blank}\subseteq\set{\mathbbold{0},\mathbbold{1},\blank}^{\omega}$ is $D_{n}(\mathbf{\Pi}_{1}^0)$-complete.
\end{lemma}

\prooflem{automatonAnddnclosedcomplete} This is a straightforward generalization of Lemma \ref{automatonA4dd4closedcomplete}. 
\boxlem{automatonAnddnclosedcomplete}

\begin{lemma} \label{Btildeinclass} Let $\Alphabet$ be any finite alphabet, $B\subseteq\Alphabet^{\leq\omega}$, $\blank\notin\Alphabet$, and $1\leq k,n<\omega$. If the set 
$B^{\blank}\subseteq(\Alphabet\cup\set{\blank})^{\omega}$  belongs to $D_{n}(\mathbf{\Pi}_{k}^0)$, then $(B^\sim)^{\blank}\subseteq(\Alphabet\cup\set{\tla,\blank})^{\omega}$  belongs to 
$D_{n}(\mathbf{\Pi}_{k+1}^0)$.\end{lemma}

\prooflem{Btildeinclass} We  show that $(B^\sim)^{\blank}$ belongs to the class $D_{n}(\mathbf{\Pi}_{k+1}^0)$ by providing a winning strategy for Player $\two$
in the game $\GWadgeBCone{(B^\sim)^{\blank}}{B^{\blank}}$. Recall that in this game, Player $\one$, being in charge of  $(B^\sim)^{\blank}$, plays letters in $\Alphabet\cup\set{\blank,\tla}$, while Player $\two$ plays letters in $\Alphabet\cup\set{\blank,\tlaBaire}$. 
After $\one$ and $\two$ have produced, respectively, the infinite sequences $x\in \big(\Alphabet\cup\set{\blank,\tla}\big)^{\omega}$ and $y\in \big(\Alphabet\cup\set{\blank,\tlaBaire}\big)^{\omega}$, Player $\two$ wins if and only if $\big((x_{[\,/\blank]})^{\tla}\in B\iff (y^{\tlaBaire})_{[\,/\blank]}\in B\big)$. 
This means that, in order to get $(x_{[\,/\blank]})^{\tla}$ starting from $x$, one first deletes all occurrences of the letter \quq{\blank}, then interprets every occurrence of the letter \quq{$\tla$} as an eraser. 
Whereas, in order to get $(y^{\tlaBaire})_{[\,/\blank]}$ starting from $y$, one first interprets every occurrence of the letter \quq{$\tlaBaire$} as an eraser --- which erases not only letters in $\Alphabet$ but also the letter \quq{\blank} --- then gets rid of all remaining occurrences of the letter \quq{\blank}.\medskip 

For simplicity, we describe a winning strategy for Player $\two$ in the version of the game 
$\GWadgeBCone{(B^\sim)^{\blank}}{B^{\blank}}$ where $\two$'s moves are finite sequences of the form  $y'_{m}\in(\Alphabet\cup\set{\blank,\tlaBaire})^{<\omega}$ and the finite or infinite sequence produced by $\two$ is the concatenation $y'=y'_0y'_1\ldots$. We recall that this variant of the game  is equivalent to the original one in that a player has a winning strategy in one of them if and only if the same player has a winning strategy in the other one.

\vfill\eject

 The strategy consists in, when Player $\one$ reaches the position $\langle x_0,x_1,\ldots,x_p\rangle$, and Player $\two$ --- who is actually\footnote{In this presentation, playing an empty sequence simply means that Player  $\two$ passes her turn.} in some position $ y'_0y'_1\ldots y'_{p-1}$ ---  replies with 
\begin{enumerate}
  \item $x_p$ if $x_p\in\Alphabet$,
  \item $d$ if $x_p=d$,
    \item $\langle\underbrace{\tlaBaire,\ldots,\tlaBaire}_{q},\blank\rangle$ if $x_p=\tla$, and $q$ is the \emph{least integer} such that 
   \[
    \Big( \big(y'_0y'_1\ldots y'_{p-1}\langle\underbrace{\tlaBaire,\ldots,\tlaBaire}_{q}\rangle\big)^{\tlaBaire}  \Big)_{[\,/\blank]}=\big(\langle x_0,x_1,\ldots,x_p\rangle_{[\,/\blank]}\big)^{\tla}.
    \] 
\end{enumerate}
In words, this means that Player $\two$ copies what Player $\one$ does except when $\one$ plays an eraser. In this case, Player $\two$ considers which   letter  $x_i$ from $\Alphabet$ (if any) this eraser is going to delete after one gets  rid of all the occurrences of the letter \quq{\blank} from $\one$'s actual position $\langle x_0,x_1,\ldots,x_p\rangle$. 
\begin{enumerate}
  \item If such a letter exists, it is the last letter of the non-empty word $\big(\langle x_0,x_1,\ldots,x_{p-1}\rangle_{[\,/\blank]}\big)^{\tla}$. In this case, by construction, Player $\two$  finds herself in some position  $y'_0y'_1\ldots y'_{p-1}$ which satisfies 
  \[
  \Big( \big(y'_0y'_1\ldots y'_{p-1}\big)^{\tlaBaire}  \Big)_{[\,/\blank]}=\big(\langle x_0,x_1,\ldots,x_{p-1}\rangle_{[\,/\blank]}\big)^{\tla}\mbox{,}
  \]
 so that, for some non-zero integer $q$, we have 
$\langle x_i, \underbrace{\blank,\ldots,\blank}_{q-1}\rangle$ is a suffix of 
$\big(y'_0y'_1\ldots y'_{p-1}\big)^{\tlaBaire}$. So, Player $\two$ then plays  
$y'_{p}=\langle\underbrace{\tlaBaire,\ldots,\tlaBaire}_{q},\blank\rangle$ in order to make the sequence 
$\langle\underbrace{\tlaBaire,\ldots,\tlaBaire}_{q}\rangle$  erase exactly the sequence 
$\langle x_i, \underbrace{\blank,\ldots,\blank}_{q-1}\rangle$, and the letter \quq{\blank} that she then plays serves to ensure that, in the event that she should return indefinitely to this same position, would nevertheless produce a sequenc $y$ such that $y^{\tlaBaire}$ is infinite.

   \item If no such  letter exists, this means that $\big(\langle x_0,x_1,\ldots,x_{p-1}\rangle_{[\,/\blank]}\big)^{\tla}=\emptysequence$, in which case by construction, Player $\two$  finds herself in some position  $y'_0y'_1\ldots y'_{p-1}$ which satisfies 
  \[
\big(y'_0y'_1\ldots y'_{p-1}\big)^{\tlaBaire}=\langle \blank,\ldots,\blank \rangle.
  \]
 Then, $\two$ simply plays    a letter \quq{\blank} so that if Player $\one$ ultimately produces a sequence $x$ that satisfies $\big(x_{[\,/\blank]}\big)^{\tla}=\emptysequence$, then the sequence she will have produced will satisfy $y^{\tlaBaire}=d^{\omega}$.
     \end{enumerate}
     
Note that this strategy makes sure that when Player $\one$ produces $x\in \big(\Alphabet\cup\set{\blank,\tla}\big)^{\omega}$, Player $\two$ produces $y\in \big(\Alphabet\cup\set{\blank,\tlaBaire}\big)^{\omega}$ that satisfies both
$
y^{\tlaBaire}\text{ is infinite and }\big(y^{\tlaBaire}\big)_{[\,/\blank]}=\big(x_{[\,/\blank]}\big)^{\tla}
$. This guarantees both that Player $\two$ wins the play and also, by Lemma \ref{lemgameBaireClassone}, that the mapping induced by $\two$'s winning strategy is Baire class 1.
Therefore, $(B^\sim)^{\blank}$ belongs to the class $D_{n}(\mathbf{\Pi}_{k+1}^0)$ since it is the inverse image by a Baire class 1 function of  the set $B^{\blank}$ which belongs to $D_{n}(\mathbf{\Pi}_{k}^0)$.
\boxlem{Btildeinclass}

\begin{definition}[Question-tree] \label{definitionquestiontree}

Given any non-empty pruned tree  $\T$ --- regarded as some  player's arena ---  on any given non-empty alphabet $\Tvar{\Alphabet}$, and a sequence $(\Tvar{Q_{n}})_{n<\omega}$ of subtrees of\;$\T$ ($\Tvar{Q_{n}}\subseteq \T$ holds for each integer $n$) we describe $\tildeT$ a non-empty pruned tree  on the alphabet  
$$\Tvar{\ddot{\Alphabet}}=\Tvar{\Alphabet}\cup \set{\textcolor{\colorforyes}{yes}}\cup\set{(\textcolor{\colorforno}{no},\Tvar{u})\mid \Tvar{u}\in \T}$$
that we call the \emph{Question-tree} induced by both $\T$ and $(\Tvar{Q_{n}})_{n<\omega}$ as an arena for the same player, by indicating which moves are the legal ones for that same player. To begin with, we list the properties that $\tildeT$ will satisfy.
\begin{itemize}
  \item Each infinite branch  $\tildex=\langle\xvar{0},\avar{0} , \xvar{1}, \avar{1}, \xvar{2}, \avar{2},\ldots\ldots\rangle$ of the tree $\tildeT$ can be decomposed into two subsequences:\begin{itemize}
  \item[$\circ$] $\x=\langle\xvar{0}, \xvar{1},\xvar{2},\ldots\ldots\rangle=\langle\tildexvar{2n}/n<\omega\rangle$ known as the \textbf{main run} of $\tildex$, 
  \item[$\circ$] $\anovar =\langle\avar{0} ,  \avar{1},\avar{2},\ldots\ldots\rangle=\langle\tildexvar{2n+1}/n<\omega\rangle$ which represents the sequence of \textbf{auxiliary moves} of $\tildex$.
   
\end{itemize}

  \item Given any   infinite branch  $\tildex=\langle\xvar{0},\avar{0} , \xvar{1}, \avar{1}, \xvar{2}, \avar{2},\ldots\ldots\rangle$ of $\tildeT$, 
  \begin{itemize}
    \item[$\circ$]  its main run $\x=\langle\xvar{0}, \xvar{1},\xvar{2},\ldots\ldots\rangle$ is an infinite branch of $\T$,
    \item[$\circ$] the sequence $\anovar =\langle\avar{0} ,  \avar{1},\avar{2},\ldots\ldots\rangle$ of its auxiliary moves is uniquely defined with respect to the main run $\x$. i.e., if 
 $\tildex=\langle\xvar{0},\avar{0} , \xvar{1}, \avar{1}, \xvar{2}, \avar{2},\ldots\ldots\rangle$ and $\tildexprime=\langle\xvar{0},\avarprime{0} , \xvar{1}, \avarprime{1}, \xvar{2}, \avarprime{2},\ldots\ldots\rangle$ are both infinite branches of $\tildeT$ with the same main run $\x=\langle\xvar{0}, \xvar{1},\xvar{2},\ldots\ldots\rangle$, then they both share the  exact same auxiliary moves  ($\avar{i}=\avarprime{i}$ holds for all integers $i$), which means that $\tildex=\tildexprime$.

\end{itemize} 
 \item Given any   infinite branch  $\x=\langle\xvar{0}, \xvar{1},\xvar{2},\ldots\ldots\rangle$ of $\T$, there exists a unique infinite branch $\tildex=\langle\xvar{0},\avar{0} , \xvar{1}, \avar{1}, \xvar{2}, \avar{2},\ldots\ldots\rangle$ of $\tildeT$ whose main run is $\x$.

%\medskip
%  \medskip
%  \medskip
%  \medskip
%  $\x=\langle\xvar{0}, \xvar{1},\xvar{2},\ldots\ldots\rangle=\langle\Tvar{\tildex_{(2n)}}/n<\omega\rangle$ known as the \textbf{main run} of $\tildex$ and 
%  $\langle\avar{0} ,  \avar{1},\avar{2},\ldots\ldots\rangle$
% 
%  yields  a unique infinite branch $\x=\langle\xvar{0}, \xvar{1},\xvar{2},\ldots\ldots\rangle=\langle\Tvar{\tildex_{(2n)}}/n<\omega\rangle$ of $\T$ known as the \textbf{main run} of $\tildex$. 
%%
%%
% \item Each infinite branch  $\tildex=\langle\xvar{0},\avar{0} , \xvar{1}, \avar{1}, \xvar{2}, \avar{2},\ldots\ldots\rangle$ of the tree $\tildeT$ yields  a unique infinite branch $\x=\langle\xvar{0}, \xvar{1},\xvar{2},\ldots\ldots\rangle=\langle\Tvar{\tildex_{(2n)}}/n<\omega\rangle$ of $\T$ known as the \textbf{main run} of $\tildex$. 
%   \item Each infinite branch  $\x=\langle\xvar{0}, \xvar{1},\xvar{2},\ldots\ldots\rangle$ of $\T$ yields a unique infinite branch of $\tildeT$  which is of the form $\tildex=\langle\xvar{0},\avar{0} , \xvar{1}, \avar{1}, \xvar{2}, \avar{2},\ldots\ldots\rangle$. 
%   i.e.,  given $\x=\langle\xvar{0}, \xvar{1},\xvar{2},\ldots\ldots\rangle\in[\T]$, there exists a unique  infinite sequence $\langle\avar{0} ,  \avar{1},\avar{2},\ldots\ldots\rangle$ such that the infinite branch $\tildex=\langle\xvar{0},\avar{0} , \xvar{1}, \avar{1}, \xvar{2}, \avar{2},\ldots\ldots\rangle\in[\tildeT]$.
%  
\end{itemize}

%%%%%%%%%%%%%%%%%%%%%%%%%%%%%%%%%%%%%%%%%%%%%%%%%%%%%
%%%%%%%%%%%%%%%%%%%%%%%%%%%%%%%%%%%%%%%%%%%%%%%%%%%%%
%%%%%%%%%%%%%%%%%%%%%%%%%%%%%%%%%%%%%%%%%%%%%%%%%%%%%
%%%%%%%%%%%%%%%%%%%%%%%%%%%%%%%%%%%%%%%%%%%%%%%%%%%%%
%%%%%%%%%%%%%%%%%%%%%%%%%%%%%%%%%%%%%%%%%%%%%%%%%%%%%
%%%%%%%%%%%%%%%%%%%%%%%%%%%%%%%%%%%%%%%%%%%%%%%%%%%%%

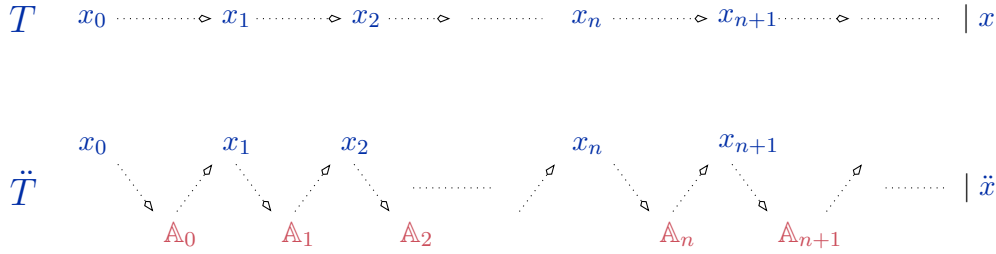
\begin{figure}[h]
\begin{center}
\begin{tikzpicture}[->,scale=.625]
 \node at (-2,0)  (tildet) {{\Large $\T$}};
   \draw[>={Kite[open,width=3pt,length=5pt]}, dotted] (0,0) node[anchor= east]{$\xvar{0}$} --  (2,0) node[anchor= west]{$\xvar{1}$};
 \draw[>={Kite[open,width=3pt,length=5pt]}, dotted] (2.95,0)   -- (4.75,0) node[anchor= west]{$\xvar{2}$};
 \draw[>={Kite[open,width=3pt,length=5pt]}, dotted] (5.75,0)  -- (7.05,0) ;
  %%%%%
   \draw[>={}, dotted]  (7.5,0) -- (9,0); 
   \draw[>={}, dotted]  (15.75,0) -- (17.5,0) node[anchor= west]{$\;|\;\xvar{}$};
     %%%%%
%    \draw[>={Kite[open,width=3pt,length=5pt]}, dotted] (10.5,-1)  -- (11.25,0);
\draw[>={Kite[open,width=3pt,length=5pt]}, dotted] (10.5,0) node[anchor= east]{$\xvar{n}$}  --  (12.5,0) node[anchor= west]{$\xvar{n+1}$};
 \draw[>={Kite[open,width=3pt,length=5pt]}, dotted] (14,0)  -- (15.5,0) ;
 %    \draw[>={Kite[open,width=3pt,length=5pt]}, dotted] (17,-1)  -- (17.75,0) ;
\end{tikzpicture}

\end{center}
\medskip
\begin{center}
\begin{tikzpicture}[->,scale=.625]
 \node at (-2,-0.5)  (tildet) {{\Large $\tildeT$}};
   \draw[>={Kite[open,width=3pt,length=5pt]}, dotted] (0,0) node[anchor=south east]{$\xvar{0}$} -- (0.75,-1) node[anchor=north west]{$\avar{0}$};
  \draw[>={Kite[open,width=3pt,length=5pt]}, dotted] (1.25,-1)  -- (2,0) node[anchor=south west]{$\xvar{1}$};
 \draw[>={Kite[open,width=3pt,length=5pt]}, dotted] (2.5,0)  -- (3.25,-1) node[anchor=north west]{$\avar{1}$};
   \draw[>={Kite[open,width=3pt,length=5pt]}, dotted] (3.75,-1)  -- (4.5,0) node[anchor=south west]{$\xvar{2}$};
 \draw[>={Kite[open,width=3pt,length=5pt]}, dotted] (5,0)  -- (5.75,-1) node[anchor=north west]{$\avar{2}$};
  %%%%%
   \draw[>={}, dotted]  (6.25,-.5) -- (8,-.5); 
   \draw[>={}, dotted]  (16.25,-.5) --  (17.5,-.5) node[anchor= west]{$\;|\;\tildevar{x}$};
     %%%%%
    \draw[>={Kite[open,width=3pt,length=5pt]}, dotted] (8.5,-1)  -- (9.25,0);
\draw[>={Kite[open,width=3pt,length=5pt]}, dotted] (10.5,0) node[anchor=south east]{$\xvar{n}$}  -- (11.25,-1) node[anchor=north west]{$\avar{n}$};
   \draw[>={Kite[open,width=3pt,length=5pt]}, dotted] (11.75,-1)  -- (12.5,0) node[anchor=south west]{$\xvar{n+1}$};
 \draw[>={Kite[open,width=3pt,length=5pt]}, dotted] (13,0)  -- (13.75,-1) node[anchor=north west]{$\avar{n+1}$};
     \draw[>={Kite[open,width=3pt,length=5pt]}, dotted] (15,-1)  -- (15.75,0) ;
\end{tikzpicture}
\end{center}
\caption{\small{An infinite play $\Tvar{x}$ inside $\T$ and the corresponding play $\tildex$ in $\tildeT$\label{TandTtilde}}}
\end{figure}

%%%%%%%%%%%%%%%%%%%%%%%%%%%%%%%%%%%%%%%%%%%%%%%%%%%%%
%%%%%%%%%%%%%%%%%%%%%%%%%%%%%%%%%%%%%%%%%%%%%%%%%%%%%
%%%%%%%%%%%%%%%%%%%%%%%%%%%%%%%%%%%%%%%%%%%%%%%%%%%%%

From the viewpoint of a player  making its way through the tree $\tildeT$, the moves from the \emph{main run} take place in the tree $\T$. The other moves --- known as \emph{auxiliary moves} --- consist in answering questions about the future of the main run with regards to the trees $(\Tvar{Q_{n}})_{n<\omega}$. 
When playing the auxiliary move $\avar{n}$, the player has several options.

\begin{description}
  \item[Option \textcolor{\colorforyes}{yes}:] if the player takes that option, it means this player agrees that the main run of its play will remain for ever inside $\Tvar{Q_{n}}$, ensuring therefore, that at the end the player will have produced $\x=\langle\xvar{0}, \xvar{1},\xvar{2},\ldots\ldots\rangle\in[\T]$. From now on, the tree $\tildeT$ is 
  designed so that its main run is restricted to $\Tvar{Q_{n}}$, which means that no position along the main run that exits the tree $\Tvar{Q_{n}}$ is legal.
  
  \item[Options (\textcolor{\colorforno}{no},$\Tvar{u}$):] the player takes such an option if this player does not agree  that the main run of its play will remain for ever inside $\Tvar{Q_{n}}$. By choosing (\textcolor{\colorforno}{no},$\Tvar{u}$), the player indicates that its main run either already has, or will reach some position 
 $\Tvar{u}\in \T\smallsetminus \Tvar{Q_{n}}$,  indicating this way how the main run of its play will exit the tree $\Tvar{Q_{n}}$. So, from now on, the tree $\tildeT$ is 
  designed so that the main run of every branch that extends the actual position is committed to either reaching the position $\Tvar{u}$, or having already reached it. This means that if $|\Tvar{u}|=k$, then   $\langle\xvar{0}, \xvar{1},\xvar{2},\ldots\xvar{k-1}\rangle=\Tvar{u}$.

\medskip

 Moreover,  the options  (\textcolor{\colorforno}{no},$\Tvar{u}$) available at this stage are exactly all (\textcolor{\colorforno}{no},$\Tvar{u}$) such that both $u\upharpoonright |\Tvar{u}|-1\in \Tvar{Q_{n}}$ but $\Tvar{u}\notin \Tvar{Q_{n}}$ --- which means that the position $\Tvar{u}$ exits $\Tvar{Q_{n}}$ for the first time since all strict prefixes of $\Tvar{u}$ were still inside $\Tvar{Q_{n}}$. 
\end{description}

Given  any set $\Tvar{A}\subseteq[\T]$ of infinite branches of  $\T$, we also define   $\tildevar{A}\subseteq[\tildeT]$ as the set of all infinite branches of $\tildeT$ whose main runs belong to $\Tvar{A}$. Namely, 
\[
\tildevar{A}=\setbig{
 \tildex\in[\tildeT]\mid (\tildexvar{2n})_{n<\omega}\in \Tvar{A}
 }
.\]

\end{definition}

 One can view the succession of  options taken as a commitment to keeping the  main run  within increasingly smaller subtrees since options of the form \textcolor{\colorforyes}{yes} 
restrict the main run to some $\Tvar{Q_{n}}$ and options of the form (\textcolor{\colorforno}{no},$\Tvar{u}$) restrict the play to $\setbig{\Tvar{v}\in\T\mid\Tvar{u}\subseteq \Tvar{v}}$. If  $\Tvar{F_{n}}$ denotes either the closed set $[\Tvar{Q_{n}}]$ if the player takes Option \textcolor{\colorforyes}{yes}, or the clopen set $\setbig{\Tvar{x}\in[\T]\mid\Tvar{u}\subseteq \Tvar{x}}$ if the player takes Option (\textcolor{\colorforno}{no},$\Tvar{u}$), the state of play after $2n$ moves is that the player  has committed to the whole infinite sequence of the main run ending in the closed subspace 
$\bigcap_{i\leq n}\Tvar{F_i}$.

\medskip

\emph{Question-trees} provide a solution to the main challenge of reducing the topological complexity of Borel sets of finite ranks.

\begin{lemma} \label{questiontreefordowngrading} Let $\Alphabet$ be any non-empty finite alphabet, 
$\T$ be any non-empty pruned tree on $\Alphabet$, $\Tvar{A}\subseteq[\T]$, and $n,k\geq 1$ be integers. If $\Tvar{A}\subseteq[\T]$ belongs to  $D_{n}(\mathbf{\Pi}_{k+1}^0)$, then there exists some (non-empty pruned) Question-tree $\tildeT$ on the alphabet  
$\Tvar{\ddot{\Alphabet}}=\Tvar{\Alphabet}\cup \set{\textcolor{\colorforyes}{yes}}\cup\set{(\textcolor{\colorforno}{no},\Tvar{u})\mid \Tvar{u}\in \T}$ induced by $\T$ such that 
$\tildevar{A}\subseteq[\tildeT]$ belongs to $D_{n}(\mathbf{\Pi}_{k}^0)$.\end{lemma}

\prooflem{questiontreefordowngrading}  Take a $\supseteq$-decreasing sequence 
$\Tvar{G_0}\supseteq \Tvar{G_1}\supseteq \ldots\supseteq \Tvar{G_{n-1}}\supseteq \Tvar{G_{n}}=\Tvar{\emptyset}$ of  $\mathbf{\Pi}_{k+1}^0$ subsets of $[\T]$ such that 
$\Tvar{A}=\bigcup_{j<n,~j\text{ even}}~\Tvar{G_j}\smallsetminus \Tvar{G_{j+1}}$.\medskip

For each integer $j< n$, there exist countably many closed subsets of $[\T]$ such that 

\begin{itemize}
  \item when $k$ is even, 
% \[
%  \Tvar{G_j}=\bigcap_{i_{0}<\omega}\,\bigcup_{i_{1}<\omega}\ldots\bigcap_{i_{k-1}<\omega}\,\bigcup_{i_k<\omega}[\Tvar{T_{(j,i_0,i_1,i_2,\ldots,i_k)}}]
%  \]
%   new:
 \[
  \Tvar{G_j}=\bigcap_{i_{0}<\omega}\,\bigcup_{i_{1}<\omega}\ldots\,\bigcap_{i_{k-2}<\omega}\bigcup_{i_{k-1}<\omega}[\Tvar{T_{(j,i_0,i_1,i_2,\ldots,i_{k-1})}}]
  \]

  \item when $k$ is odd, 
%  \[
%  \Tvar{G_j}=\bigcap_{i_0<\omega}\,\bigcup_{i_1<\omega}\ldots\bigcup_{i_{k-1}<\omega}\,\bigcap_{i_k<\omega}\big([\T]\smallsetminus[\Tvar{T_{(j,i_0,i_1,\ldots,i_k)}}]\big).
%  \]
%  new:
 \[
  \Tvar{G_j}=\bigcap_{i_0<\omega}\,\bigcup_{i_1<\omega}\ldots\bigcup_{i_{k-2}<\omega}\,\bigcap_{i_{k-1}<\omega}\big([\T]\smallsetminus[\Tvar{T_{(j,i_0,i_1,\ldots,i_{k-1})}}]\big).
  \]

\end{itemize}
We let $\code{\phantom{a}}:\set{0, 1,\ldots,n-1}\times \N^{k} \xrightarrow{\textit{bij.}}\N$ be any bijection, and $\codeinv{\ }$
 its inverse, so we have 
 \[
 \code{j,i_0,i_1,\ldots,i_{k-1}}=m\text{ ~~if only if~~ }\codeinv{m}=(j,i_0,i_1,\ldots,i_{k-1}).
 \]
We  let also $\Tvar{Q_m}=\Tvar{T_{\codeinv{m}}}$, and finally the \emph{Question-tree} $\tildeT$ be the one induced by both $\T$ and $(\Tvar{Q_m})_{m<\omega}$. We let  $\tildevar{A}\subseteq[\tildeT]$ be the set of all infinite branches of $\tildeT$ whose main run belongs to  $\Tvar{A}$. For each integer $j\leq n$, we let 
$\tildevar{G_j}\subseteq[\tildeT]$ be the set of all infinite branches of $\tildeT$ whose main run belongs to  $\Tvar{G_j}$.
We have

\begin{itemize}
  \item 
  when $k$ is even:\medskip
  
 \centerline{$\begin{array}{lll}
   \tildevar{x}\in\tildevar{G_j}&\iff&\Tvar{x}\in \Tvar{G_j}\\
&\iff&\Tvar{x}\in\bigcap_{i_0<\omega}\bigcup_{i_1<\omega}\ldots\bigcap_{i_{k-2}<\omega}\bigcup_{i_{k-1}<\omega}[\Tvar{T_{(j,i_0,i_1,\ldots,i_{k-1})}}]\\
&\iff&\forall_{i_0<\omega}\;\exists_{i_1<\omega}\ldots\forall_{i_{k-2}<\omega}\;\exists_{i_{k-1}<\omega}~~\Tvar{x}\in[\Tvar{T_{(j,i_0,i_1,\ldots,i_{k-1})}}
]\\
&\iff&\underbrace{\forall_{i_0<\omega}\;\exists_{i_1<\omega}\ldots\forall_{i_{k-2}<\omega}\;\underbrace{\exists_{i_{k-1}<\omega}~~\tildevar{x}(2\cdot\code{j,i_0,i_1,\ldots,i_{k-1}}+1)=\textcolor{\colorforyes}{yes}}_{\mathbf{\Sigma}^{0}_{1}}}_{\mathbf{\Pi}^0_{k}}
\end{array}$}\medskip

%OLD:
%
% when $k$ is even, we have 
% \[ \begin{array}{lll}
%   \tildevar{x}\in\tildevar{G_j}&\iff&\Tvar{x}\in \Tvar{G_j}\\
%&\iff&\Tvar{x}\in\bigcap_{i_0<\omega}\bigcup_{i_1<\omega}\ldots\bigcap_{i_{k-1}<\omega}\bigcup_{i_k<\omega}[\Tvar{T_{(j,i_0,i_1,\ldots,i_{n})}}]\\
%&\iff&\forall_{i_0<\omega}\exists_{i_1<\omega}\ldots\forall_{i_{k-1}<\omega}\exists_{i_k<\omega}~~\Tvar{x}\in[\Tvar{T_{(j,i_0,i_1,\ldots,i_{n})}}]\\
%&\iff&\underbrace{\forall_{i_0<\omega}\exists_{i_1<\omega}\ldots\forall_{i_{k-1}<\omega}\underbrace{\exists_{i_k<\omega}~~\tildevar{x}(2\cdot\code{j,i_0,i_1,\ldots,i_{n}}+1)=\textcolor{\colorforyes}{yes}}_{\mathbf{\Sigma}^{0}_{1}}}_{\mathbf{\Pi}^0_{k+1}}
%\end{array}
%\]
%

  \item when $k$ is odd:\medskip
  
\centerline{$\begin{array}{lll}
   \tildevar{x}\in\tildevar{G_j}&\iff&\Tvar{x}\in \Tvar{G_j}\\
&\iff&\Tvar{x}\in\bigcap_{i_0<\omega}\bigcup_{i_1<\omega}\ldots\bigcup_{i_{k-2}<\omega}\bigcap_{i_{k-1}<\omega}\big([\T]\smallsetminus[\Tvar{T_{(j,i_0,i_1,\ldots,i_{k-1})}}]\big)\\
&\iff&\forall_{i_0<\omega}\;\exists_{i_1<\omega}\ldots\exists_{i_{k-2}<\omega}\;\forall_{i_{k-1}<\omega}~~\Tvar{x}\notin[\Tvar{T_{(j,i_0,i_1,\ldots,i_{k-1})}}]\\
&\iff&\underbrace{\forall_{i_0<\omega}\;\exists_{i_1<\omega}\ldots\exists_{i_{k-2}<\omega}\;\underbrace{\forall_{i_{k-1}<\omega}~~\tildevar{x}(2\cdot\code{j,i_0,i_1,\ldots,i_{k-1}}+1)\neq\textcolor{\colorforyes}{yes}}_{\mathbf{\Pi}^{0}_{1}}}_{\mathbf{\Pi}^0_{k}}.
\end{array}$}\medskip

%OLD
%
%when $k$ is odd, we have 
%  \[ \begin{array}{lll}
%   \tildevar{x}\in\tildevar{G_j}&\iff&\Tvar{x}\in \Tvar{G_j}\\
%&\iff&\Tvar{x}\in\bigcap_{i_0<\omega}\bigcup_{i_1<\omega}\ldots\bigcup_{i_{k-1}<\omega}\bigcap_{i_k<\omega}\big([\T]\smallsetminus[\Tvar{T_{(j,i_0,i_1,\ldots,i_{n})}}]\big)\\
%&\iff&\forall_{i_0<\omega}\exists_{i_1<\omega}\ldots\exists_{i_{k-1}<\omega}\forall_{i_k<\omega}~~\Tvar{x}\notin[\Tvar{T_{(j,i_0,i_1,\ldots,i_{n})}}]\\
%&\iff&\underbrace{\forall_{i_0<\omega}\exists_{i_1<\omega}\ldots\exists_{i_{k-1}<\omega}\underbrace{\forall_{i_k<\omega}~~\tildevar{x}(2\cdot\code{j,i_0,i_1,\ldots,i_{n}}+1)\neq\textcolor{\colorforyes}{yes}}_{\mathbf{\Pi}^{0}_{1}}}_{\mathbf{\Pi}^0_{k+1}}.
%\end{array}
%\]

\end{itemize}
Thus, regardless of the parity of $k$,  each set $\tildevar{G_j}$ turns out to be some $\mathbf{\Pi}^0_{k}$-subset of $\tildevar{T}$. Since we also have 
both $\tildevar{G_0}\supseteq \tildevar{G_1}\supseteq \ldots\supseteq \tildevar{G}_{\Tvar{n-1}}$ and  $\tildevar{A}=\bigcup_{j<n,~j\text{ even}}~\tildevar{G_j}\smallsetminus \Tvar{\tildevar{G}_{j+1}}$, we then have  $\tildevar{A}\subseteq[\tildeT]$ belongs to $D_{n}(\mathbf{\Pi}_{k}^0)$. 
\boxlem{questiontreefordowngrading}

\begin{lemma} \label{fromquestiontreetotilde} Let $\Alphabet$ be any non-empty finite alphabet, 
$\T$ be any non-empty pruned tree on $\Alphabet$, $\Tvar{A}\subseteq[\T]$, and 
$(\Tvar{Q_{n}})_{n<\omega}$ be any sequence  of subtrees of $\T$. We also let $\tildeT$ be the (non-empty pruned) Question-tree $\tildeT$ on the alphabet 
$\Tvar{\ddot{\Alphabet}}=\Tvar{\Alphabet}\cup \set{\textcolor{\colorforyes}{yes}}\cup\set{(\textcolor{\colorforno}{no},\Tvar{u})\mid \Tvar{u}\in \T}$  induced by both $\T$ and 
$(\Tvar{Q_{n}})_{n<\omega}$ and  $\tildevar{A}\subseteq[\tildeT]$ be the set of all infinite branches of 
$\tildeT$ whose main run belongs to  $\Tvar{A}$. For any $B\subseteq\Alphabet^{\leq\omega}$, we have\medskip

\centerline{$\tildevar{A}\winf B^{\blank}\Longrightarrow\Tvar{A}\winf (B^{\sim})^{\blank}.$}
\end{lemma}

\prooflem{fromquestiontreetotilde} Let $\tildevar{\sigma}$ be some winning strategy for Player $\two$ in the game $\GWadge{\tildevar{A}}{B^{\blank}}$
 We build a  winning strategy $\Tvar{\sigma}$ for Player $\two$ in the game $\GWadge{\Tvar{A}}{(B^{\sim})^{\blank}}$. The idea in designing $\Tvar{\sigma}$ is to follow the strategy 
 $\tildevar{\sigma}$ where $\one$'s main run is exactly the run played by $\one$ in $\GWadge{\Tvar{A}}{(B^{\sim})^{\blank}}$, and as for  auxiliary moves,  Player $\two$  simulates  that Player $\one$ always answers \textcolor{\colorforyes}{yes}, thus committing to remain in each of the different trees submitted to her.\medskip
 
 In the game $\GWadge{\Tvar{A}}{(B^{\sim})^{\blank}}$, in case  Player   $\one$ reaches a position 
$\Tvar{u}$ which contradicts a previous commitment (some $\avar{k}=\textcolor{\colorforyes}{yes}$) then this means that $\one$'s play in  $\GWadge{\Tvar{A}}{(B^{\sim})^{\blank}}$ just exited the tree 
$\Tvar{Q_k}$. So, Player $\two$ needs to go  back and modify the answer 
$\avar{k}=\textcolor{\colorforyes}{yes}$ by turning it into  
$\avar{k}=($\textcolor{\colorforno}{no},$\Tvar{u})$ which also forces $\tildevar{\sigma}$ to make Player $\two$ play some series of \quq{$\tla$} in order to go back to the position that she was in right before 
$\two$ simulated $\avar{k}=\textcolor{\colorforyes}{yes}$ in the game 
$\GWadge{\tildevar{A}}{B^{\blank}}$.\medskip 
 
 Recall  that the occurrences of the letter \quq{\blank} do not count when deciding on how many 
\quq{$\tla$} needs to be played in order to reach a previously established position (up to the removal of all \quq{\blank}). Notice also that when some $\avar{k}=\textcolor{\colorforyes}{yes}$ is turned into   some $\avar{k}=($\textcolor{\colorforno}{no},$\Tvar{u})$, the value of $\avar{k}$ will then never change again because $\one$'s move being already located outside $\Tvar{Q_k}$, will always remain that way in the future.\boxlem{fromquestiontreetotilde}

\begin{lemma} \label{Acompleteandtilde} Let $\Alphabet$ be any non-empty finite alphabet, 
$B\subseteq\Alphabet^{\leq\omega}$, and $1\leq n,k<\omega$. If the set 
$B^{\blank}\subseteq(\Alphabet\cup\set{\blank})^{\omega}$  is $D_{n}(\mathbf{\Pi}_{k}^0)$-complete, then $(B^\sim)^{\blank}\subseteq(\Alphabet\cup\set{\tla,\blank})^{\omega}$ is 
$D_{n}(\mathbf{\Pi}_{k+1}^0)$-complete.\end{lemma}

\prooflem{Acompleteandtilde} By Lemma \ref{Btildeinclass}, $(B^\sim)^{\blank}$  belongs to $D_{n}(\mathbf{\Pi}_{k+1}^0)$. For completeness, notice that 
 by Lemma \ref{questiontreefordowngrading}, for all $\Tvar{A}\subseteq[\T]$  belonging to  $D_{n}(\mathbf{\Pi}_{k+1}^0)$,  there exists $\tildeT$ such that $\tildevar{A}\subseteq[\tildeT]$ belongs to $D_{n}(\mathbf{\Pi}_{k}^0)$. Therefore, the inequality $\tildevar{A}\winf B^{\blank}$ holds
  since  $B^{\blank}$  is $D_{n}(\mathbf{\Pi}_{k}^0)$-complete. By Lemma
 \ref{fromquestiontreetotilde}, $\Tvar{A}\winf (B^{\sim})^{\blank}$, which shows that $(B^\sim)^{\blank}$  is $D_{n}(\mathbf{\Pi}_{k+1}^0)$-complete.\boxlem{Acompleteandtilde}

%%%%%%%%%%%%%%%%%%%%%%%%%%%%%%%%%%%%%%%%%%%%%%%%%%%%%%%%%%%%%%%%%%%%%%%%%%%%%%%%%%%%
%%%%%%%%%%%%%%%%%%%%%%%%%%%%%%%%%%%%%%%%%%%%%%%%%%%%%%%%%%%%%%%%%%%%%%%%%%%%%%%%%%%%
%%%%%%%%%%%%%%%%%%%%%%%%%%%%%%%%%%%%%%%%%%%%%%%%%%%%%%%%%%%%%%%%%%%%%%%%%%%%%%%%%%%%
%%%%%%%%%%%%%%%%%%%%%%%%%%%%%%%%%%%%%%%%%%%%%%%%%%%%%%%%%%%%%%%%%%%%%%%%%%%%%%%%%%%%
 %%%%%%%%%%%%%% %%%%%%%%%%%%%%%%%%%%          preliminarylemmaforcorollaries            %%%%%%%%%%%%%%%%% %%%%%%%%
%%%%%%%%%%%%%%%%%%%%%%%%%%%%%%%%%%%%%%%%%%%%%%%%%%%%%%%%%%%%%%%%%%%%%%%%%%%%%%%%%%%%
%%%%%%%%%%%%%%%%%%%%%%%%%%%%%%%%%%%%%%%%%%%%%%%%%%%%%%%%%%%%%%%%%%%%%%%%%%%%%%%%%%%%
%%%%%%%%%%%%%%%%%%%%%%%%%%%%%%%%%%%%%%%%%%%%%%%%%%%%%%%%%%%%%%%%%%%%%%%%%%%%%%%%%%%%
%%%%%%%%%%%%%%%%%%%%%%%%%%%%%%%%%%%%%%%%%%%%%%%%%%%%%%%%%%%%%%%%%%%%%%%%%%%%%%%%%%%%
%%%%%%%%%%%%%%%%%%%%%%%%%%%%%%%%%%%%%%%%%%%%%%%%%%%%%%%%%%%%%%%%%%%%%%%%%%%%%%%%%%%%

      \begin{lemma} \label{preliminarylemmaforcorollaries} Let  $\T$ be any non-empty pruned tree on some non-empty countable alphabet $\Sigma$, and $\blank\notin\Alphabet$. Let   $\Tvar{T}^{\blank}=\setbig{\Tvar{s}\in (\Alphabet\cup\set{\blank})^{<\omega}\mid\Tvar{s}_{[\,/\blank]}\in \T}$ denote the set of all finite sequences $\Tvar{s}$  that belong to $\Tvar{T}$ once all occurences of the letter $\blank$ have been erased.
      
\vfill\eject
      
       Let $\Tvar{A}\subseteq[\T]$,  and $
\Tvar{A}^{\blank}=\setbig{ 
\Tvar{x}\in [\Tvar{T}^{\blank}]\mid \Tvar{x}_{[\,/\blank]}\in \Tvar{A}
}
$
 denote the set of all the infinite branches of $\Tvar{T}^{\blank}$ that belong to $\Tvar{A}$ once all occurences of the letter $\blank$ have been removed.
  Let $\mathbf{\Gamma}$ be any Wadge class closed under intersections with $\mathbf{\Pi}_{2}^0$-sets.
\begin{center}
      If $\Tvar{A}$  is $\mathbf{\Gamma}$-complete, then $\Tvar{A}^{\blank}$  is $\mathbf{\Gamma}$-complete.
\end{center}
            \end{lemma}

\prooflem{preliminarylemmaforcorollaries} Clearly, $\Tvar{A}\winf \Tvar{A}^{\blank}$ holds which proves that $\Tvar{A}^{\blank}$ is $\mathbf{\Gamma}$-hard. To show that  $\Tvar{A}^{\blank}$ belongs to $\mathbf{\Gamma}$, we let   $[{\Tvar{T}^{\blank}}]^{\proptoinverse}=\setbig{\Tvar{x}\in[\T^{\blank}]\mid \forall m\exists n>m~~\Tvar{x}(n)\neq \blank}.
$ denote the subset of $[{\Tvar{T}^{\blank}}]$ formed 
 of all the infinite branches of ${\Tvar{T}^{\blank}}$ containing infinitely many letters different from the letter $\quq{\blank}$. 
 Clearly, $[{\Tvar{T}^{\blank}}]^{\proptoinverse}$ is a $\mathbf{\Pi}_{2}^0$-subset of $[{\Tvar{T}^{\blank}}]$, and the mapping 
\[\begin{array}{cccc}
 f:&[{\Tvar{T}^{\blank}}]^{\proptoinverse}&\xrightarrow{cont.}&[\Tvar{T}]     \\
      &   \Tvar{x} & \mapsto &\Tvar{x}_{[\,/\blank]}\\
\end{array}
\]
 is continuous, so that we have $f^{-1}[\Tvar{A}]$ is a $\mathbf{\Gamma}$-subset of $[{\Tvar{T}^{\blank}}]^{\proptoinverse}$. Since $f^{-1}[\Tvar{A}]$ is no other than $\Tvar{A}^{\blank}$, we obtain that $\Tvar{A}^{\blank}$ is some  
  $\mathbf{\Gamma}$-subset of $[{\Tvar{T}^{\blank}}]^{\proptoinverse}$. Being the intersection of some $\mathbf{\Gamma}$-subset of $[{\Tvar{T}^{\blank}}]$ and a  $\mathbf{\Pi}_{2}^0$-subset of  $[{\Tvar{T}^{\blank}}]$, the set $\Tvar{A}^{\blank}$ belongs to $\mathbf{\Gamma}$.
\boxlem{preliminarylemmaforcorollaries}  
    
 %%%%%%%%%%%%%% %%%%%%%%%%%%%%%%%%%%          preliminarylemmaforcorollaries            %%%%%%%%%%%%%%%%% %%%%%%%%

      \begin{corollary} \label{preliminarycorollary} Let  $\T$ be any non-empty pruned tree on some non-empty countable alphabet $\Sigma$,  $\Tvar{A}\subseteq[\T]$, and $\blank\notin\Alphabet$. Let 
$n\geq 1$ and $k\geq 2$ be integers.

      \begin{enumerate}
  \item \label{preliminarycorollarya} If $\Tvar{A}\subseteq[{\Tvar{T}}]$  is $D_{n}(\mathbf{\Pi}_{k}^0)$-complete, then $\Tvar{A}^{\blank}\subseteq[{\Tvar{T}^{\blank}}]$  is $D_{n}(\mathbf{\Pi}_{k}^0)$-complete.

  \item \label{preliminarycorollaryb} If $\Tvar{A}\subseteq[{\Tvar{T}}]$ is $\checked{D}_{n}(\mathbf{\Pi}_{k+1}^0)$-complete, then $\Tvar{A}^{\blank}\subseteq[{\Tvar{T}^{\blank}}]$ is $\checked{D}_{n}(\mathbf{\Pi}_{k+1}^0)$-complete.
  \end{enumerate}
      
   \end{corollary}

\proofcor{preliminarycorollary} In order to obtain the result, it is enough to show that these classes are closed under intersections with $\mathbf{\Pi}_{2}^0$ sets and apply  Lemma \ref{preliminarylemmaforcorollaries}
\begin{enumerate}
  \item We show that the class $D_{n}(\mathbf{\Pi}_{k}^0)$ is closed under intersections with 
$\mathbf{\Pi}_{2}^0$-sets. In order to prove this, consider any $D_{n}(\mathbf{\Pi}_{k}^0)$-set 
$\Tvar{A}\subseteq[\T]$. Since $\Tvar{A}\in D_{n}(\mathbf{\Pi}_{k}^0)$ we can find a $\supseteq$-decreasing sequence 
$\Tvar{G_0}\supseteq \Tvar{G_1}\supseteq \ldots\supseteq \Tvar{G_{n-1}}\supseteq \Tvar{G_{n}}=\Tvar{\emptyset}$ of $\mathbf{\Pi}_{k}^0$ subsets of $[\T]$ such that 
$\Tvar{A}=\bigcup_{j\text{ even }<n}~\Tvar{G_j}\smallsetminus \Tvar{G_{j+1}}$. Let 
$\Tvar{B}\subseteq[\T]$ be any $\mathbf{\Pi}_{2}^0$-set. Since $k\geq 2$, the class 
$\mathbf{\Pi}_{k}^0$ is closed under intersections with $\mathbf{\Pi}_{2}^0$-sets, so we have\medskip
  
\scalebox{0.70}{\centerline{$\Tvar{A}\cap\Tvar{B} 
 =\parenthesesBig{\bigcup_{j\text{ even }<n}~\Tvar{G_j}\smallsetminus \Tvar{G_{j+1}}}\cap\Tvar{B}
 %
% =\bigcup_{j\text{ even}<n}~\parenthesesbig{\parentheses{\Tvar{G_j}\smallsetminus \Tvar{G_{j+1}}}\cap\Tvar{B}}
  %
 =\bigcup_{j\text{ even }<n}~\parenthesesbig{\underbrace{\Tvar{G_j}\cap\Tvar{B}}_{\mathbf{\Pi}_{k}^0}}\smallsetminus \parenthesesbig{\underbrace{\Tvar{G_{j+1}\cap\Tvar{B}}}_{\mathbf{\Pi}_{k}^0}}.$}}\medskip
 
\noindent Notice that $\Tvar{G_{0}}\cap\Tvar{B}\supseteq \Tvar{G_1}\cap\Tvar{B}\supseteq \Tvar{G_2}\cap\Tvar{B}\supseteq \ldots$ forms a $\supseteq$-decreasing sequence , and 
  since $k\geq 2$ holds, all sets  $\Tvar{G_j}\cap\Tvar{B}$ (any $0\leq j\leq n$) belong to $\mathbf{\Pi}_{k}^0$, which shows that $\Tvar{A}\cap\Tvar{B}$  belongs to $\checked{D}_{n}(\mathbf{\Pi}_{k}^0)$.

    \item   We show that the class $\checked{D}_{n}(\mathbf{\Pi}_{k+1}^0)$ is closed under intersections with $\mathbf{\Pi}_{2}^0$-sets. For this, notice that any set in $\checked{D}_{n}(\mathbf{\Pi}_{k+1}^0)$ is of the form $[\T]\smallsetminus \Tvar{A}$ with  $\Tvar{A}\subseteq[\T]$ belonging to 
$D_{n}(\mathbf{\Pi}_{k+1}^0)$. So, there exists a $\supseteq$-decreasing sequence 
$\Tvar{G_0}\supseteq \Tvar{G_1}\supseteq \ldots\supseteq \Tvar{G_{n-1}}\supseteq \Tvar{G_{n}}=\Tvar{\emptyset}$ of $\mathbf{\Pi}_{k+1}^0$ subsets of $[\T]$ such that
$\Tvar{A}=\bigcup_{j\text{ even }<n}~\Tvar{G_j}\smallsetminus \Tvar{G_{j+1}}$. Let 
$\Tvar{B}\subseteq[\T]$ be any $\mathbf{\Pi}_{2}^0$-set. We have\medskip

\scalebox{0.70}{\centerline{$\begin{array}{ccl}
  \parentheses{[\T]\smallsetminus \Tvar{A}}\cap\Tvar{B} &=&\parenthesesbig{
   \parentheses{[\T]\smallsetminus \Tvar{G_{0}}}
   \cup
      \parentheses{\Tvar{G_1}\smallsetminus \Tvar{G_{2}}}
         \cup\ldots
  }\cap\Tvar{B}\\
 &=&
 \parenthesesBig{
   \parenthesesbig{\parentheses{[\T]\smallsetminus \Tvar{G_{0}}}\cap\Tvar{B}}
   \cup
      \parenthesesbig{\parentheses{\Tvar{G_1}\smallsetminus \Tvar{G_{2}}}\cap\Tvar{B}}
         \cup\ldots
  }\\
  &=&
 \parenthesesBig{
 \parenthesesBig{[\T]\smallsetminus \parenthesesBig{\underbrace{\Tvar{G_{0}}\cup\Tvar{B}^{\complement}}_{\mathbf{\Pi}_{k+1}^0}}}
   \cup
      \parenthesesbig{\parenthesesbig{\underbrace{\Tvar{G_1}\cap\Tvar{B}}_{\mathbf{\Pi}_{k+1}^0}}\smallsetminus \parenthesesbig{\underbrace{\Tvar{G_{2}\cap\Tvar{B}}}_{\mathbf{\Pi}_{k+1}^0}}}
         \cup\ldots
  }\\
 \end{array}$}}\medskip
 
\noindent Notice that $\Tvar{G_{0}}\cup\Tvar{B}^{\complement}\supseteq \Tvar{G_1}\cap\Tvar{B}\supseteq \Tvar{G_2}\cap\Tvar{B}\supseteq \ldots$ forms a $\supseteq$-decreasing sequence , and 
  since $k+1\geq 3$ holds, the set   $\Tvar{G_{0}}\cup\Tvar{B}^{\complement}$ together with all $\Tvar{G_j}\cap\Tvar{B}$ (any $1\leq j\leq n$),  all belong to $\mathbf{\Pi}_{k+1}^0$, which shows that $\parentheses{[\T]\smallsetminus \Tvar{A}}\cap\Tvar{B}$  belongs to $\checked{D}_{n}(\mathbf{\Pi}_{k+1}^0)$.
\end{enumerate}
This finishes the proof.\boxcor{preliminarycorollary}  

\vfill\eject

We recall Theorem \ref{thm:thedupthree}, which we are now ready to prove.    

\thedupthree*

%
%stated that, given integers $n\geq 1$ and $k\geq 2$, 
% if $A\subseteq\Alphabet^{\omega}$ is $D_{n}(\mathbf{\Pi}_{k}^0)$-complete, then $A^\sim$ is $D_{n}(\mathbf{\Pi}_{k+1}^0)$-complete.
%
%
%
%AAAAAAAA

%%%
%%%%First appearance
%%%\begin{theorem}
%%%  \label{thm:a}
%%%  \thmtext
%%%\end{theorem}
%%%
%%%%Second appearance
%%%\begin{theorem}
%%%  \thmtext
%%%\end{theorem}

%\begin{oneshot}{Theorem~\ref{thm:thedupthree}} 
%Let $\Alphabet$ be any finite alphabet, $A\subseteq\Alphabet^{\omega}$, and 
%$n\geq 1$, $k\geq 2$ be integers. If $A$ is $D_{n}(\mathbf{\Pi}_{k}^0)$-complete, then $A^\sim$ is $D_{n}(\mathbf{\Pi}_{k+1}^0)$-complete.\end{oneshot}
%

\proofthm{thm:thedupthree} \label{thispageproofthedupthree}Since  $A$  is $D_{n}(\mathbf{\Pi}_{k}^0)$-complete, by Corollary \ref{preliminarycorollary} (\ref{preliminarycorollarya}), the set $A^{\blank}$  is $D_{n}(\mathbf{\Pi}_{k}^0)$-complete as well, and by Lemma \ref{Acompleteandtilde}, $(A^\sim)^{\blank}$  is $D_{n}(\mathbf{\Pi}_{k+1}^0)$-complete. To conclude, we notice that $(A^\sim)^{\blank}\wequ A^\sim$, since, while in charge of $A^\sim$ playing some letter from $\Alphabet$ and immediately erasing it is equivalent to playing the letter \quq{\blank} while  in charge of $(A^\sim)^{\blank}$.
\boxthm{thm:thedupthree}

    %%%%%%%%%%%%%%%%%%%%%%%%%%%%%%%%%%%%%%%%%%%%%%%%%%%%%%%%%%%%%%%%%%%%%%%%%%%%%%%%%%%%%%

 We now turn our attention to the automaton $\checked{\mathcal{A}}_{{n}^{\blank}}$ as shown in Figure \ref{therealautomatonBnd}.
 
\begin{figure}[h]
\begin{center}
\tikzset{
%->,  % makes the edges directed
>=stealth', % makes the arrow heads bold
node distance=20mm, % specifies the minimum distance between two nodes. Change if necessary.
every state/.style={thick, fill=gray!14}, % sets the properties for each ?state? node
initial text=$ $, % sets the text that appears on the start arrow
}
\scalebox{0.85}{\begin{tikzpicture}%[shorten >=1pt,node distance=3cm,on grid]
\node[state, accepting, initial] (q4) {$q_{2l}$};
\node[left of=q4] (def) {$\checked{\mathcal{A}}_{2l^{\blank}}:=$};
\node[state, right of=q4] (q3) {${\scriptstyle q_{2l-1}}$};
\node[rectangle, right of=q3, ] (q3bis) {};
\node[rectangle, right of=q3bis, ] (q3ter) {};
\node[state, accepting, right of=q3ter] (q2) {$q_2$};
\node[state, right of=q2] (q1) {$q_1$};
\node[state, accepting, right of=q1] (q0) {$q_0$};
\draw  
(q4) edge[loop above] node{$\mathbbold{0}$,\blank} (q4)
(q3) edge[loop above] node{$\mathbbold{0}$,\blank} (q3)
(q2) edge[loop above] node{$\mathbbold{0}$,\blank} (q2)
(q1) edge[loop above] node{$\mathbbold{0}$,\blank} (q1)
(q0) edge[loop above] node{$\mathbbold{0},\mathbbold{1}$,\blank} (q0)
(q4) edge[->,above] node{$\mathbbold{1}$} (q3)
(q3) edge[->,above] node{$\mathbbold{1}$} (q3bis)
(q3ter) edge[->,above] node{$\mathbbold{1}$} (q2)
(q2) edge[->,above] node{$\mathbbold{1}$} (q1)
(q1) edge[->,above] node{$\mathbbold{1}$} (q0)
;
\draw[dotted]  (q3bis) -- (q3ter);
\end{tikzpicture}}
\tikzset{
%->,  % makes the edges directed
>=stealth', % makes the arrow heads bold
node distance=20mm, % specifies the minimum distance between two nodes. Change if necessary.
every state/.style={thick, fill=gray!14}, % sets the properties for each ?state? node
initial text=$ $, % sets the text that appears on the start arrow
}
\scalebox{0.85}{\begin{tikzpicture}%[shorten >=1pt,node distance=3cm,on grid]
\node[state, initial] (q4) {${\scriptstyle q_{2l+1}}$};
\node[left of=q4] (def) {$\checked{\mathcal{A}}_{2l+1^{\blank}}:=~~~~~$};
\node[state, accepting, right of=q4] (q3) {$q_{2l}$};
\node[rectangle, right of=q3, ] (q3bis) {};
\node[rectangle, right of=q3bis, ] (q3ter) {};
\node[state, accepting, right of=q3ter] (q2) {$q_2$};
\node[state, right of=q2] (q1) {$q_1$};
\node[state, accepting, right of=q1] (q0) {$q_0$};
\draw  
(q4) edge[loop above] node{$\mathbbold{0}$,\blank} (q4)
(q3) edge[loop above] node{$\mathbbold{0}$,\blank} (q3)
(q2) edge[loop above] node{$\mathbbold{0}$,\blank} (q2)
(q1) edge[loop above] node{$\mathbbold{0}$,\blank} (q1)
(q0) edge[loop above] node{$\mathbbold{0},\mathbbold{1}$,\blank} (q0)
(q4) edge[->,above] node{$\mathbbold{1}$} (q3)
(q3) edge[->,above] node{$\mathbbold{1}$} (q3bis)
(q3ter) edge[->,above] node{$\mathbbold{1}$} (q2)
(q2) edge[->,above] node{$\mathbbold{1}$} (q1)
(q1) edge[->,above] node{$\mathbbold{1}$} (q0)
;
\draw[dotted]  (q3bis) -- (q3ter);
\end{tikzpicture}}
\end{center}
\caption{\small{The automaton $\checked{\mathcal{A}}_{{n}^{\blank}}$\label{therealautomatonBnd}, depending on whether $n$ is even ($n=2l$) or odd ($n=2l+1$)}}
\end{figure}

 %%%%%%%%%%%%%%%%% %%%%%%%%%%%%%%%%% %%%%%%%%%%%%%%%%% %%%%%%%%%%%%%%%%% %%%%%%%%%%%%%%%%%%
 %%%%%%%%%%%%%%%%% %%%%%%%%%%%%%%%%% %%%%%%%%%%%%%%%%% %%%%%%%%%%%%%%%%% %%%%%%%%%%%%%%%%%%
 %%%%%%%%%%%%%%%%% %%%%%%%%%%%%%%%%% %%%%%%%%%%%%%%%%% %%%%%%%%%%%%%%%%% %%%%%%%%%%%%%%%%%%
 %%%%%%%%%%%%%%%%% %%%%%%%%%%%%%%%%% %%%%%%%%%%%%%%%%% %%%%%%%%%%%%%%%%% %%%%%%%%%%%%%%%%%%
 %%%%%%%%%%%%%%%%% %%%%%%%%%%%%%%%%% %%%%%%%%%%%%%%%%% %%%%%%%%%%%%%%%%% %%%%%%%%%%%%%%%%%%

\begin{proposition} \label{automatonBnddnclosedcompletecheck} Let $n\geq 1$ be any integer, 
$\checked{\mathcal{A}}_{{n}^{\blank}}$ be the automaton described in Figure \ref{therealautomatonBnd}, and $\#_\mathbbold{1}(x)$ represent the number of occurrences of the letter $\quq{\mathbbold{1}}$ the infinite word $x$. The $\omega$-language $\Linfinite(\checked{\mathcal{A}}_{{n}^{\blank}})=\setbig{x\in \set{\mathbbold{0},\mathbbold{1},\blank}^{\omega}\mid \#_\mathbbold{1}(x)\geq n \textit{ or }\textit{parity}\big(\#_\mathbbold{1}(x)\big)=\textit{parity}(n)}$ is $\checked{D}_{n}(\mathbf{\Pi}_{1}^0)$-complete.\end{proposition}

\proofpro{automatonBnddnclosedcompletecheck} Clearly, we have $\Linfinite(\checked{\mathcal{A}}_{{n}^{\blank}})=\set{\mathbbold{0},\mathbbold{1},\blank}^{\omega}\smallsetminus \big(\Lboth(\mathcal{A}_{n})\big)^{\blank}$ where $\mathcal{A}_{n}$ is the automaton described in Figure \ref{therealautomatonAnplusonewithnoblanks}. By Lemma \ref{automatonAnddnclosedcomplete}, the language $\big(\Lboth(\mathcal{A}_{n})\big)^{\blank}\subseteq\set{\mathbbold{0},\mathbbold{1},\blank}^{\omega}$ is $D_{n}(\mathbf{\Pi}_{1}^0)$-complete, henceforth 
$\Linfinite(\checked{\mathcal{A}}_{{n}^{\blank}})$ is $\checked{D}_{n}(\mathbf{\Pi}_{1}^0)$-complete.
\boxpro{automatonBnddnclosedcompletecheck}

    Our next goal is to prove the following result.
   
\begin{restatable}{lemma}{lemmadupfour} \label{thm:lemmadupfour}\label{lemmadup4} Let 
$\Alphabet$ be any finite alphabet, $A\subseteq\Alphabet^{\omega}$, and $n\geq 1$ be any integer. 
\begin{center}
If $A$ is $\checked{D}_{n}(\mathbf{\Pi}_{2}^0)$-complete, then $A^\sim\subseteq(\Alphabet\cup\set{\tla})^{\omega}$ is $D_{n+1}(\mathbf{\Pi}_{3}^0)$-complete.
\end{center}
\end{restatable}

Prior to proving Lemma \ref{lemmadup4}, we need the  following preliminary result for which we introduce in Figure \ref{therealautomatonBndonfinitealso} the automaton $\checked{\mathcal{A}}_{n}$.

\begin{figure}[h]

\begin{center}
\tikzset{
%->,  % makes the edges directed
>=stealth', % makes the arrow heads bold
node distance=16mm, % specifies the minimum distance between two nodes. Change if necessary.
every state/.style={thick, fill=gray!10}, % sets the properties for each ?state? node
initial text=$ $, % sets the text that appears on the start arrow
}
\scalebox{0.85}{\begin{tikzpicture}%[shorten >=1pt,node distance=3cm,on grid]
\node[state, accepting, initial] (q4) {$q_{2l}$};
\node[left of=q4] (def) {$\checked{\mathcal{A}}_{2l}:=$};
\node[state, right of=q4] (q3) {${\scriptstyle  q_{2l-1}}$};
\node[rectangle, right of=q3, ] (q3bis) {};
\node[rectangle, right of=q3bis, ] (q3ter) {};
\node[state, accepting, right of=q3ter] (q2) {$q_2$};
\node[state, right of=q2] (q1) {$q_1$};
\node[state, accepting, right of=q1] (q0) {$q_0$};
\draw  
(q4) edge[loop above] node{$\mathbbold{0}$} (q4)
(q3) edge[loop above] node{$\mathbbold{0}$} (q3)
(q2) edge[loop above] node{$\mathbbold{0}$} (q2)
(q1) edge[loop above] node{$\mathbbold{0}$} (q1)
(q0) edge[loop above] node{$\mathbbold{0},\mathbbold{1}$} (q0)
(q4) edge[->,above] node{$\mathbbold{1}$} (q3)
(q3) edge[->,above] node{$\mathbbold{1}$} (q3bis)
(q3ter) edge[->,above] node{$\mathbbold{1}$} (q2)
(q2) edge[->,above] node{$\mathbbold{1}$} (q1)
(q1) edge[->,above] node{$\mathbbold{1}$} (q0)
;
\draw[dotted]  (q3bis) -- (q3ter);
\end{tikzpicture}}
\tikzset{
%->,  % makes the edges directed
>=stealth', % makes the arrow heads bold
node distance=16mm, % specifies the minimum distance between two nodes. Change if necessary.
every state/.style={thick, fill=gray!10}, % sets the properties for each ?state? node
initial text=$ $, % sets the text that appears on the start arrow
}
\scalebox{0.85}{\begin{tikzpicture}%[shorten >=1pt,node distance=3cm,on grid]
\node[state, initial] (q4) {${\scriptstyle q_{2l+1}}$};
\node[left of=q4] (def) {$\checked{\mathcal{A}}_{2l+1}:=~~~~~$};
\node[state, accepting, right of=q4] (q3) {$q_{2l}$};
\node[rectangle, right of=q3, ] (q3bis) {};
\node[rectangle, right of=q3bis, ] (q3ter) {};
\node[state, accepting, right of=q3ter] (q2) {$q_2$};
\node[state, right of=q2] (q1) {$q_1$};
\node[state, accepting, right of=q1] (q0) {$q_0$};
\draw  
(q4) edge[loop above] node{$\mathbbold{0}$} (q4)
(q3) edge[loop above] node{$\mathbbold{0}$} (q3)
(q2) edge[loop above] node{$\mathbbold{0}$} (q2)
(q1) edge[loop above] node{$\mathbbold{0}$} (q1)
(q0) edge[loop above] node{$\mathbbold{0},\mathbbold{1}$} (q0)
(q4) edge[->,above] node{$\mathbbold{1}$} (q3)
(q3) edge[->,above] node{$\mathbbold{1}$} (q3bis)
(q3ter) edge[->,above] node{$\mathbbold{1}$} (q2)
(q2) edge[->,above] node{$\mathbbold{1}$} (q1)
(q1) edge[->,above] node{$\mathbbold{1}$} (q0)
;
\draw[dotted]  (q3bis) -- (q3ter);
\end{tikzpicture}}
\end{center}
\caption{\small{The automaton $\checked{\mathcal{A}}_{n}$\label{therealautomatonBndonfinitealso}, depending on whether $n$ is even ($n=2l$) or odd ($n=2l+1$)}}
%\caption{The automaton $\checked{\mathcal{A}}_{2n+1}^{\blank}$\label{automatonB2nplusoned}}
\end{figure}
    
\begin{lemma} \label{preliminarylemmadup4} Let $\Alphabet$ be any finite alphabet, 
$A\subseteq\Alphabet^{\omega}$, and $n\geq 1$ be any integer. 
\begin{center}
If $A$ is $\checked{D}_{n}(\mathbf{\Pi}_{2}^0)$-complete, then $A^{\blank}\subseteq(\Alphabet\cup\set{\blank})^{\omega}$ is $D_{n+1}(\mathbf{\Pi}_{2}^0)$-complete.
\end{center}
\end{lemma}

\prooflem{preliminarylemmadup4}
%We write $\Lboth(\checked{\mathcal{A}}_{n})\subseteq\set{0,1}^{\leq\omega}$ for the language of finite and infinite words accepted by the automaton $\checked{\mathcal{A}}_{n}$ represented on Figure \ref{therealautomatonBndonfinitealso} with the following acceptance condition: a  word is accepted if after processing the input, the automaton either ends in an accepting state or visit infinitely often an (necessarily unique here!) accepting state.
 Note that the graph of the automaton $\checked{\mathcal{A}}_{n}$ in Figure \ref{therealautomatonBndonfinitealso}  is nothing but  the graph of the automaton 
$\checked{\mathcal{A}}_{n^{\blank}}$ in Figure \ref{therealautomatonBnd} deprived of its 
transitions labeled \quq{$\blank$} (which  only consist in trivial loops on each state). Since  
$x\in \Linfinite(\checked{\mathcal{A}}_{n^{\blank}})\iff 
x_{[\,/\blank]}\in \Lboth(\checked{\mathcal{A}}_{n})$ for all $x\in \set{\mathbbold{0},\mathbbold{1},\blank}^{\omega}$, we also immediately have 
$\parenthesesbig{\Lboth(\checked{\mathcal{A}}_{n})}^{\blank}=
\Linfinite(\checked{\mathcal{A}}_{n^{\blank}})$. 
 We let $\mathcal{A}_{n}$ be the automaton represented in Figure \ref{therealautomatonAnplusonewithnoblanks}.
 
\vfill\eject
 
  Note that the automaton $\mathcal{A}_{n}$ is obtained from   the automaton $\checked{\mathcal{A}}_{n}$ in Figure \ref{therealautomatonBndonfinitealso} by simply swapping the accepting states and the rejecting states. So,  we have both $\Lboth(\checked{\mathcal{A}}_{n})=\set{\mathbbold{0},\mathbbold{1}}^{\leq\omega}\smallsetminus\Lboth(\mathcal{A}_{n})$, and $\Linfinite(\mathcal{A}^{\blank}_{n})=\parenthesesbig{\Lboth(\mathcal{A}_{n})}^{\blank}$.
This gives us:
\[\begin{array}{ccl}
\parenthesesBig{{\Lboth(\checked{\mathcal{A}}_{n})\,}^{\sim}}^{\blank}&=&  
\parenthesesBig{\parenthesesbig{\set{\mathbbold{0},\mathbbold{1}}^{\leq\omega}\smallsetminus \Lboth(\mathcal{A}_{n})}^{\sim}}^{\blank} \\
&=&\parenthesesBig{\set{\mathbbold{0},\mathbbold{1}\tla}^{\leq\omega}\smallsetminus {\Lboth(\mathcal{A}_{n})\,}^{\sim}}^{\blank}\\
&=&\set{\mathbbold{0},\mathbbold{1}\tla,\blank}^{\omega}\smallsetminus \parenthesesbig{{\Lboth(\mathcal{A}_{n})\,}^{\sim}}^{\blank}\\
&=&\parenthesesBig{\parenthesesbig{{\Lboth(\mathcal{A}_{n})\,}^{\sim}}^{\blank}}^{\macroforcomplement}.\\
\end{array}
\]

By Lemma  \ref{automatonAnddnclosedcomplete}  the language $\big(\Lboth(\mathcal{A}_{n})\big)^{\blank}\subseteq\set{\mathbbold{0},\mathbbold{1},\blank}^{\omega}$ is $D_{n}(\mathbf{\Pi}_{1}^0)$-complete.
 By Lemma \ref{Acompleteandtilde}, we obtain that 
the language  $\parenthesesbig{{\Lboth(\mathcal{A}_{n})\,}^{\sim}}^{\blank}$ is $D_{n}(\mathbf{\Pi}_{2}^0)$-complete, 
which shows that the $\omega$-language 
$\parenthesesBig{\parenthesesbig{{\Lboth(\mathcal{A}_{n})\,}^{\sim}}^{\blank}}^{\macroforcomplement}=\parenthesesBig{{\Lboth(\checked{\mathcal{A}}_{n})\,}^{\sim}}^{\blank}$ is 
$\checked{D}_{n}(\mathbf{\Pi}_{2}^0)$-complete.

\medskip

Since $A$ is $\checked{D}_{n}(\mathbf{\Pi}_{2}^0)$-complete, we have $A\wequ \parenthesesBig{{\Lboth(\checked{\mathcal{A}}_{n})\,}^{\sim}}^{\blank}$. We are going to show that $A^{\blank}\wequ 
\parenthesesBig{{\Lboth(\mathcal{A}_{n+1})\,}^{\sim}}^{\blank}$ which will give the result since we already know by  Lemma \ref{Acompleteandtilde} that $\parenthesesBig{{\Lboth(\mathcal{A}_{n+1})\,}^{\sim}}^{\blank}$ is  
$D_{n+1}(\mathbf{\Pi}_{2}^0)$-complete. In order to get the result, we prove both  directions of the Wadge reduction: $A^{\blank}\winf \parenthesesBig{{\Lboth(\mathcal{A}_{n+1})\,}^{\sim}}^{\blank}$ and $A^{\blank}\wsup \parenthesesBig{{\Lboth(\mathcal{A}_{n+1})\,}^{\sim}}^{\blank}$.
 
\begin{description}
  \item[($\bm{\winf}$):] First, we pick a winning strategy $\sigma$ for Player $\two$ in the game  
  $\GWadge{A}{\parenthesesBig{{\Lboth(\checked{\mathcal{A}}_{n})\,}^{\sim}}^{\blank}}$. Without loss of generality, we may assume that $\sigma$ satisfies both
  
  \begin{enumerate}
  \item Player $\two$ never plays the letter $\quq{\mathbbold{0}}$, and
  \item every position $v\in \set{\mathbbold{1},\tla, d}^{<\omega}$ reached by $\two$ according to the strategy $\sigma$ satisfies $(v_{[\,/\blank]})^{\tla}=\mathbbold{1}^{k}$ for some $k\leq n$.
  \end{enumerate}   %
  This means that 
  the strategy $\sigma$  only commands $\two$ to   play according to the following three options: 
  \begin{enumerate}
  \item either play  the letter $\quq{\mathbbold{1}}$, 
  \item or erase the letter $\quq{\mathbbold{1}}$,
  \item or play the letter $\quq{\blank}$ (which stands for passing, henceforth does not interact with the final result).
\end{enumerate}
\medskip

As indicated above, we can assume that $\sigma$ satisfies the two conditions above for the following reason:
 given any strategy $\sigma'$ for Player $\two$ in the  game  $\GWadge{A}{\parenthesesBig{{\Lboth(\checked{\mathcal{A}}_{n})\,}^{\sim}}^{\blank}}$, the strategy $\sigma$ , it is easy to construct a  strategy $\sigma$ that produces the same outcome as $\sigma'$, in the sense that given any play $x\in\Alphabet^{\omega}$ by Player $\one$, Player $\two$'s responses $\sigma*x$  and $\sigma'*x$, in accordance with $\sigma$ and $\sigma'$ respectively, satisfy the same membership condition with respect to the winning set for Player $\two$:
 \[
 \sigma*x\in \parenthesesBig{{\Lboth(\checked{\mathcal{A}}_{n})\,}^{\sim}}^{\blank}\iff \sigma'*x\in\parenthesesBig{{\Lboth(\checked{\mathcal{A}}_{n})\,}^{\sim}}^{\blank}.
 \]
 
Indeed, $\sigma$ is constructed as follows. Assume, after $r$ rounds of moves, Player $\one$ is in position $u_{r}\in\Alphabet^{r}$, and Player $\two$'s responses, in accordance with $\sigma$ and 
$\sigma'$ respectively, are $v_{r}=\sigma*u$  and $v'_{r}=\sigma'*u$ (at the beginning, we have 
$u_{0}=v_{0}=v'_{0}=\emptysequence$). We describe Player $\two$'s next move in accordance with 
$\sigma$, only by considering  $\two$'s next move in accordance with $\sigma'$. Assume Player 
$\one$ plays the letter $a\in \Alphabet$ (so that Player $\one$ is now in position $u_{r+1}:=u_{r}a$), and Player $\two$'s response according to $\sigma'$ is $v'_{r+1}:=v'_{r}b'$, with either
$b'=\skiponesturn$, or $b'$ in $\set{\mathbbold{0},\mathbbold{1},\tla,\blank}$, we describe Player $\two$'s response according to $\sigma$ which is $v_{r+1}:=v_{r}b$, with either $b=\skiponesturn$, or $b\in\set{\mathbbold{1},\tla,\blank}$:

\begin{itemize}
  \item if $b':=\skiponesturn$, then $b:=\skiponesturn$,
  \item if $b':=\blank$, then  $b:=\blank$,
  \item if $b':=\mathbbold{0}$, then  $b:=\blank$,
  \item if $b':=\mathbbold{1}$, then 
\begin{enumerate}[label=(\emph{\greek*})]
  \item if $({v_{r}}_{[\,/\blank]})^{\tla}=\mathbbold{1}^{k}$ for some $k< n$, then $b:=\mathbbold{1}$,
  \item if $({v_{r}}_{[\,/\blank]})^{\tla}=\mathbbold{1}^{k}$ for some $k\geq n$, then $b:=\blank$,

   \end{enumerate}

  \item if $b'=\tla$, then 

\begin{enumerate}[label=(\emph{\greek*})]
  \item if $({v'_{r}}_{[\,/\blank]})^{\tla}:=\emptysequence$, then $b:=\blank$,
  \item if $({v'_{r}}_{[\,/\blank]})^{\tla}\in \set{\mathbbold{0},\mathbbold{1}}^{*}\mathbbold{0}$, then $b:=\blank$,
  \item if $({v'_{r}}_{[\,/\blank]})^{\tla}\in \set{\mathbbold{0},\mathbbold{1}}^{*}\mathbbold{1}$, then 

\begin{enumerate}[label=(\emph{\roman*})]
  \item if $({v_{r}}_{[\,/\blank]})^{\tla}=\mathbbold{1}^{k}$ for some $k< n$, then $b:=\tla$
    \item if $({v_{r}}_{[\,/\blank]})^{\tla}=\mathbbold{1}^{k}$ for some $k\geq n$, then $b:=\blank$.
       \end{enumerate}
     \end{enumerate}
\end{itemize}
We leave it to the reader to verify that the strategy $\sigma$ meets the requirements.

\medskip

For any  $y\in \set{\mathbbold{1},\tla, d}^{\omega}$ played by $\two$ according to the strategy $\sigma$, and any integer $i$ we write 
  $w_i=\parenthesesbig{(y\upharpoonright i)_{[\,/\blank]}}^{\tla}$. We have 
  \begin{itemize}
  \item $w_i=\mathbbold{1}^{k_i}$ for some integer $k_i\leq n$, and 
  \item $y\in \parenthesesbig{{\Lboth(\checked{\mathcal{A}}_{n})\,}^{\sim}}^{\blank}$ if and only if the least integer $k$ such that $w_i=\mathbbold{1}^{k}$ holds for infinitely many integers $i$ has the property that $n-k$ is even (or equivalently $\textit{parity}(k)=\textit{parity}(n)$).
 
\end{itemize}

  Secondly, we  describe a winning strategy $\tau$ for $\two$ in $\GWadge{A^{\blank}}{\parenthesesBig{{\Lboth(\mathcal{A}_{n+1})\,}^{\sim}}^{\blank}}$. Notice first that 
  \begin{itemize}
  \item for all integers $k\leq n$, we have  $\mathbbold{1}^k\in \Lboth(\mathcal{A}_{n+1}) \iff 
  \mathbbold{1}^k\in \Lboth(\checked{\mathcal{A}}_{n})$, and 
  
    \item for all integers $k> n$, we have $\mathbbold{1}^{k}\notin \Lboth(\mathcal{A}_{n+1})$ but 
    $\mathbbold{1}^{k}\in \Lboth(\checked{\mathcal{A}}_{n})$.
 \end{itemize}

\vfill\eject
  
  The strategy $\tau$ works as follows:
  
  \begin{enumerate}
  \item As long as Player $\one$ does not play the letter $\quq{\blank}$, $\tau$ commands Player $\two$ to apply the strategy $\sigma$.
  \item\label{thenwegobacktotwo} If Player $\one$ plays the letter $\quq{\blank}$,  $\tau$ commands Player $\two$ to  play $\mathbbold{1}^{n+1}$ followed by the same number of letters  
   $\quq{\blank}$ that Player $\one$ plays after this first $\quq{\blank}$. (Playing $\mathbbold{1}^{n+1}$ makes sure that the play  reaches  a position\footnote{See Figure \ref{therealautomatonAnplusonewithnoblanks}.}  that would be rejected if Player $\one$ keeps on playing the letter $\quq{\blank}$ for ever.)  
  \item If Player $\one$ stops playing  the letter $\quq{\blank}$ and resumes playing letters from $\Alphabet$, then  Player $\two$ plays $\tla^{n+1}$ (in order to erase the sequence $\mathbbold{1}^{n+1}$ she previously played), then she resumes applying the strategy $\sigma$ while discounting the occurrences of the letter $\quq{\blank}$ previously played by $\one$; and she does so as long as Player $\one$ does not play anymore $\quq{\blank}$. If Player $\one$ plays $\quq{\blank}$ again, then the strategy goes back to  case \ref{thenwegobacktotwo}.
\end{enumerate}
  
  Assume that Player $\one$  has produced $x\in (\Alphabet\cup\set{\blank})^{\omega}$ and Player $\two$, according to $\tau$, has answered  with $y\in \set{\mathbbold{0},\mathbbold{1},\tla,\blank}^{\omega}$.
  Note that 
  \begin{itemize}
  \item  if $x$ contains only finitely many letters from $\Alphabet$, then we have both $x\notin A$, and  $\mathbbold{1}^{n+1}$ is a suffix of the finite sequence  ${y_{[\,/\blank]}}$, henceforth also a suffix of 
    ${y_{[\,/\blank]}}^\tla$, which shows that ${y_{[\,/\blank]}}^\tla\notin \Lboth(\mathcal{A}_{n+1})$ and therefore that  
    $y\notin  \parenthesesBig{{\Lboth(\mathcal{A}_{n+1})\,}^{\sim}}^{\blank}$, which awards Player $\two$ a victory.
  
   \item   if $x$ contains infinitely many letters from $\Alphabet$, then by construction, apart from sequences of the form $\mathbbold{1}^{n+1}\tla^{n+1}$ 
   (which cancel themselves out when interpreting the letter $\quq{\tla}$ as an eraser), the sequence ${y_{[\,/\blank]}}$  is nothing but the application of the strategy $\sigma$ to $x_{[\,/\blank]}$, which shows again that in this configuration Player $\two$ wins.
 \end{itemize} 

  \item[($\bm{\wsup}$)]:
 We describe a winning strategy $\tau$ for Player $\two$ in the underlying game   
$$\GWadge{\parenthesesBig{{\Lboth(\mathcal{A}_{n+1})\,}^{\sim}}^{\blank}}{A^{\blank}}.$$
For exactly the same reasons as explained in the previous case, without loss of generality, we may assume that \begin{enumerate}
  \item Player $\one$ never plays the letter $\quq{\mathbbold{0}}$.
  \item Every position $u\in \set{\mathbbold{1},\tla, d}^{<\omega}$ that Player $\one$ reaches satisfies $(u_{[\,/\blank]})^{\tla}=\mathbbold{1}^{k}$ for some $k\leq n+1$.
  \end{enumerate}   %
We fix a winning strategy $\sigma$ for $\two$ in $\GWadge{\parenthesesbig{{\Lboth(\checked{\mathcal{A}}_{n})\,}^{\sim}}^{\blank}}{A}$. 
The strategy $\tau$ works as follows.

\begin{enumerate}
  \item\label{thenwegobacktopositionone} As long as all the successive  positions $u\in \set{\mathbbold{1},\tla, d}^{<\omega}$ reached by Player $\one$  satisfy  $(u_{[\,/\blank]})^{\tla}=\mathbbold{1}^{k}$ for some $k\leq n$, 
  Player $\two$ keeps applying  strategy $\sigma$.
  \item\label{thenwegobacktopositiontwo} Each time Player $\one$ finds himself in a position $u\in \set{\mathbbold{1},\tla, d}^{<\omega}$ with the property that $(u_{[\,/\blank]})^{\tla}=\mathbbold{1}^{n+1}$, Player $\two$ plays the letter $\quq{\blank}$.
  \item When Player $\one$ quits being in a position $u\in \set{\mathbbold{1},\tla, d}^{<\omega}$ as in case \ref{thenwegobacktopositiontwo}  and goes back to some position as in  case 
  \ref{thenwegobacktopositionone} (i.e.,  Player $\one$ was in some position $u$ satisfying $(u_{[\,/\blank]})^{\tla}=\mathbbold{1}^{n+1}$ and just played $\quq{\tla}$ so that he finds himself  in the new position $u\tla$ which satisfies $(u\tla_{[\,/\blank]})^{\tla}=\mathbbold{1}^{n}$), then Player $\two$ resumes applying $\sigma$ to $\one$'s play deprived of the moves that lead to positions as in   case 
  \ref{thenwegobacktopositiontwo}. This means that $\two$ removes from $\one$'s play all the subwords of the form $\mathbbold{1}d\ldots d\tla$ that lead to reaching then leaving 
  positions as described in  case   \ref{thenwegobacktopositiontwo}, and replaces each of them by the simple letter $\quq{\blank}$ before applying the strategy $\sigma$. 
  \end{enumerate}
To show that  strategy $\tau$ is winning for Player $\two$, assume that Player $\one$  has produced $x$ in $\set{\mathbbold{0},\mathbbold{1},\tla,\blank}^{\omega}$ and Player $\two$ has answered according to $\tau$ with 
$y\in (\Alphabet\cup\set{\blank})^{\omega} $.
We distinguish between whether ${x_{[\,/\blank]}}^{\tla}=\mathbbold{1}^{n+1}$ and ${x_{[\,/\blank]}}^{\tla}\neq\mathbbold{1}^{n+1}$.

  \begin{itemize}
  \item If  ${x_{[\,/\blank]}}^{\tla}=\mathbbold{1}^{n+1}$, then $\exists i~\forall j\geq i~~ \parenthesesbig{(x\upharpoonright j)_{[\,/\blank]}}^{\tla}=\mathbbold{1}^{n+1}$. 
  So, necessarily $\blank^{\omega}$ is a suffix of $x$. 
  More precisely, we   have $x=u\blank^{\omega}$ with ${u_{[\,/\blank]}}^{\tla}=\mathbbold{1}^{n+1}$. Then, by construction, $y$ is also of the form $y=v\blank^{\omega}$, so that $y_{[\,/\blank]}$ is finite.
   So we end up with both \begin{itemize}
  \item ${x_{[\,/\blank]}}^{\tla}=\mathbbold{1}^{n+1}\notin \Lboth(\mathcal{A}_{n+1})$, which yields  $x\notin \parenthesesBig{{\Lboth(\mathcal{A}_{n+1})\,}^{\sim}}^{\blank}$, and

  \item $y_{[\,/\blank]}\notin A$, which yields $y\notin A^{\blank}$,\medskip
   
\end{itemize} 

which shows that Player $\two$ wins.
     
  \item If ${x_{[\,/\blank]}}^{\tla}\neq\mathbbold{1}^{n+1}$, then $\forall i~\exists j\geq i~~ \parenthesesbig{(x\upharpoonright j)_{[\,/\blank]}}^{\tla}\neq\mathbbold{1}^{n+1}$.\medskip 
       
       Firstly, since  $(x_{[\,/\blank]})^{\tla}=\mathbbold{1}^{k}$ holds for some $k\leq n$, we  have 
    \[
    (x_{[\,/\blank]})^{\tla}\in \Lboth(\mathcal{A}_{n+1})\iff  (x_{[\,/\blank]})^{\tla}\in \Lboth(\checked{\mathcal{A}}_{n}).
    \]
    
    Secondly, by construction,  $\blank^{\omega}$ is not a suffix of $y$, hence $y_{[\,/\blank]}$ is infinite, so  $y$ contains infinitely many letters from $\Alphabet$. We let $x'$ be the sequence obtained  from $x$ by removing all  occurrences of subwords of the from $\mathbbold{1}\blank\ldots\blank\tla$ (which  correspond to Player $\one$ reaching some position as in case  \ref{thenwegobacktopositiontwo}  by playing \quq{$\mathbbold{1}$}, then staying at the same position for a while by playing \quq{$\blank$}, for finally leaving that same position by playing \quq{$\tla$}). More precisely, this  means that if $x=u\mathbbold{1}\blank\ldots\blank\tla z$, and ${u_{[\,/\blank]}}^{\tla}=\mathbbold{1}^{n}$, then 
  $\mathbbold{1}\blank\ldots\blank\tla$ is deleted and replaced with the single letter $\quq{\blank}$.
  We clearly have\medskip 
  
\begin{itemize}
  \item $x'$ is infinite,
  \item ${x_{[\,/\blank]}}^{\tla}={x'_{[\,/\blank]}}^{\tla}$
  \item $y_{[\,/\blank]}=\sigma*x'$ (the result of applying the strategy $\sigma$ to $x'$).
\end{itemize}\medskip

So, we end up with 
\[
 \arraycolsep=1.4pt\def\arraystretch{1.5}
\begin{array}{ccl}
   x\in \parenthesesBig{{\Lboth(\mathcal{A}_{n+1})\,}^{\sim}}^{\blank}   & \iff & {x_{[\,/\blank]}}^{\tla}\in \Lboth(\mathcal{A}_{n+1})  \\
   & \iff &   {x_{[\,/\blank]}}^{\tla}\in\Lboth(\checked{\mathcal{A}}_{n})\\
   & \iff &   {x'_{[\,/\blank]}}^{\tla}\in \Lboth(\checked{\mathcal{A}}_{n})\\
   & \iff &   x'\in \parenthesesbig{{\Lboth(\checked{\mathcal{A}}_{n})\,}^{\sim}}^{\blank} \\
   & \iff & y_{[\,/\blank]}\in  A\\
   & \iff & y\in  A^{\blank}.\\
   \end{array}
\]
  Which shows once again that Player $\two$ wins.
 \end{itemize}
\end{description}
This finishes the proof.\boxlem{preliminarylemmadup4}  

We recall Lemma \ref{thm:lemmadupfour}, which we are now ready to prove.    
\lemmadupfour*
\medskip

\prooflem{lemmadup4} By Lemma \ref{preliminarylemmadup4}, if  $A\subseteq\Alphabet^{\omega}$  is $\checked{D}_{n}(\mathbf{\Pi}_{2}^0)$-complete, then $A^{\blank}$ is 
 $D_{n+1}(\mathbf{\Pi}_{2}^0)$-complete. By Lemma \ref{Acompleteandtilde}  $(A^\sim)^{\blank}\subseteq(\Alphabet\cup\set{\tla,\blank})^{\omega}$  is $D_{n+1}(\mathbf{\Pi}_{3}^0)$-complete.
So, it is enough to show that $(A^\sim)^{\blank}\wequ A^\sim$ to get the result. 
The reduction $A^\sim\winf (A^\sim)^{\blank}$ is obvious since the identity provides a winning strategy for Player $\two$ in the underlying Wadge game. 
To see that $(A^\sim)^{\blank}\winf A^\sim$ holds it is enough to notice that any  player in charge of a set 
of the form $A^\sim$ can always mimic passing (or equivalently playing the letter $\quq{\blank}$) 
by simply playing successively $\quq{a\tla}$ (any letter $a\in\Alphabet$). So, Player $\two$ has an obvious winning strategy in the game $\GWadge{(A^\sim)^{\blank}}{A^\sim}$ which consists in copying 
Player $\one$'s moves as long as they are letters from $\Alphabet$ and replacing every letter $\quq{\blank}$ by $\quq{a\tla}$ (any letter $a\in\Alphabet$).
\boxlem{lemmadup4}  

\medskip
%

 %%%%%%%%%%%%%%%%% %%%%%%%%%%%%%%%%% %%%%%%%%%%%%%%%%% %%%%%%%%%%%%%%%%% %%%%%%%%%%%%%%%%%%
 %%%%%%%%%%%%%%%%% %%%%%%%%%%%%%%%%% %%%%%%%%%%%%%%%%% %%%%%%%%%%%%%%%%% %%%%%%%%%%%%%%%%%%

      \begin{lemma} \label{lemmadup5} Let $\Alphabet$ be any finite alphabet, $A\subseteq\Alphabet^{\omega}$, 
$n\geq 1$ and $k\geq 3$ be integers. If $A$ is $\checked{D}_{n}(\mathbf{\Pi}_{k}^0)$-complete, then $A^\sim\subseteq(\Alphabet\cup\set{\tla})^{\omega}$ is $\checked{D}_{n}(\mathbf{\Pi}_{k+1}^0)$-complete.\end{lemma}

\prooflem{lemmadup5} 
By Corollary \ref{preliminarycorollary} (\ref{preliminarycorollaryb}), if  $A$ is $\checked{D}_{n}(\mathbf{\Pi}_{k}^0)$-complete, then 
$A^{\blank}\subseteq(\Alphabet\cup\set{\blank})^{\omega}$ is $\checked{D}_{n}(\mathbf{\Pi}_{k}^0)$-complete as well. 
So, $(A^{\blank})^\macroforcomplement=(\Alphabet\cup\set{\blank})^{\omega}\smallsetminus A^{\blank}$ is $D_{n}(\mathbf{\Pi}_{k}^0)$-complete. 
Note that $(A^{\blank})^\macroforcomplement=(A^\macroforcomplement\cup\Alphabet^{<\omega})^{\blank}=\parenthesesbig{(\Alphabet^{\omega}\smallsetminus A)\cup\Alphabet^{<\omega}}^{\blank}$. 
By Lemma \ref{Acompleteandtilde}, 
$$\parenthesesbig{(A^\macroforcomplement\cup\Alphabet^{<\omega})^\sim}^{\blank}\subseteq(\Alphabet\cup\set{\tla,\blank})^{\omega}$$ 
is $D_{n}(\mathbf{\Pi}_{k+1}^0)$-complete. Thus, 
$\parenthesesBig{\parenthesesbig{(A^\macroforcomplement\cup\Alphabet^{<\omega})^\sim}^{\blank}}
^\macroforcomplement=\parenthesesbig{A^\sim}^{\blank}$ is $\checked{D}_{n}(\mathbf{\Pi}_{k+1}^0)$-complete. It remains to show that $\parenthesesbig{A^\sim}^{\blank}\wequ A^\sim$. The reduction $A^\sim\winf\parenthesesbig{A^\sim}^{\blank}$ is obvious. 
For $\parenthesesbig{A^\sim}^{\blank}\winf A^\sim$, notice that playing some letter from $\Alphabet$ and immediately erasing it is equivalent to playing the letter \quq{\blank}.
\boxlem{lemmadup5}  

\medskip

    The next theorem  sums up all the results obtained so far concerning the topological complexity of $A^\sim$ with regard to the complexity of $A$.

    \begin{theorem} \label{resumetheorem} Let $\Alphabet$ be any finite alphabet with at least two different letters, $A\subseteq\Alphabet^{\omega}$,  \quq{$\tla$} some letter not in $\Alphabet$,   and $n\geq 1$, $k\geq 2$ be integers. 
\begin{enumerate}[label=(\textit{\alph*})]

     \item\label{resumetheorem1} If $A$ is $D_{n}(\mathbf{\Pi}_{k}^0)$-complete, then $A^\sim\subseteq(\Alphabet\cup\set{\tla})^{\omega}$ is $D_{n}(\mathbf{\Pi}_{k+1}^0)$-complete
 
   \item\label{resumetheorem3} If $A$ is $\checked{D}_{n}(\mathbf{\Pi}_{2}^0)$-complete, then $A^\sim\subseteq(\Alphabet\cup\set{\tla})^{\omega}$ is $D_{n+1}(\mathbf{\Pi}_{3}^0)$-complete
 
  \item\label{resumetheorem2}  If $A$ is $\checked{D}_{n}(\mathbf{\Pi}_{k+1}^0)$-complete, then $A^\sim\subseteq(\Alphabet\cup\set{\tla})^{\omega}$ is $\checked{D}_{n}(\mathbf{\Pi}_{k+2}^0)$-complete.

\end{enumerate}
 \end{theorem}

\proofthm{resumetheorem} \begin{multicols}{3}\begin{itemize}
  \item[\ref{resumetheorem1}]  is Theorem \ref{thm:thedupthree}, 
     \item[\ref{resumetheorem3}]  is Lemma \ref{lemmadup4},
 \item[\ref{resumetheorem2}]  is Lemma \ref{lemmadup5}. 
 
   \end{itemize}
\end{multicols}
%Case \ref{resumetheorem1} is Theorem \ref{thm:thedupthree}, case \ref{resumetheorem2} is Lemma \ref{lemmadup5}, and case \ref{resumetheorem3} is Lemma \ref{lemmadup4}. 
\noindent This finishes the proof.\boxthm{resumetheorem}  

%\end{document}
    %%%%%%%%%%%%%%%%%%%%%%%%%%%%%%%%%%%%%%%%%%%%%%%%%%%%%%%%%%%%%%%%%%%%%%%%%%%%%%%%%%%%%%
    %%%%%%%%%%%%%%%%%%%%%%%%%%%%%%%%%%%%%%%%%%%%%%%%%%%%%%%%%%%%%%%%%%%%%%%%%%%%%%%%%%%%%%

    %%%%%%%%%%%%%%%%%%%%%%%%%%%%%%%%%%%%%%%%%%%%%%%%%%%%%%%%%%%%%%%%%%%%%%%%%%%%%%%%%%%%%%
    %%%%%%%%%%%%%%%%%%%%%%%%%%%%%%%%%%%%%%%%%%%%%%%%%%%%%%%%%%%%%%%%%%%%%%%%%%%%%%%%%%%%%%
    %%%%%%%%%%%%%%%%%%%%%%%%%%%%%%%%%%%%%%%%%%%%%%%%%%%%%%%%%%%%%%%%%%%%%%%%%%%%%%%%%%%%%%
    %%%%%%%%%%%%%%%%%%%%%%%%%%%%%%%%%%%%%%%%%%%%%%%%%%%%%%%%%%%%%%%%%%%%%%%%%%%%%%%%%%%%%%
    %%%%%%%%%%%%%%%%%%%%%%%%%%%%%%%%%%%%%%%%%%%%%%%%%%%%%%%%%%%%%%%%%%%%%%%%%%%%%%%%%%%%%%
    %%%%%%%%%%%%%%%%%%%%%%%%%%%%%%%%%%%%%%%%%%%%%%%%%%%%%%%%%%%%%%%%%%%%%%%%%%%%%%%%%%%%%%

 We now recall a  slightly modified variant of the operation $\sim$, introduced in \cite{Fin01a}, and which is particularly well suited for deducing the properties of $\omega$-powers. 
  
\begin{definition}
Let  $\Alphabet_A$ be any finite alphabet, $\strongtla\notin\Alphabet_A$ be any letter, $\Alphabet:=\Alphabet_A\cup\{\strongtla\}$, and $A\subseteq\Alphabet_A^{\leq\omega}$. Then, we set 
$A^\approx:=\{ x\in\Alphabet^{\leq\omega}\mid x^{\strongtlascripstyle}\in A\}$, where $x^{\strongtlascripstyle}$ is inductively defined as follows.

     \begin{itemize}
     \item if $x=\emptysequence$, then $x^{\strongtlascripstyle}:=\emptysequence$,

     \item if $x=ua$  for  some $u\in\Alphabet^{<\omega}$ and $a\in\Alphabet$, then 
     \begin{enumerate}[label=(\textit{\alph*})]

     \item  if  $u^{\strongtlascripstyle}$ is defined, then 
$\left\{\!\!\!\!
\begin{array}{lllll}
%\text{ if }a=\strongtla,&\text{and}&u^{\strongtlascripstyle}\text{ is undefined }, &\text{then} &x^{\strongtlascripstyle}\text{ is undefined}\\
\text{ if }a=\strongtla,&\text{and}&|u^{\strongtlascripstyle}|= 0, &\text{then} &x^{\strongtlascripstyle}:=\text{ is undefined}\\
\text{ if }a=\strongtla,&\text{and}&|u^{\strongtlascripstyle}|> 0, &\text{then} &x^{\strongtlascripstyle}:=u^{\strongtlascripstyle}\upharpoonright_{(|u^{\strongtlascripstyle}|-1)}\\
\text{ if }a\neq\tla,&& &\text{then}&x^{\strongtlascripstyle}:= u^{\strongtlascripstyle} a\\
\end{array}
\right.
$

 \item if  $u^{\strongtlascripstyle}$ is undefined, then $x^{\strongtlascripstyle}$ is also undefined;
    
\end{enumerate}

\vfill\eject

   \item   if $x\in\Alphabet^{\omega}$, then 
   
  \begin{enumerate}[label=(\textit{\alph*})]
   \item  if for some $n\in\omega$, $(x\upharpoonright n)^{\strongtlascripstyle}$ is undefined, then $x^{\strongtlascripstyle}$ is undefined;
  
   \item otherwise, 
   $x^{\strongtlascripstyle}:=\lim_{n\in\omega}~(x\upharpoonright n)^{\strongtlascripstyle}$, where given any word $w\in\Alphabet_A^{<\omega}$, we have 
  \[
w\subseteq\mbox{lim}_{n\in\omega}~(x\upharpoonright n)^{\strongtlascripstyle}\Longleftrightarrow\exists p\in\omega\,
\forall n> p\,~~w\subseteq(x\upharpoonright n)^{\strongtlascripstyle}.
\]

\end{enumerate}

\end{itemize}

%\bigskip\bigskip\bigskip\bigskip\bigskip
%
%
%
%
%- $\emptysequence^\tla:=\emptysequence$.\smallskip
%
%- For a finite word $u\in\Alphabet^{<\omega}$, 
%$\left\{
%\begin{array}{ll}
%& (ua)^\tla:= u^\tla a\mbox{ if }a\in\Alphabet_A\mbox{,}\cr
%& (u\tla)^\tla:= u^\tla\mbox{ with its last letter removed if }|u^\tla| > 0\mbox{,}\cr
%& (u\tla)^\tla\mbox{ is undefined if }|u^\tla|= 0.
%\end{array}
%\right.$\smallskip
%
%- For an infinite word $\sigma$, $\sigma^\tla:=\mbox{lim}_{n\in\omega}~(\sigma\upharpoonright n)^\tla$.

\end{definition}

 The only difference between $A^{\approx}$ and $A^{\sim}$ relies in here $(u\strongtla)^{\strongtlascripstyle}$ is undefined if $|u^{\strongtlascripstyle}|= 0$, while 
 $(u\tla)^{\tla}$ remains always defined, no matter the length of $u^{\tla}$. In fact, in all but one case, $A^{\approx}$ and $A^{\sim}$ are Wadge equivalent, so we easily obtain the following version of Theorem \ref{thm:thedupthree}. 

%
%\begin{theorem}
%Let  $\Alphabet_A$ be any finite alphabet, $n\geq 1$ and $k\geq 2$ be natural numbers. If 
%$A\subseteq\Alphabet_A^{\omega}$ is $D_{n}(\mathbf{\Pi}_k^0)$-complete, then $A^\approx$ is 
%$D_{n}(\mathbf{\Pi}_{k+1}^0)$-complete.
%\end{theorem}
% 

  \begin{theorem} \label{thedup2} Let $\Alphabet$ be any finite alphabet with at least two different letters, $A\subseteq\Alphabet^{\omega}$,  \quq{$\strongtla$} some letter not in $\Alphabet$,   and $n\geq 1$ and $k\geq 2$ be integers. 
\begin{enumerate}[label=(\textit{\alph*})]

     \item If $A$ is $D_{n}(\mathbf{\Pi}_{k}^0)$-complete, then $A^{\approx}\subseteq(\Alphabet\cup\set{\strongtla})^{\omega}$ is $D_{n}(\mathbf{\Pi}_{k+1}^0)$-complete.
 
   \item  If $A$ is $\checked{D}_{n}(\mathbf{\Pi}_{2}^0)$-complete, then $A^{\approx}\subseteq(\Alphabet\cup\set{\strongtla})^{\omega}$ is $D_{n+1}(\mathbf{\Pi}_{3}^0)$-complete.
 
  \item  If $A$ is $\checked{D}_{n}(\mathbf{\Pi}_{k+1}^0)$-complete, then $A^{\approx}\subseteq(\Alphabet\cup\set{\strongtla})^{\omega}$ is $\checked{D}_{n}(\mathbf{\Pi}_{k+2}^0)$-complete.

\end{enumerate}
 \end{theorem}
 
\proofthm{thedup2}
  It is easy to see that if $A\subseteq\Alphabet_A^{\omega}$ is  Borel but not the whole space (i.e., 
$\thecomplement{A}\neq\emptyset$), then $A^\approx$ is Wadge equivalent to $A^\sim$ (see \cite[page 690]{Fin01a}). So, the results follow directly from Theorem \ref{resumetheorem}.
\boxthm{thedup2}

%because:
%\begin{enumerate}
%\ite[a)] In the  Wadge game $W(A^\sim, A^\approx)$ 
 % the player in charge of 
%$A^\approx$ has clearly a winning strategy 
%which consists in copying the play of the other player except if player 1 writes the eraser
%$\tla$ but he has nothing to erase. In this case player 2 writes for example a letter 
%$a\in X_A$ and the eraser $\tla$ at the next step of the play.
%Now  if, in $\omega$ steps, player 1 has written the $\omega$-word $\alpha$ and 
%player 2 has written the $\omega$-word $\beta$, it is easy to see that 
%$[\alpha^\tla = \beta^\tla]$ and then $\alpha \in  A^\sim$ iff $\beta \in A^\approx$. 
%Thus player 2 has 
%a winning strategy in the Wadge game $W(A^\sim, A^\approx)$.
 
%\ite[b)] Consider now the Wadge game $W(A^\approx , A^\sim)$. 
%The only extra  
%possibility for the player in charge of $A^\approx$ is to get out of the set $A^\approx$ by 
%writing the eraser $\tla$ when in fact there is not any letter of his previous play to erase.
%But then his final play is surely outside $A^\approx$.
%\nl If $A \neq X_A^{\omega}$, i.e. $A^-\neq \emptyset$, then 
% the player in charge of the set 
%$A^\sim$  may write a word in $(A^\sim)^-$, 
%(by playing the letter $\tla$ to erase his previous play and then writing 
%a word in $A^-$) then 
%player 2 has a winning strategy in the Wadge game $W(A^\approx , A^\sim)$.  

%\end{enumerate}

\section{Differences of $\mathbf{\Pi}_{k}^0$-complete $\omega$-powers}\label{differences-of-complete-powers}

 We will begin by describing the language $\mathbb{L}$, which was first introduced under the name $L_3$  in \cite{Fin01a}.
  
  \begin{notation}[the language $\mathbb{L}$]\label{notationforLthree}
\rm\ Let $\Alphabet_A$ be any finite alphabet, $\strongtla\notin\Alphabet_A$ any letter, and 
$\Alphabet$ be $\Alphabet_A\cup\{\strongtla\}$. The   context-free language over $\Alphabet$ generated by the context-free grammar with the following production rules is called $\mathbb{L}$:
%\begin{figure}[h]
\[
\begin{array}{lllll}
\circ~~S&\longrightarrow&aS\strongtla\, S&\text{with}&a\in\Alphabet_A,\\
\circ~~S&\longrightarrow&a\strongtla\, S&\text{with}&a\in\Alphabet_A,\\
\circ~~S&\longrightarrow&{\emptysequence}.\\
\end{array}
\]
%\caption{Production rules of the grammar generating $\mathbb{L}$}
%\end{figure} 
\end{notation}

\noindent (Readers who are unfamiliar with basic concepts about grammars can refer to \cite{HopcroftMotwaniUllman2001}.) This language $\mathbb{L}$ corresponds to the words where every letter of $\Alphabet_A$ has been removed after using the backspace operation. It is easy to see that $\mathbb{L}$ is a deterministic one-counter language, i.e., $\mathbb{L}$ is accepted by a deterministic one-counter automaton. Moreover,  for each $a\in\Alphabet_A$, the language $\mathbb{L}a$ is also accepted by a deterministic one-counter automaton.
\medskip

 We can now state the following result. 
 
\begin{lemma}\label{Lemma8-2-in-Fin01a} Let $A\subseteq\Alphabet_A^{\omega}$, 
$\strongtla\notin\Alphabet_A$, and $\Alphabet=\Alphabet_A\cup\set{\strongtla}$.\smallskip

The $\omega$-language    $A^\approx\subseteq\Alphabet^{\omega}$ is obtained by replacing every letter $a\in\Alphabet_A$ in every word of $A$ with $\mathbb{L}a$.\end{lemma}

\prooflem{Lemma8-2-in-Fin01a} See \cite[Lemma 8.2]{Fin01a}.
\boxlem{Lemma8-2-in-Fin01a}

Note that an $\omega$-word $x\in A^\approx$ may be considered as an $\omega$-word $x^{\strongtla}\in A$ to which we may add a finite word belonging to the context-free (finitary) language $\mathbb{L}$, either before the first letter $x^{\strongtla} (0)$ of $x^{\strongtla}$, or  between  any two consecutive letters $x^{\strongtla} (n)$ and $x^{\strongtla} (n+ 1)$ of $x^{\strongtla}$.

\begin{corollary} \label{FLpresop}
Let $L_A\subseteq\Alphabet_A^{<\omega}$ be any finitary language, and $A:= L_A^\infty$ be the $\omega$-power of $L_A$. Then, 
\begin{enumerate}
  \item there exists some (finitary) language $E_A$ such that $A^\approx= E_A^\infty$;

  \item providing $L_A$ belongs to the class $OCL(k)$ for some natural number $k$, the language $E_A$ can be chosen to be contained within the class $OCL(k+ 1)$.

\end{enumerate}\end{corollary}

\proofcor{FLpresop} We reproduce the proof that first appeared in \cite{Fin01b}. Let $\mathbb{L}$ be the context free  language defined above, and $h :\Alphabet_A\rightarrow 2^{\Alphabet_A^{<\omega}}$ be the substitution defined by $a\mapsto\mathbb{L}a$. Then, it is easy to see that  now $A^\approx$ is obtained by substituting in $A$ the language $\mathbb{L}a$ for each letter $a\in\Alphabet_A$. Thus $E_A = h(L_A)$ satisfies the statement of the theorem.\boxcor{FLpresop}
\medskip

We can now easily state the following result, which will be  improved in the sequel. 

 %   We now  recall the following result, proved in \cite{Fin01a}. %????????? C'est une erreur n'est-ce pas ?

\begin{proposition}\label{FLpin} 
For each natural numbers $n,k\geq 1$, 

\begin{enumerate}
  \item There are context-free languages $U_{n,k}$ in the subclass of iterated counter languages such that $U_{n,k}^\infty$ is $D_{2n-1}(\mathbf{\Pi}_k^0)$-complete.

  \item There are context-free languages $V_{n,1}$ and $V_{n,2}$ in the subclass of iterated counter languages such that $V_{n,1}^\infty$ and $V_{n,2}^\infty$ are respectively  $\checked{D}_{2n}(\mathbf{\Pi}_1^0)$-complete and $\checked{D}_{2n}(\mathbf{\Pi}_2^0)$-comple-te. Actually these languages can be chosen to be regular, by 
\cite{Fink-Lec5, DBLP:journals/tcs/FinkelGD26}

\item Providing there exist context-free languages $V_{n,3}$ in the subclass of iterated counter languages such that $V_{n,3}^\infty$ are  $\checked{D}_{2n}(\mathbf{\Pi}_{3}^0)$-complete,  there would also exist context-free languages $V_{n,k+3}$ in the subclass of iterated counter languages such that $V_{n,k+3}^\infty$ are   $\checked{D}_{2n}(\mathbf{\Pi}_{k+3}^0)$-complete.

%\item Assuming there are context-free languages $V_{n,k+2}$ in the subclass of iterated counter languages such that $V_{n,k+2}^\infty$ are  $\checked{D}_{2n}(\mathbf{\Pi}_{k+2}^0)$-complete, then there are also context-free languages $V_{n,k+3}$ in the subclass of iterated counter languages such that $V_{n,k+3}^\infty$ are   $\checked{D}_{2n}(\mathbf{\Pi}_{k+3}^0)$-complete.

\end{enumerate}\end{proposition}

\proofpro{FLpin}

\begin{enumerate}
\item \label{FLpina} The main theorem in \cite{Fink-Lec5} provides regular languages $U_{n,1}\subseteq 2^{<\omega}$ such that $U_{n,1}^\infty$ is $D_{2n-1}(\mathbf{\Pi}_1^0)$-complete. 
Theorem 2 in \cite{DBLP:journals/tcs/FinkelGD26} provides regular languages 
$U_{n,2}\subseteq 2^{<\omega}$ with the properties that $U_{n,2}^\infty$ is 
$D_{2n-1}(\mathbf{\Pi}_2^0)$-complete. In particular, both $U_{n,1}$ and $U_{n,2}$ are in the class 
$\mbox{OCL}(0)$.\medskip

\ \ We now consider the substitution 
$h:\{ \mathbbold{0},\mathbbold{1}\}\to 2^{(\{ \mathbbold{0},\mathbbold{1}\}\cup\{\tla\})^{<\omega}}$ from the proof of Corollary \ref{FLpresop}, and set $U_{n,3}:= h(U_{n,2})$, which is a context-free language in the class 
$\mbox{OCL}(1)$. Note that the set 
${U_{n,3}^\infty= h(U_{n,2})^\infty= (U_{n,2}^\infty )^\approx}$ is $D_{2n-1}(\mathbf{\Pi}_3^0)$-complete, by Theorem \ref{thm:thedupthree}.\medskip

\ \ Iterating this method $k$ many times ($k\geq 1$), we easily obtain a context-free language 
$U_{n,k+2}\in\mbox{OCL}(k)$ such that $U_{n,k+2}^\infty$ is $D_{2n-1}(\mathbf{\Pi}_{k+2}^0)$-complete.

    \item The main theorem in \cite{Fink-Lec5} provides regular languages 
 $V_{n,1}\subseteq 2^{<\omega}$ such that $V_{n,1}^\infty$ is $\checked{D}_{2n}(\mathbf{\Pi}_1^0)$-complete. Theorem 3  in \cite{DBLP:journals/tcs/FinkelGD26} provides regular languages $V_{n,2}\subseteq 2^{<\omega}$ with the properties that $V_{n,2}^\infty$ is $\checked{D}_{2n}(\mathbf{\Pi}_2^0)$-complete. In particular, both $V_{n,1}$ and $V_{n,2}$ are in the class $\mbox{OCL}(0)$.

  \item Granted with the assumption that there exists some context-free languages $V_{n,3}$, we proceed  as in \ref{FLpina}., making use of the same substitution  $h:\{ \mathbbold{0},\mathbbold{1}\}\to 2^{(\{ \mathbbold{0},\mathbbold{1}\}\cup\{\tla\})^{<\omega}}$ in order to obtain $V_{n,4}:= h(V_{n,3})$, and iterating this method accordingly. Besides,  if $V_{n,3}$ would belong to $\mbox{OCL}(1)$, then $V_{n,k+2}$ would belong to $\mbox{OCL}(k)$.

\end{enumerate}

\noindent This finishes the proof.\boxpro{FLpin}

 Note that all languages $U_{n,1},V_{n,1},U_{n,2}$ and $V_{n,2}$ are regular, hence accepted by some (real-time deterministic) finite automata (without any counter).
 
\vfill\eject
 
  On the other hand, the languages $\mathbb{L}a$, for $a\in\Alphabet_A$, are one-counter languages. Moreover we can easily see that, for each 
$a\in\Alphabet_A$,  the language $\mathbb{L}a$ is accepted by a real-time one-counter automaton 
$\mathcal{A}$, such that $\mathcal{A}$ accepts a finite word $w$ iff there is a run on $w$ ending in an accepting state \emph{and} with empty stack. This implies that the languages $U_{n,3}$   are  also accepted by a real-time one-counter automaton, by final states \emph{and} empty stack. Hypothetically speaking, if the languages $V_{n,3}$  exist as described above,  then they would be accepted by a real-time one-counter automaton, by final states \emph{and} empty stack as well. Now is the time to  state the following  proposition.

\begin{proposition} \label{sub} 
Let $A\subseteq\Alphabet_A^{<\omega}$ be any finitary language accepted by a real-time one-counter automaton $\mathcal{A}$ accepting words by final states \emph{and} empty stack. Let also  
$h:\Alphabet_A\to 2^{\Alphabet^{<\omega}}$ be the substitution defined by $a\mapsto \mathbb{L}a$, where $\mathbb{L}$ is the one-counter language introduced above in   Notation \ref{notationforLthree}. Then, the language $h(A)$ is in $\mbox{OCL}(2)$, and it is accepted by a real-time two-iterated counter automaton $\mathcal{B}$ accepting words by final states \emph{and} empty stack. 
\end{proposition}
    
    \proofpro{sub} 
    We provide informal proof by simply explaining the idea behind the construction of a real-time two-iterated counter automaton $\mathcal{B}$ from the automaton $\mathcal{A}$.
 The stack alphabet of $\mathcal{A}$ is of the form $\{Z_0, z_0\}$, and the stack alphabet of 
$\mathcal{B}$ is of the form $\{Z_0, z_0, z_1\}$. The automaton $\mathcal{B}$ starts the reading of a word over the alphabet $\Alphabet$     as the one-counter automaton $\mathcal{A}$ accepting the language $A$. Then, at any point during the computation, it may guess (using non-determinism) that it is reading a finite segment $w$ of $\mathbb{L}$ that will later be erased (using the eraser $\tla$). In which case, it reads the word $w$ using the additional stack letter $z_1$ which permits to simulate a one-counter automaton at the top of the stack while keeping the memory of the stack of $\mathcal{A}$ (which is actually a counter). 
 Then, after  reading the word $w$, $\mathcal{B}$ simulates again the one-counter automaton 
$\mathcal{A}$, and so on. The automaton $\mathcal{B}$ accepts words by final states (corresponding to final states of $\mathcal{A}$) \emph{and} empty stack. 
    \boxpro{sub} 

We now state one of our main technical results. 
%
%\begin{proposition} \label{maintechnicalresult} 
%Let $A\subseteq\Alphabet^{<\omega}$ be any finitary language accepted by a real-time two-iterated counter automaton $\mathcal{A}$ accepting words by final states  \emph{and} empty stack, and such that the 
%$\omega$-power $A^\infty$ is $D_{2n-1}(\mathbf{\Pi}_k^0)$-complete  (respectively, 
%$\checked{D}_{2n}(\mathbf{\Pi}_k^0)$-complete) for some natural numbers $n\geq 1$ and $k\geq 3$. Then, we can find a finite alphabet $Y$ and a finitary language $B\subseteq Y^{<\omega}$ such that \begin{itemize}
%  \item $B$ is the language accepted by a real-time one-counter automaton $\mathcal{B}$ accepting words by final states 
%\emph{and} empty stack, 
%  \item   $B^\infty$ is $D_{2n-1}(\mathbf{\Pi}_k^0)$-complete (respectively, 
%$\checked{D}_{2n}(\mathbf{\Pi}_k^0)$-complete),
% \item  Furthermore, one can even take $Y=\set{0, 1}$.
%
%   
%\end{itemize}\end{proposition}
%

\begin{proposition} \label{maintechnicalresult} 
Let $A\subseteq\Alphabet^{<\omega}$ be any finitary language accepted by a real-time \underline{two-iterated} counter automaton $\mathcal{A}$ accepting words by final states  \emph{and} empty stack, and such that the 
$\omega$-power $A^\infty$ is $\mathbf{\Gamma}$-complete for some non-self-dual Wadge class $\mathbf{\Gamma}$ that satisfies $\mathbf{\Gamma}=\mathbf{\Delta}^{0}_{3}\mbox{-PU}(\mathbf{\Gamma})$. Then, we can find a finite alphabet $Y$ and a finitary language $B\subseteq Y^{<\omega}$ such that \begin{itemize}
  \item $B$ is the language accepted by a real-time \underline{one-counter} automaton $\mathcal{B}$ accepting words by final states 
\emph{and} empty stack, 
  \item   $B^\infty$ is $\mathbf{\Gamma}$-complete.   
\end{itemize}
Furthermore, one can even take $Y=\set{\mathbbold{0},\mathbbold{1}}$.
 \end{proposition}

\proofpro{maintechnicalresult} Assume first that $\emptyset\notin {\bf\Gamma}$. Then $A^\infty =\Sigma^\omega$, since otherwise there is $x_0\in\Sigma^\omega\setminus A^\infty$, and the map constantly equal to $x_0$ continuously reduces $\emptyset$ to $A^\infty$, and thus $\emptyset\in {\bf\Gamma}$. If $A^\infty =\Sigma^\omega$, then $B=Y$ is as desired. So in the sequel we may and will assume that $\emptyset\in {\bf\Gamma}$.\medskip

 Note first that we have already seen in Section \ref{section-automata} that a (real-time) two-iterated counter automaton (accepting words by final states  \emph{and} empty stack) may be seen as a (real-time) two-counter automaton (accepting words by final states \emph{and} counters having the value zero). The idea is to code the content of two counters. We will need the following notion. Let $m\geq 1$ be any natural number, and $n,p,q$ be natural numbers such that neither $2$ nor $3$ divides $n\geq 1$, and $m= n\cdot 2^p\cdot 3^q$. Then, we set $M_2(m):= p$ and $M_3(m)= q$. So $2^{M_2(m)}$ is the greatest power of $2$ which divides $m$, and $3^{M_3(m)}$ is the greatest power of $3$ which divides $m$.
\medskip 

 Let then $\mathcal{A}:= (Q,\Alphabet,q_0,\Delta ,F)$ be any (real-time) two-iterated counter automaton accepting the language $A= \Lfinite(\mathcal{A})\subseteq\Alphabet^{<\omega}$, by final states  \emph{and} empty stack.
 
\vfill\eject
 
  We define the finitary language $\mathcal{L}$ as the set of finite words over the alphabet 
 $\Alphabet\cup\{\mathbbold{0},\mathbbold{1},\mathbbold{2}\}$, where  $\mathbbold{0},\mathbbold{1},\mathbbold{2}$ are new letters not in $\Alphabet$, of the form ${{}^\smallfrown}_{i<n}~v_i~a_i~\mathbbold{1}~w_i~z_i~\mathbbold{2}~u_{i+1}$, where $|v_0|= 1$, 
 $n\geq 1$, $v_i, w_i\in 1^+$, $z_i,u_i\in 1^{<\omega}$, $a_i\in\Alphabet$, $|u_{i+1}|= |z_i|$ and we can find a sequence $(q_i)_{i\leq n}$ of states of $Q$ and integers $l_i,l'_i\in\{ -1,0,1\}$ such that, for each $i< n$,
\[
a_i:\big( q_i,M_2(|v_i|),M_3(|v_i|)\big)\mapsto_{\mathcal{A}}
\big( q_{i+1},M_2(|v_i|)+ l_i,M_3(|v_i|)+ l'_i\big).
\]
and $|w_i|= |v_i|\cdot 2^{l_i}\cdot 3^{l'_i}$. Moreover, the state $q_{n}$ is a final state of 
$\mathcal{A}$, i.e., $q_{n}\in F$, and $M_2(|w_{n-1}|)= 0$, $M_3(|w_{n-1}|)= 0$. Note that the state $q_0$ of the sequence $(q_i)_{i\leq n}$ is also the initial state of $\mathcal{A}$.\medskip 

%\noindent\bf Claim 1\label{claimnumberone}\it\ 
\begin{claim}\label{claimnumberone} 
%The language $\mathcal{L}$ is accepted by a one-counter automaton $\mathcal{C}$.
There exists some  one-counter automaton $\mathcal{C}$ such that $\mathcal{L}=\Lfinite(\mathcal{C})$.
\end{claim}
%\rm\medskip
%%%%%%%%%%%%%%%%%%%%%%%%%%%%%%%%%%%%%%%%%%%%%%%%%%
%%%%%%%%%%%%%%%%%%%%%%%%%%%%%%%%%%%%%%%%%%%%%%%%%%
%%%%%%%%%%%%%%%%%%%%%%%%%%%%%%%%%%%%%%%%%%%%%%%%%%
%%%%%%%%%%%%%%%%%%%%%%%%%%%%%%%%%%%%%%%%%%%%%%%%%%
%%%%%%%%%%%%%%%%%%%%%%%%%%%%%%%%%%%%%%%%%%%%%%%%%%
%%%%%%%%%%%%%%%%%%%%%%%%%%%%%%%%%%%%%%%%%%%%%%%%%%

 \proofclaim{claimnumberone}
 We reproduce here the proof found in \cite[Proposition 17, Claim 1]{Fink-Lec4}, for sake of completeness and in order to keep the paper as self-contained as possible. This proof   gives an informal explanation of the way of a one-counter automaton $\mathcal{C}$ accepts the finitary language $\mathcal{L}$.  
 We first consider the reading of a word $w\in (\Si\cup\{ \mathbbold{0},\mathbbold{1},\mathbbold{2}\})^{<\omega}$ of the form 
\[
({{}^\frown}_{i<n}~ \mathbbold{0}^{p_i}~a_i~\mathbbold{1}~ \mathbbold{0}^{m_i}~\mathbbold{2})~\mathbbold{0}^{p_n}\mbox{,}
\]

\noindent where each $p_i$ and $m_i$ are  positive integers, and every $a_i$ belongs to $\Si$.\medskip

 Using the finite control, (i.e.,  a finite set of states together with a set of transitions involving only these states, together with the input letters that are read; this corresponds to the behavior of a finite state automaton)   the automaton $\mathcal{C}$ first checks that the first three letters of $w$ form an initial segment $\mathbbold{0}~a_0~\mathbbold{1}$, for some letter $a_0\in\Si$. Moreover, when reading the $p_0= 1$ letter $\mathbbold{0}$ before $a_0$, the automaton $\mathcal{C}$, using the finite control, checks that $p_0 > 0$ and determines whether $M_2(p_0)= 0$, and whether $M_3(p_0)= 0$. Here we actually have $M_2(p_0)= 0$ and $M_3(p_0)= 0$. Moreover the counter content of $\mathcal{C}$ is increased by one for each letter $\mathbbold{0}$ read.\medskip
  
  The automaton $\mathcal{C}$ now reads the letter $a_0$ and it guesses a transition of $\mathcal{A}$ leading to  
\[
a_0 :\big( q_0,M_2(p_0),M_3(p_0)\big)\mapsto_{\mathcal{A}}\big( q_1,M_2(p_0) + l_0,M_3(p_0) + l'_0\big).
\]
We set $v_0:= \mathbbold{0}^{p_0}$ and $w_0:= \mathbbold{0}^{p_0.2^{l_0}.3^{l'_0}}$.\medskip
 
 The counter value is now equal to $p_0$ and, when reading the letters $\mathbbold{0}$ following $a_0$, the automaton $\mathcal{C}$ checks that $m_0\geq p_0.2^{l_0}.3^{l'_0}$ in such a way that the counter value becomes zero after having read the 
$p_0.2^{l_0}.3^{l'_0}$ letters $\mathbbold{0}$ following the first letter $\mathbbold{1}$. For instance, if $l_0= l'_0= 1$, then 
$|w_0|= |v_0|.6$ so this can be done by decreasing the counter content by one each time six letters $\mathbbold{0}$ are read. Other cases are handled in a similar manner. Details are left to the reader's discretion.\medskip

Please also note that the automaton $\mathcal{C}$ has retained the value of state $q_1$ in its finite control.\medskip

 We now set $\mathbbold{0}^{m_0}:= w_0.z_0$. We have seen that, after having read $w_0$, the counter value of the automaton 
$\mathcal{C}$ is equal to zero. Now when reading $z_0$ the counter content is increased by one for each letter read so that it becomes $|z_0|$ after having read $z_0$. The automaton $\mathcal{C}$ now reads a letter $\mathbbold{2}$ and next decreases its counter by one for each letter $\mathbbold{0}$ read until the counter content is equal to zero. We set $\mathbbold{0}^{p_1}:= u_1.v_1$ with 
$u_1= z_0$. The automaton $\mathcal{C}$ now reads the segment $v_1$. Using the finite control, it checks that 
$|v_1| > 0$ and determines whether $M_2(|v_1|)= 0$, and whether $M_3(|v_1|)= 0$. Moreover the counter content is increased by one for each letter $\mathbbold{0}$ read. The automaton $\mathcal{C}$ now reads the letter $a_1$ and it guesses a transition of $\mathcal{A}$ leading to  
\[
a_1 :\big( q_1,M_2(|v_1|),M_3(|v_1|)\big)\mapsto_{\mathcal{A}}
\big( q_2,M_2(|v_1|) + l_1,M_3(|v_1|) + l'_1\big).
\]

\noindent We set $w_1:= \mathbbold{0}^{|v_1|.2^{l_1}.3^{l'_1}}$. The counter value is now equal to $|v_1|$. The automaton $\mathcal{C}$ now reads the second letter $\mathbbold{1}$ and, when reading the $m_1$ letters $\mathbbold{0}$ following this letter $\mathbbold{1}$, the automaton 
$\mathcal{C}$ checks that $m_1\geq |v_1|.2^{l_1}.3^{l'_1}$ in such a way that the counter value becomes zero after having read the $|v_1|.2^{l_1}.3^{l'_1}$ letters $\mathbbold{0}$ following the second letter $\mathbbold{1}$.
\medskip

 For instance, if $l_1= 0$ and $l'_1= -1$, then $|w_1|= |v_1|.3^{-1}$, so this can be done by decreasing the counter content by three each time one letter $\mathbbold{0}$ is read. And if $l_1= -1$ and $l'_1= -1$, then $|w_1|= |v_1|.2^{-1}.3^{-1}= |v_1|.6^{-1}$ so this can be done by decreasing the counter content by six each time one letter $\mathbbold{0}$ is read. The other cases are handled in a similar manner. Details are left to the reader's discretion.\medskip

 Note that these different cases can be treated using $\lambda$-transitions, in such a way that there will be at most 5 consecutive 
$\lambda$-transitions during a run of $\mathcal{C}$ on $w$. This will be an important and useful fact later on.\medskip
 
Please also note that the automaton $\mathcal{C}$ has retained the value of state $q_2$ in its finite control.\medskip
 
 The reading of $w$ by $\mathcal{C}$ continues in a similar manner. An acceptance condition by final states and empty stack  can be used to ensure that  $q_n\in F$, $M_2(|w_{n-1}|)= 0$, $M_3(|w_{n-1}|)= 0$, and $|z_{n-1}|= p_n$.\medskip 
 
 In order to complete the proof, we can remark that 
$\mathcal{R}= (\mathbbold{0}^{<\omega}.\Si.\mathbbold{1}.\mathbbold{0}^{<\omega}.\mathbbold{2}.\mathbbold{0}^{<\omega})^{<\omega}$ is a regular language,  so we have only considered the reading by $\mathcal{C}$ of words $w\in\mathcal{R}$. Indeed, if the language $L(\mathcal{C})$ was not included into 
$\mathcal{R}$, then we could replace it with $L(\mathcal{C})\cap\mathcal{R}$ because the class $OCL$ is closed under intersection with regular languages (by a classical construction of product of automata, the language $\mathcal{R}$ being accepted by a deterministic finite-state automaton).\medskip  

% We will now use a coding of $\omega$-words over $\Si$ given by the map $g_{N,l} :\Sio\ra (\Si\cup\{ \mathbbold{0},\mathbbold{1},\mathbbold{2}\})^\omega$, where $(N,l)\in\mathcal{P} :=\{ (N,l)\in (\omega\setminus\{ 0\} )\times\omega\mid 6\mbox{ does not divide }N\}$, and 
%\[
%g_{N,l}(\sigma ):= \mathbbold{0}~({{}^\frown}_{i\in\omega}~\sigma (i)~\mathbbold{1}~\mathbbold{0}^{N.6^{l+i}}~\mathbbold{2}~\mathbbold{0}^{N.6^{l+i}}).
%\]

  This proves Claim 1.\boxclaim{claimnumberone}

% \hfill{$\diamond$}\medskip  
%%%%%%%%%%%%%%%%%%%%%%%%%%%%%%%%%%%%%%%%%%%%%%%%%%
%%%%%%%%%%%%%%%%%%%%%%%%%%%%%%%%%%%%%%%%%%%%%%%%%%
%%%%%%%%%%%%%%%%%%%%%%%%%%%%%%%%%%%%%%%%%%%%%%%%%%
%%%%%%%%%%%%%%%%%%%%%%%%%%%%%%%%%%%%%%%%%%%%%%%%%%
%%%%%%%%%%%%%%%%%%%%%%%%%%%%%%%%%%%%%%%%%%%%%%%%%%
%%%%%%%%%%%%%%%%%%%%%%%%%%%%%%%%%%%%%%%%%%%%%%%%%%

 We then use a coding of $\omega$-words over $\Alphabet$ given by the mapping 
$g_{N,l}:\Alphabet^{\omega}\to (\Alphabet\cup\{ \mathbbold{0},\mathbbold{1},\mathbbold{2}\})^{\omega}$, where $(N,l)\in\mathcal{P}:=
\{ (N,l)\in (\omega\smallsetminus\{ 0\} )\times\omega\mid 6\mbox{ does not divide }N\}$, and 
\[
g_{N,l}(x ):= \mathbbold{0}~({{}^\smallfrown}_{i\in\omega}~
x (i)~\mathbbold{1}~\mathbbold{0}^{N\cdot 6^{l+i}}~\mathbbold{2}~\mathbbold{0}^{N\cdot 6^{l+i}}).
\]

\begin{claim}\label{claimnumbertwo} 
The equality 
$\big( \Lfinite(\mathcal{A})\big)^\infty= g_{N,l}^{-1}(\mathcal{L}^\infty )$ holds, i.e., 
 for all $x\in\Alphabet^{\omega}$ we have 
 \[
 g_{N,l}(x )\in\mathcal{L}^\infty\Leftrightarrow
x\in\big( \Lfinite(\mathcal{A}\big)^\infty.
\]
\end{claim}
%\rm\medskip
%%%%%%%%%%%%%%%%%%%%%%%%%%%%%%%%%%%%%%%%%%%%%%%%%%%%%%%%%%%%%%%%%%%%%%%%%%%%%%%%%%%%%%%%%%%%
%%%%%%%%%%%%%%%%%%%%%%%%%%%%%%%%%%%%%%%%%%%%%%%%%%%%%%%%%%%%%%%%%%%%%%%%%%%%%%%%%%%%%%%%%%%%
%%%%%%%%%%%%%%%%%%%%%%%%%%%%%%%%%%%%%%%%%%%%%%%%%%%%%%%%%%%%%%%%%%%%%%%%%%%%%%%%%%%%%%%%%%%%
%%%%%%%%%%%%%%%%%%%%%%%%%%%%%%%%%%%%%%%%%%%%%%%%%%%%%%%%%%%%%%%%%%%%%%%%%%%%%%%%%%%%%%%%%%%%
%%%%%%%%%%%%%%%%%%%%%%%%%%%%%%%%%%%%%%%%%%%%%%%%%%%%%%%%%%%%%%%%%%%%%%%%%%%%%%%%%%%%%%%%%%%%
%%%%%%%%%%%%%%%%%%%%%%%%%%%%%%%%%%%%%%%%%%%%%%%%%%%%%%%%%%%%%%%%%%%%%%%%%%%%%%%%%%%%%%%%%%%%
%%%%%%%%%%%%%%%%%%%%%%%%%%%%%%%%%%%%%%%%%%%%%%%%%%%%%%%%%%%%%%%%%%%%%%%%%%%%%%%%%%%%%%%%%%%%
%%%%%%%%%%%%%%%%%%%%%%%%%%%%%%%%%%%%%%%%%%%%%%%%%%%%%%%%%%%%%%%%%%%%%%%%%%%%%%%%%%%%%%%%%%%%
%%%%%%%%%%%%%%%%%%%%%%%%%%%%%%%%%%%%%%%%%%%%%%%%%%%%%%%%%%%%%%%%%%%%%%%%%%%%%%%%%%%%%%%%%%%%

 \proofclaim{claimnumbertwo}Once again, we reproduce here the proof found in \cite[Proposition 17, Claim 2]{Fink-Lec4}.
   We let $\mathcal{A}$ be a real-time two-iterated counter automaton accepting finite words over $\Si$, by final states 
{\it and} empty stack, and $\mathcal{L}\subseteq (\Si\cup\{ \mathbbold{0},\mathbbold{1},\mathbbold{2}\})^{<\omega}$ be defined as above.\medskip

 Let $x\in\Si^{\omega}$ be an $\omega$-word such that $g_{N,l}( x )\in\mathcal{L}^\infty$. Recall that $g_{N,l}( x )$ can be written 
\[
\mathbbold{0}~({{}^\frown}_{i\in\omega}~x (i)~\mathbbold{1}~\mathbbold{0}^{N.6^{l+i}}~\mathbbold{2}~\mathbbold{0}^{N.6^{l+i}}).
\]
As $g_{N,l}( x )\in\mathcal{L}^\infty$, we can also write 
$g_{N,l}( x )= {^\frown}_{j\in\omega}~({{}^\frown}_{i<n_j}~v_{i,j}~a_{i,j}~\mathbbold{1}~w_{i,j}~z_{i,j}~\mathbbold{2}~u_{i+1,j})$, where, for each natural number $j$, $|v_{0,j}|= 1$, $n_j\geq 1$, $v_{i,j},w_{i,j}\in 1^+$, $z_{i,j},u_{i,j}\in 1^{<\omega}$, 
$a_{i,j}\in\Si$, $|u_{i+1,j}|= |z_{i,j}|$ and we can find a sequence $(q_{i,j})_{i\leq n_j}$ of states of $Q$ such that 
$q_{0,j}= q_0$ is the initial state of $\mathcal{A}$, and integers $l_{i,j}, l'_{i,j}\in\{-1,0,1\}$ such that, for each $i< n_j$, 
\[
a_{i,j} :\big( q_{i,j},M_2(|v_{i,j}|),M_3(|v_{i,j}|)\big)\mapsto_{\mathcal{A}}
\big( q_{i+1,j},M_2(|v_{i,j}|) + l_{i,j},M_3(|v_{i,j}|) + l'_{i,j}\big).
\]

\noindent and $|w_{i,j}|= |v_{i,j}|.2^{l_{i,j}}.3^{l'_{i,j}}$. Moreover, the state $q_{n_j,j}$ is a final state, $M_2(|w_{n_j-1,j}|)= 0$, and 
$M_3(|w_{n_j-1,j}|)= 0$.\medskip 

 In particular, $|v_{0,0}|= 1= 2^0.3^0$. We will prove, by induction on $i< n_0$, that 
\[
|w_{i,0}|= 2^{M_2(|w_{i,0}|)}.3^{M_3(|w_{i,0}|)}\mbox{,}
\]
 
\noindent and $|w_{i,0}|= |v_{i+1,0}|$ if $i< n_0 - 1$. Moreover, setting $c_0^i= M_2(|v_{i,0}|)$ and $c_1^i= M_3(|v_{i,0}|)$, we will prove that, for each $i< n_0 - 1$, 
$a_{i,0} : (q_{i,0}, c_0^i,c_1^i)\mapsto_{\mathcal{A}} (q_{i+1,0},c_0^{i+1},c_1^{i+1})$.\medskip

 We have already seen that $|v_{0,0}|= 1= 2^0.3^0$. By hypothesis we can find a state $q_{1,0}\in Q$ and integers 
$l_{0,0},l'_{0,0}\in\{-1,0,1\}$ such that 
\[
a_{0,0} :\big( q_0,M_2(|v_{0,0}|),M_3(|v_{0,0}|)\big)\mapsto_{\mathcal{A}} 
\big( q_{1,0},M_2(|v_{0,0}|) + l_{0,0},M_3(|v_{0,0}|) + l'_{0,0}\big)\mbox{,}
\]

\noindent i.e., $a_{0,0} : (q_0,0,0)\mapsto_{\mathcal{A}} (q_{1,0},l_{0,0},l'_{0,0})$. Then, 
$|w_{0,0}|= |v_{0,0}|.2^{l_{0,0}}.3^{l'_{0,0}}= 2^{l_{0,0}}.3^{l'_{0,0}}$. Now note that 
$|w_{0,0}.z_{0, 0}|= |u_{1,0}.v_{1,0}|= \mathbbold{0}^{N.6^l}$ and $|u_{1,0}|= |z_{0,0}|$. Thus 
$|v_{1,0}|= |w_{0,0}|= 2^{l_{0,0}}.3^{l'_{0,0}}$. Setting $c_0^0:= 0$, $c_1^0:= 0$, 
$c_0^1:= l_{0,0}:= M_2(|v_{1,0}|)$ and $c_1^1:= l'_{0,0}:= M_3(|v_{1,0}|)$, it holds that 
\[
a_{0,0} : (q_0,c_0^0,c_1^0)\mapsto_{\mathcal{A}} (q_{1,0},c_0^1, c_1^1).
\]

\noindent Assume now that, for each $i< n_0 - 1$, it holds that 
$|w_{i,0}|= |v_{i+1,0}|= 2^{M_2(|w_{i,0}|)}.3^{M_3(|w_{i,0}|)}$ and 
$a_{i,0} : (q_{i,0},c_0^i,c_1^i)\mapsto_{\mathcal{A}} (q_{i+1,0}, c_0^{i+1},c_1^{i+1})$. We know that we can find a state $q_{n_0,0}\in Q$ and integers $l_{n_0-1,0},l'_{n_0-1,0}\in\{-1,0,1\}$ such that\medskip
 
\noindent $a_{n_0-1,0} :\big( q_{n_0-1,0},M_2(|v_{n_0-1,0}|),M_3(|v_{n_0-1,0}|)\big)\mapsto_{\mathcal{A}}$\smallskip
 
\centerline{$~~~~~~~~~~~~~~~~~~~~~~~~~~~~~~~~~~~~~~~~~~~~~~~~~~~~~~~~~~
\big( q_{n_0,0},M_2(|v_{n_0-1,0}|) + l_{n_0-1,0},M_3(|v_{n_0-1,0}|) + l'_{n_0-1,0}\big)\mbox{,}$}\medskip

\noindent i.e., $a_{n_0-1,0} : (q_{n_0-1,0},c_0^{n_0-1},c_1^{n_0-1})\mapsto_{\mathcal{A}} 
(q_{n_0},c_0^{n_0-1} + l_{n_0-1,0},c_1^{n_0-1} + l'_{n_0-1,0})$.\medskip

 Then, $|w_{n_0-1, 0}|= |v_{n_0-1,0}|.2^{l_{n_0-1,0}}.3^{l'_{n_0-1,0}}= 
2^{c_0^{n_0-1} + l_{n_0-1,0}}.3^{c_1^{n_0-1} + l'_{n_0-1,0}}= 1$ since, moreover, by hypothesis, 
$M_2(|w_{n_0-1,0}|)= M_3(|w_{n_0-1,0}|)= 0$.\medskip

 Finally, we proved the stated assertion by induction, and this shows that $a_{0,0}a_{1,0}\ldots a_{n_0-1,0}$ is accepted by 
$\mathcal{A}$, by final states and empty stack. On the other hand, 
$$|w_{n_0-1,0}.z_{n_0-1,0}|= N.6^{l+n_0-1}\mbox{,}$$ 
$|w_{n_0-1,0}|= 1$ and $|u_{n_0}|= |z_{n_0-1,0}|$, thus ${|u_{n_0}|= N.6^{l+n_0-1} - 1}$. As above, we can prove, by induction on $j$, that, for every $j\in\omega$, the finite word $a_{0,j}a_{1,j}\ldots a_{n_j-1,j}$ is in $\Lfinite(\mathcal{A})$, and thus 
$ x $ is in $\big( \Lfinite(\mathcal{A})\big)^\infty$.\medskip

 Conversely it is easy to see that if $x\in \big( \Lfinite(\mathcal{A})\big)^\infty$, then $g_{N,l}( x )\in\mathcal{L}^\infty$.
  \boxclaim{claimnumbertwo}%\hfill{$\diamond$}\medskip

 We now come back to the proof of Proposition \ref{maintechnicalresult}. By Claim 1, $\mathcal{L}$ is accepted by a one-counter automaton 
$\mathcal{C}$, which further satisfies that  there are never  more than 5 consecutive $\emptysequence$-transitions during any run of $\mathcal{C}$ on any finite word $w$. 
 The alphabet $Y$ will be $\Alphabet\cup\{ \mathbbold{0},\mathbbold{1},\mathbbold{2}\}$. 
\medskip

 We first define the language $B'$ which will be of the form $\mu\cup\mathcal{L}$, where $\mu$ is a finitary language over $Y$.
    In addition, we will ensure that $\mu$ is accepted by a one-counter automaton, (for which any run  on  finite words involves at most 5 consecutive $\emptysequence$-transitions) by finite states \emph{and} empty stack, so that it will also be the case of $B'$, by non-determinism. The set $\mathcal{L}^\infty$ will look like $A^\infty$ on some compact set $K_{1,0}$. We actually define, for any natural numbers $N,l$ such that  $6$ does not  divide $N\geq 1$, some compact sets $K_{N,l}$. The goal is to be able,  on each  $K_{N,l}$ space, to control the complexity of $B'^\infty$ (this will mainly be that of the complex $\mathcal{L}^\infty$).  Outside of the 
$K_{N,l}$'s spaces,  the complexity of $\mathcal{L}^\infty$ remains unknown to us. This is why we introduced $\mu$ in the first place. The set $\mu^\infty$ will be simple. It will hide the possible complexity of 
$\mathcal{L}^\infty$ out of the union of the $K_{N,l}$'s, but that will not hide the complexity of 
$\mathcal{L}^\infty$ on the $K_{N,l}$'s.

\vfill\eject

 So, we set 
\begin{itemize}
  \item $K_{N,l}:= g_{N,l}[\Alphabet^{\omega}]$, 

  \item $\mu:=\set{w\in Y^{<\omega}\mid \varphi(w)}$ where $\varphi(w)$ is the following formula:
\[
\varphi(w):=\begin{array}{l}\exists n\geq 2\\
\exists (a_i)_{i<n}\in\Alphabet^n\\
\exists (P_i)_{i<n}\in (\omega\smallsetminus\{ 0\} )^n\\
\exists (Q_i)_{i<n}\in\omega^n  \end{array}
 \left( \begin{array}{c}w= \mathbbold{0}~({{}^\smallfrown}_{i<n}~a_i~\mathbbold{1}~\mathbbold{0}^{P_i}~\mathbbold{2}~\mathbbold{0}^{Q_i})\\
  \bigwedge\\
(P_{n-2}\neq Q_{n-2}~\vee ~P_{n-1}\neq 6\cdot P_{n-2})\\
  \end{array}\right).
\]
\end{itemize}

\noindent Note that since $g_{N,l}$ is a homeomorphism onto its range, $K_{N,l}$ is a compact set. By Claim 2, 
$g_{N,l}^{-1}(\mathcal{L}^\infty\cap K_{N,l})= A^\infty$ holds for each $l\in\omega$.
 Note also  that all the words in $B'$ have the same form: $\mathbbold{0}~({{}^\smallfrown}_{i<n}~a_i~\mathbbold{1}~\mathbbold{0}^{P_i}~\mathbbold{2}~\mathbbold{0}^{Q_i})$. Finally, note that any finite concatenation of words of this form retains this form. We set 
\[
S:=\setbig{\mathbbold{0}({{}^\smallfrown}_{i\in\omega}~a_i~\mathbbold{1}~\mathbbold{0}^{P_i}~\mathbbold{2}~\mathbbold{0}^{Q_i})\mid (a_i)_{i\in\omega}\in\Alphabet^{\omega} ~\wedge ~(P_i)_{i\in\omega}\in (\omega\smallsetminus\{ 0\} )^{\omega} ~\wedge ~
(Q_i)_{i\in\omega}\in\omega^{\omega}}.
\]
We note that $\mu^\infty$  belongs to  $\mathbf{\Pi}^{0}_{2} (Y^{\omega} )$ for it can be written as
\[
\mu^\infty=\setbig{\gamma\in Y^{\omega}\mid\forall l\in\omega~\exists t\in\mu^l~~
{^\smallfrown}_{i<l}\ t(i)\subseteq\gamma}.
\]

 %Let $\mathbf{\Gamma}\in\setbig{D_{2n-1}(\mathbf{\Pi}^{0}_{k}),\checked{D}_{2n}(\mathbf{\Pi}^{0}_{k})}$. 
\noindent Note that 
$B'^\infty\cap K_{1,0}=\mathcal{L}^\infty\cap K_{1,0}$ satisfies $g_{1,0}^{-1}(\mathcal{L}^\infty\cap K_{1,0})= A^\infty$. Since $A^\infty$   
is $\mathbf{\Gamma}$-complete, and $\mathbf{\Gamma}$ is non-self-dual, we have $A^\infty$ is not a $\checked{\mathbf{\Gamma}}$-subset of $K_{1,0}$.
 Since $g_{1,0}$ is continuous, we also have $B'^\infty\cap K_{1,0}$ is not a $\checked{\mathbf{\Gamma}}$-subset of $K_{1,0}$. Finally, since the range of 
 $g_{1,0} :\Alphabet^{\omega}\to Y^{\omega}$ is none other than  $K_{1,0}$, we obtain that $B'^\infty$ is not a $\checked{\mathbf{\Gamma}}$-subset of $Y^{\omega}$. So, it only remains to show that $B'^\infty$ is a $\bf\Gamma$-subset of $Y^{\omega}$ in order to show that  $B'^\infty$ is $\mathbf{\Gamma}$-complete.
 
%
%is not a $\checked{\mathbf{\Gamma}}$-subset of $K_{1,0}$  since $g_{1,0}^{-1}(\mathcal{L}^\infty\cap K_{1,0})= A^\infty$, and $A^\infty$   
%being $\mathbf{\Gamma}$-complete does not belong\footnote{See \cite[Section 22.B]{kechris1995classical} and recall that $D_{2n-1}(\mathbf{\Pi}^{0}_{k})$ and $\checked{D}_{2n}(\mathbf{\Pi}^{0}_{k})$ are a non-self-dual classes.}  to the class $\checked{\mathbf{\Gamma}}$, so that $B'^\infty$ is not a 
%$\checked{\mathbf{\Gamma}}$-subset of $Y^{\omega}$. Since $\mathbf{\Gamma}$ and $\checked{\mathbf{\Gamma}}$ are non-self-dual classes,  it remains to see that $B'^\infty$ is a $\bf\Gamma$-subset of $Y^{\omega}$ in order to show that  $B'^\infty$ is $\mathbf{\Gamma}$-complete.
 %
\medskip

 We define $F: S\smallsetminus\mu^\infty\to (\{\emptysequence\}\cup\mu )\times\omega\times\omega$ as follows. Let 
 \begin{multicols}{2}
 \begin{itemize}
  \item $\gamma:=  \mathbbold{0}\,({{}^\smallfrown}_{i\in\omega}\,a_i\,\mathbbold{1}\,\mathbbold{0}^{P_i}\,\mathbbold{2}\,\mathbbold{0}^{Q_i})\in 
 S\smallsetminus\mu^\infty$
  \item $(N,l)\in\mathcal{P}\text{ with }P_0= N\cdot 6^l$
 \end{itemize}
 \end{multicols}
 
%
%\[\gamma:=  \mathbbold{0}\,({{}^\smallfrown}_{i\in\omega}\,a_i\,\mathbbold{1}\,\mathbbold{0}^{P_i}\,\mathbbold{2}\,\mathbbold{0}^{Q_i})\in 
% S\smallsetminus\mu^\infty~~\text{ and }~~ (N,l)\in\mathcal{P}\text{ with }P_0= N\cdot 6^l.
% \]
  \begin{enumerate}[(a)]
  \item If ${\gamma\in K_{N,l}}$,  then we set $F(\gamma ):= (\emptysequence ,N,l)$.
  \item If ${\gamma\notin K_{N,l}}$, then there is some maximal integer $i_0$ for which $P_{i_0}\neq Q_{i_0}$ or $P_{i_0+1}\neq 6\cdot P_{i_0}$.\medskip 

  We let $(N',l')\in\mathcal{P}$ with $P_{i_0+1}= N'\cdot6^{l'}$, and set
\[
F(\gamma ):=\big( \mathbbold{0}~({{}^\smallfrown}_{i\leq i_0}~a_i\,\mathbbold{1}\,\mathbbold{0}^{P_i}\,\mathbbold{2}\,\mathbbold{0}^{Q_i})\,
a_{i_0+1}\,\mathbbold{1}\,\mathbbold{0}^{P_{i_0+1}}\,\mathbbold{2}\,\mathbbold{0}^{Q_{i_0+1}-1},N',l'+ 1\big).
\]

\end{enumerate} 
Consider any $\gamma\in B'^\infty\smallsetminus\mu^\infty$. Note that   $\gamma\in S\smallsetminus\mu^\infty$, so that $F(\gamma):=(t,N,l)$ is defined,   with $t\subseteq\gamma$, and  $\gamma-t\in K_{N,l}$.
%\begin{enumerate}[i.]
%
%  \item  $(t,N,l):= F(\gamma )$ is defined,   \item $t\subseteq\gamma$   \item  $\gamma-t\in K_{N,l}$.
% 
%\end{enumerate} 
 For $(t,N,l)\in \mathrm{range}(F)$, we set  
\begin{itemize}
  \item $P_{t,N,l}:=\setbig{\gamma\in Y^{\omega}\mid t\subseteq\gamma ~\wedge ~\gamma -  t\in K_{N,l}}$ 
\item $A_{t,N,l}:=\setbig{\gamma\in P_{t,N,l}\mid\gamma -  t\in\mathcal{L}^\infty\cap K_{N,l}}$ .
\end{itemize}
 Note that $P_{t,N,l}$ is compact, contained in $S\smallsetminus\mu^\infty$, and $F(\gamma )= (t,N,l)$ if $\gamma\in P_{t,N,l}$. This shows that the $P_{t,N,l}$'s are pairwise disjoint and disjoint from 
$\mu^\infty$. Note also that $A_{t,N,l}$ belongs to $\bf\Gamma$. The previous discussion shows that 
\[
B'^\infty=\mu^\infty\cup\bigcup_{(t,N,l)\in \mathrm{range}(F)}~A_{t,N,l}.
\]
 In particular, $B'^\infty$ belongs to 
$\mathbf{\Delta}^{0}_{3}\mbox{-PU}(\mathbf{\Gamma})$ which is none other than $\mathbf{\Gamma}$, according to the hypothesis.

%
%It follows, therefore, from Proposition \ref{closureunderpartitionedunions} that $B'^\infty$ also belongs to  $\mathbf{\Gamma}$.
 %
\medskip

To conclude this part of the proof, we first note that it is easy to see that, as was the case for the finite language $\mathcal{L}$,  the finitary language $\mu$ is accepted by a non-deterministic one-counter automaton, for which  there are also at most 5 consecutive $\emptysequence$-transitions during any run  on any finite word,  with acceptance by final states \emph{and} empty counter. We leave the details to the reader's discretion.
\medskip

 From this, it is also easy to see that the language $B'=\mu\cup\mathcal{L}$ is  accepted by a  non-deterministic one-counter automaton,   by final states \emph{and} empty counter, for which   there are also at most 5 consecutive $\emptysequence$-transitions during any run  on any finite word. In order to find the language $B$ from the language $B'$ we use a simple morphism which is a very particular case of a substitution.  If the alphabet of $B'$ is $Y:=\{ a_1,a_2,\cdots, a_k\}$ for some integer $k\geq 1$, then we add a letter $c$ to $Y$, set $Z:= Y\cup\{ c\}$, and consider the morphism 
$h: Y\to Z^{<\omega}$ defined by $h(a_i)= a_i c^6$ for each integer $i\in [1, k]$. 
This morphism naturally extends to words, then to languages, and we set $B= h[B']$. 
A word of $B$ is simply obtained from a word $w$ of $B'$ by adding the word \quq{$cccccc$} after each letter of $w$. It is then easy to see that the language $B$ is accepted by a  non-deterministic {\it  real-time} one-counter automaton $\mathcal{A}$ by final states \emph{and} empty counter. This automaton is simply obtained from a  non-deterministic one-counter automaton  $\mathcal{B}$, for which   there are also at most 5 consecutive $\emptysequence$-transitions during a run  on a finite word, accepting $B'$ by final states \emph{and} empty counter. 
The simple idea is that $\emptysequence$-transitions from $\mathcal{B}$ now occur while $\mathcal{A}$ reads the letters $c$ in a   word of $B$.
\medskip

 Furthermore, $B^\infty$ is $\bf\Gamma$-complete, for the reason that 
${B'}^\infty\subseteq Y^\omega$ and $B^\infty\subseteq (Y\cup\{ c\} )^\omega$ are Wadge equivalent. Indeed, ${B'}^\infty\leq_\mathbb{W}B^\infty$ holds since $B=h[B']$. In particular, 
$B^\infty$ is $\bf\Gamma$-hard. For the other direction, let 
$$P=\{ x\in (Y\cup\{ c\} )^\omega\mid\forall n\in\omega ~~x(7n)\in Y\wedge\forall p<6~~x(7n+p)=c\} .$$ Then $B^\infty$ is contained in the closed subset $P$ of $(Y\cup\{ c\} )^\omega$. [Kec95, 2.8] provides a continuous map $r:(Y\cup\{ c\} )^\omega\rightarrow P$ such that $r(x)=x$ for each 
$x\in P$. The map $g:P\rightarrow Y^\omega$ defined by $g(x)=x_{[/c]}$ is continuous and satisfies 
$B^\infty =g^{-1}({B'}^\infty)$. Now note that 
$$B^\infty =\big( P\cap (g\circ r)^{-1}({B'}^\infty )\big)\cup 
\Big(\big( (Y\cup\{ c\} )^\omega\setminus P\big)\cap\emptyset\Big)$$ 
is in ${\bf\Delta}^0_3$-PU$({\bf\Gamma})={\bf\Gamma}$ since 
$(g\circ r)^{-1}({B'}^\infty ),\emptyset\in {\bf\Gamma}$. Thus $B^\infty$ is $\bf\Gamma$-complete.\medskip

%Furthermore,  $B^\infty$ is $\mathbf{\Gamma}$-complete, for the reason that $B'^\infty\subseteq Y^{\omega}$ and $B^\infty\subseteq (Y\cup\{ c\})^{\omega}$ are Wadge equivalent. 
%Indeed, $B'^\infty\winf B^\infty$ holds since $B= h[B']$. For the other direction, it is enough to show that there exists some winning strategy for Player $\two$ in the Wadge game $\GWadge{B'^\infty}{B'^\infty}$ that satisfies the following extra condition: for all positions $u\in  Y^{<\omega}$ that player $\two$ may reach, there exists some $x\in Y^{\omega}$ such that 
%$ux\notin B'^\infty$. It turns out that such a condition is necessarily fulfilled as soon as $\bigcup_{n<\omega}\mathbf{\Delta}^0_n\subsetneq\mathbf{\Gamma}$ holds (for instance, this is an easy  consequence of \cite[Proposition 17]{Duparc01}). Moreover, since $\mathbf{\Gamma}=\mathbf{\Delta}^{0}_{3}\mbox{-PU}(\mathbf{\Gamma})$ and clearly 
%$\mathbf{\Delta}^{0}_{3}\subseteq\mathbf{\Delta}^{0}_{3}\mbox{-PU}(\mathbf{\Gamma})$ holds, we have  $\mathbf{\Delta}^{0}_{3}\subseteq\mathbf{\Gamma}$, hence in particular $\bigcup_{n<\omega}\mathbf{\Delta}^0_n\subsetneq\mathbf{\Gamma}$. This guarantees that there exists a winning strategy for Player $\two$ in $\GWadge{B'^\infty}{B'^\infty}$ with the property that whatever position that Player $\two$ reaches can always be extended to an infinite word that does not belong to $B'^\infty$. 
%
%\medskip
%

 Finally, by using the  injective homomorphism  
$
 \begin{array}[t]{crcl}
 f:& Z&\to&\set{\mathbbold{0},\mathbbold{1}}^{<\omega} \\
 & a_j&\mapsto&\mathbbold{0}^j\mathbbold{1}\hspace{3em}(1\leq j\leq k)\mbox{,}\\
 & c&\mapsto&\mathbbold{0}^{k+1}\mathbbold{1}\\
\end{array}
$ 

\noindent one easily sees that the language 
$f[B]\subseteq\{\mathbbold{0},\mathbbold{1}\}^{<\omega}$ is accepted by a non-deterministic 
{\it real-time} one-counter automaton by final states \emph{and} empty counter.
\medskip

Finally, for the same reasons as those mentioned above, it is easy to see that if  $B^\infty$ is $\mathbf{\Gamma}$-complete,  then $f[B]^\infty$ is also $\mathbf{\Gamma}$-complete.
\boxpro{maintechnicalresult} 

 As an immediate consequence, we obtain the following.

\begin{corollary} \label{oldmaintechnicalresultnowcor} 
Let $A\subseteq\Alphabet^{<\omega}$ be any finitary language accepted by a real-time \underline{two-iterated} counter automaton $\mathcal{A}$ accepting words by final states  \emph{and} empty stack, and such that the 
$\omega$-power $A^\infty$ is $D_{2n-1}(\mathbf{\Pi}_k^0)$-complete  (respectively, 
$\checked{D}_{2n}(\mathbf{\Pi}_k^0)$-complete) for some natural numbers $n\geq 1$ and $k\geq 3$. Then, one can find a finite alphabet $Y$ and a finitary language $B\subseteq Y^{<\omega}$ such that \begin{itemize}

    \item $B$ is the language accepted by a real-time \underline{one-counter} automaton $\mathcal{B}$ accepting words by final states 
\emph{and} empty stack, 
  \item   $B^\infty$ is $D_{2n-1}(\mathbf{\Pi}_k^0)$-complete (respectively, $\checked{D}_{2n}(\mathbf{\Pi}_k^0)$-complete),

\end{itemize}
Furthermore, one can even take $Y=\set{\mathbbold{0},\mathbbold{1}}$.

\end{corollary}

\vfill\eject

\proofcor{oldmaintechnicalresultnowcor}By Proposition \ref{closureunderpartitionedunions}, the class  $D_{2n-1}(\mathbf{\Pi}_k^0)$ is closed under partitioned  unions with $\mathbf{\Delta}^{0}_{k}$-sets.  As for the class $\checked{D}_{2n}(\mathbf{\Pi}_k^0)$, notice that 
${\mathbf\Delta}^{0}_{k}\mbox{-PU}\big(\checked{D}_{2n}(\mathbf{\Pi}_k^0)\big)$ is equal to 
\[
 \arraycolsep=1pt\def\arraystretch{1.5}
\begin{array}{ll}
& \{\bigcup_{i\in\omega}~(P_{i}\cap A_{i}^{-})\mid 
(P_{i})_{i\in\omega}\in (\mathbf{\Delta}^{0}_{k})^{\omega}\mbox{-partition }, 
(A_{i})_{i\in\omega}\in {D}_{2n}(\mathbf{\Pi}_k^0)^\omega\}\cr
= & \{\big(\bigcup_{i\in\omega}~(P_{i}\cap A_{i})\big)^{-}\mid 
(P_{i})_{i\in\omega}\in (\mathbf{\Delta}^{0}_{k})^{\omega}\mbox{-partition }, 
(A_{i})_{i\in\omega}\in {D}_{2n}(\mathbf{\Pi}_k^0)^{\omega}\}\cr
= & \{A^{-}\mid 
A\in {D}_{2n}(\mathbf{\Pi}_k^0)\}\cr
= & {D}_{2n}^{-}(\mathbf{\Pi}_k^0).\\
\end{array}
\]
So, it results that both classes $D_{2n-1}(\mathbf{\Pi}_k^0)$ and ${D}_{2n}^{-}(\mathbf{\Pi}_k^0)$ are closed under partitioned  unions with $\mathbf{\Delta}^{0}_{k}$-sets.
 Since $k\geq 3$ holds, it follows that $D_{2n-1}(\mathbf{\Pi}_k^0)$ and ${D}_{2n}^{-}(\mathbf{\Pi}_k^0)$ are closed under partitioned  unions with $\mathbf{\Delta}^{0}_{3}$-sets. The results are then  immediate consequences of Proposition  \ref{maintechnicalresult}.\medskip

 We recall Proposition \ref{closureunderpartitionedunions} which stated that or any  natural numbers $n',k'\geq 1$, the class $D_{n'}(\mathbf{\Pi}^{0}_{k'})$ is closed under   partitioned  unions with $\mathbf{\Delta}^{0}_{k'}$-sets; namely  
$\mathbf{\Gamma} = \mathbf{\Delta}^{0}_{k'}\mbox{-PU}(\mathbf{\Gamma})$ where $\mathbf{\Gamma}:= D_{n'}(\mathbf{\Pi}^{0}_{k'})$ and 
\[
\mathbf{\Delta}^{0}_{k'}\mbox{-PU}(\mathbf{\Gamma}):=\setBig{\bigcup_{i\in\omega}~(P_{i}\cap A_{i})\mid 
(P_{i})_{i\in\omega}\in (\mathbf{\Delta}^{0}_{k'})^{\omega}\mbox{-partition }\wedge 
(A_{i})_{i\in\omega}\in \mathbf{\Gamma}^{\omega}}.
\]
\noindent By Proposition \ref{closureunderpartitionedunions}, the class  $D_{2n-1}(\mathbf{\Pi}_k^0)$ is closed under partitioned  unions with $\mathbf{\Delta}^{0}_{k}$-sets. Since $k\geq 3$ holds, it follows that the class  $D_{2n-1}(\mathbf{\Pi}_k^0)$ is closed under partitioned  unions with $\mathbf{\Delta}^{0}_{3}$-sets, finishing the proof.\boxcor{oldmaintechnicalresultnowcor}

 We now come to the proof of the main result of this section.

  \themaintheorem
   
\proofthm{thm:themaintheorem}
For $k= 1,2$, Proposition \ref{FLpin} and the discussion that follows provide, regular finitary languages  $U_{n,k},V_{n,k}$ such that $U_{n,k}^\infty$ is $D_{2n-1}(\mathbf{\Pi}_k^0)$-complete and 
$V_{n,k}^\infty$ is $\checked{D}_{2n}(\mathbf{\Pi}_k^0)$-complete. So we are done if $k\leq 2$.\medskip 

 This discussion also provides a finite alphabet $\Alphabet_3$ and a finitary language  
$U_{n,3}\subseteq\Alphabet_3^{<\omega}$, accepted by a real-time one-counter automaton, by final states \emph{and} empty stack, such that $U_{n,3}^\infty$ is $D_{2n-1}(\mathbf{\Pi}_3^0)$-complete. By an argument similar to the one used in the last paragraph of the proof of Proposition \ref{maintechnicalresult}, it is possible to get $\Alphabet_3=\{ \mathbbold{0},\mathbbold{1}\}$ with the same property.
\medskip

By Proposition \ref{sub}, if $h:\Alphabet_3\to 2^{(\Alphabet_3\cup\{\tla\} )^{<\omega}}$ is the substitution defined by $a\mapsto \mathbb{L}a$, then the language $h(U_{n,3})$ is in $\mbox{OCL}(2)$, and $h(U_{n,3})$ is accepted by a real-time two-iterated counter automaton accepting words by final states \emph{and} empty stack. The proof of Proposition \ref{FLpin} shows that $h(U_{n,3})^\infty$ is 
$D_{2n-1}(\mathbf{\Pi}_4^0)$-complete. From there,  Proposition \ref{maintechnicalresult} provides a finite alphabet 
$\Alphabet_4=\{ \mathbbold{0},\mathbbold{1}\}$ and a finitary language $U_{n,4}\subseteq\Alphabet_4^{<\omega}$, accepted by a real-time one-counter automaton, by final states \emph{and} empty stack, such that 
$U_{n,4}^\infty$ is $D_{2n-1}(\mathbf{\Pi}_4^0)$-complete. All that is left is simply to repeat this argument with 
$k\geq 4$ instead of $3$.\medskip

The argument is similar for the classes $\checked{D}_{2n}(\mathbf{\Pi}_k^0)$, on the sole condition, however, that  there exists a finitary language $V_{n,3}$ accepted by a one-counter automaton,  such that its $\omega$-power $V_{n,3}^\infty$ is $\checked{D}_{2n}(\mathbf{\Pi}_3^0)$-complete.
\medskip

 We obtained an inductive construction of languages $U_{n,k}$ accepted by one counter automata such that $U_{n,k}^\infty$ is $D_{2n-1}(\mathbf{\Pi}_k^0)$-complete. We can argue slightly differently, as follows.
\medskip

  First we can show that Proposition  \ref{maintechnicalresult} is valid if we replace in the hypothesis ``a real-time two-iterated counter automaton $\mathcal{A}$''  by ``a real-time $l$-iterated counter automaton 
$\mathcal{A}$, for some integer $l\geq 2$'';  the proof of this extension of Proposition \ref{maintechnicalresult} is very similar to the proof of Proposition \ref{maintechnicalresult}, the idea being that we have in this case to code the content of $l$ counters, but the ideas and the constructions of the proof are very similar, details are here left to the reader.  Theorem 
\ref{thm:themaintheorem}{\rm .(a)} now follows from Proposition \ref{FLpin} and from this extension of Proposition \ref{maintechnicalresult}. Note that we  only gave a detailed proof of Proposition  \ref{maintechnicalresult} in the case $l= 2$ because it is easier to explain and this case encompasses all the fundamental ideas of the proof of the extended case of an integer $l\geq 2$. 
\boxthm{thm:themaintheorem}

\section{A conditional result for the regular $\omega$-powers in ${\bf\Delta}^0_2$}\indent

The following result is proved in [FL24b].\medskip
 
\noindent\bf Theorem.\it\ Let $n$ be a natural number.\smallskip

(a) We can find a regular language $L\!\subseteq\! 2^{<\omega}$ with the property that $L^\infty$ is complete for the class $D_n({\bf\Sigma}^0_1)$, and another one for $D_n^-({\bf\Sigma}^0_1)$.\smallskip

(b) We can find a regular language $L\!\subseteq\! 2^{<\omega}$ with the property that $L^\infty$ is complete for the class $D_0({\bf\Sigma}^0_1)\oplus D^-_0({\bf\Sigma}^0_1)$, and another one for $D_{2n+1}({\bf\Sigma}^0_1)\oplus D^-_{2n+1}({\bf\Sigma}^0_1)$.\rm\medskip

 It was also noticed that only the classes of the form 
$D_{2n+2}({\bf\Sigma}^0_1)\oplus D^-_{2n+2}({\bf\Sigma}^0_1)$ were missing to get a complete result for the Wadge classes ${\bf\Gamma}\subseteq {\bf\Delta}^0_2$ having an $\omega$-regular complete set. The main result of this section conditionally fills this gap:

\begin{theorem} \label{condelta} If we can find a regular language $A\!\subseteq\! 2^{<\omega}$ with the property that $A^\infty$ is complete for the class $D_2(\boraone )\oplus\checked{D}_2(\boraone )$, then, for each natural number $n$, we can find a regular language $L\!\subseteq\! 2^{<\omega}$ with the property that $L^\infty$ is complete for the class $D_{2n+2}(\boraone )\oplus\checked{D}_{2n+2}(\boraone )$.\end{theorem}

 Our proof is based on lemmas in the style of those in [FL24b]. We use the following alternative classical definition of the differences (see [Kec95, 22E]). If $\xi$ is a countable ordinal and 
$O\! :=\! (O_\eta )_{\eta <\xi}$ is an increasing sequence of subsets of a set $X$, then\medskip

\centerline{$D_\xi (O)\! :=\!\{ x\!\in\! X\mid\exists\eta\! <\!\xi ~~\mbox{parity}(\eta )\!\not=\!\mbox{parity}(\xi )\wedge x\!\in\! O_\eta\!\setminus\! (\bigcup_{\theta <\eta}~O_\theta )\} .$}\medskip

\noindent If moreover $\bf\Gamma$ are classes of subsets of Polish spaces, then 
$$D_\xi ({\bf\Gamma})\! =\!\{ D_\xi (O)\mid X\mbox{ Polish}\wedge
\forall\eta\! <\!\xi ~~O_\eta\!\in\! {\bf\Gamma}(X)\}\mbox{,}$$
${\bf\Gamma}\oplus {\bf\Lambda}\! :=\!\{ (A\cap C)\cup 
(B\!\setminus\! C)\mid C\!\in\!\borone\wedge A\!\in\! {\bf\Gamma}\wedge B\!\in\! {\bf\Lambda}\}$, and 
${\bf\Gamma}\sqcap {\bf\Lambda}\! :=\!
\{ A\cap B\mid A\!\in\! {\bf\Gamma}\wedge B\!\in\! {\bf\Lambda}\}$. 

\begin{lemma} \label{set} Let $\bf\Gamma$ be a non self-dual Wadge class of Borel subsets of Polish spaces closed under finite intersections and unions, $k$ be a natural number, and $\varepsilon\!\in\! 2$.\smallskip

(a) $\checked{D}_k({\bf\Gamma})\sqcap {\bf\Gamma}\! =\! D_{k+1}({\bf\Gamma})$.\smallskip

(b) $\checked{D}_{2k}({\bf\Gamma})\sqcap\checked{\bf\Gamma}\! =\!\checked{D}_{2k+1}({\bf\Gamma})$.\smallskip

(c) $\checked{D}_{2k+\varepsilon}({\bf\Gamma})\sqcap\checked{D}_2({\bf\Gamma})\! =\!
\checked{D}_{2k+2+\varepsilon}({\bf\Gamma})$.\smallskip

(d) $D_{2k+1+\varepsilon}({\bf\Gamma})\sqcap\checked{D}_2({\bf\Gamma})\! =\! D_{2k+3+\varepsilon}({\bf\Gamma})$.\smallskip

(e) $\big(D_{2k+1+\varepsilon}({\bf\Gamma})\oplus\checked{D}_{2k+1+\varepsilon}({\bf\Gamma})\big)
\sqcap\checked{D}_2({\bf\Gamma})\! =\! 
D_{2k+3+\varepsilon}({\bf\Gamma})\oplus\checked{D}_{2k+3+\varepsilon}({\bf\Gamma})$.\end{lemma}

\vfill\eject

 Some items of Lemma \ref{set} were known for quite some time. Namely, they are particular cases of [\cite{Selivanov85}, §5, Proposition 8].\medskip

\prooflem{set} Fix a Polish space $X$. We start with (c) since (a), (b) are proved in 
[FL24b, Lemma 3.1].\medskip 

\noindent (c) By [FL24b, Lemma 3.1], we may assume that $\varepsilon\! =\! 1$. By (b), we may assume that $k\! >\! 0$. Let $U\! :=\! (U_\eta )_{\eta <2k+1}$ be an increasing sequence of subsets of $X$ in 
$\bf\Gamma$, $A\! :=\! X\!\setminus\! D_{2k+1}(U)$, ${V\! :=\! (V_\eta )_{\eta <2}}$ be an increasing sequence of subsets of $X$ in $\bf\Gamma$, and ${B\! :=\! X\!\setminus\! D_2(V)}$. We successively define $O_0\! :=\! U_0$, $O_1\! :=\! (V_0\cap U_1)\cup U_0$, 
$O_{2j}\! :=\! (V_1\cap U_{2j})\cup U_{2j-2}$ if $0\! <\! j\!\leq\! k$,  
$O_{2j+1}\! :=\! (V_0\cap U_{2j+1})\cup U_{2j-1}\cup (V_1\cap U_{2j})$ if $0\! <\! j\! <\! k$, 
$O_{2k+1}\! :=\! V_0\cup U_{2k-1}\cup (V_1\cap U_{2k})$ and 
$O_{2k+2}\! :=\! V_1\cup U_{2k}$, so that $(O_\eta )_{\eta <2k+3}$ is an increasing sequence of subsets of $X$ in $\bf\Gamma$. Moreover, 
$$\begin{array}{ll}
x\!\in\! A\!\cap\! B\!\!\!\!\!
& \Leftrightarrow\! (x\!\notin\! U_{2k}\vee
\exists j\! <\! k~~x\!\in\! U_{2j+1}\!\setminus\! U_{2j})\wedge 
x\!\in\!\big( V_0\cup (X\!\setminus\! V_1)\big)\cr
& \Leftrightarrow\! x\!\in\! V_0\!\setminus\! U_{2k}\vee
\exists j\! <\! k~~x\!\in\! (V_0\cap U_{2j+1})\!\setminus\! U_{2j}\vee
x\!\notin\! V_1\cup U_{2k}~\vee\cr
& \hfill{\exists j\! <\! k~~x\!\in\! U_{2j+1}\!\setminus\! (V_1\cup U_{2j})}\cr
& \Leftrightarrow\! x\!\in\! (V_0\!\cap\! U_1)\!\setminus\! U_0\vee\exists j\!\in\! (0,k)~
x\!\in\!\big( (V_0\!\cap\! U_{2j+1})\!\setminus\! U_{2j})\big)\!\cup\!
\big( U_{2j-1}\!\setminus\! (V_1\!\cup\! U_{2j-2})\big)\cr
& \hfill{\vee x\!\in\!\big( V_0\!\setminus\! U_{2k}\cup 
U_{2k-1}\!\setminus\! (V_1\cup U_{2k-2})\big)\vee
x\!\notin\! V_1\cup U_{2k}}\cr
& \Leftrightarrow\! x\!\in\! (V_0\!\cap\! U_1)\!\setminus\! U_0\vee\exists j\!\in\! (0,k)~
x\!\in\!\big( (V_0\!\cap\! U_{2j+1})\cup U_{2j-1}\big)\!\setminus\!\big( (V_1\!\cap\! U_{2j})\!\cup\! U_{2j-2}\big)\cr
& \hfill{\vee x\!\in\! (V_0\cup U_{2k-1})\!\setminus\!\big( (V_1\cap U_{2k})\cup U_{2k-2}\big)\vee 
x\!\notin\! V_1\cup U_{2k}}\cr
& \Leftrightarrow\! x\!\notin\! D_{2k+3}(O)\mbox{,}
\end{array}$$
proving that $A\cap B\!\in\!\checked{D}_{2k+3}({\bf\Gamma})$ (this is based on the fact that 
$$(C_1\!\setminus\! C_0)\cup (D_1\!\setminus\! D_0)\! =\! (C_1\cup D_1)\!\setminus\! (C_0\cap D_0)$$ 
if $C_1\!\subseteq\! D_0$ and $D_1\!\subseteq\! C_0$).\medskip

 Conversely, assume that $D_{2k+3}(O)\!\in\! D_{2k+3}({\bf\Gamma})$. We set, for $\eta\! <\! 2k\! +\! 1$, 
$U_\eta\! :=\! O_\eta$, which implies that $U$ is an increasing sequence of subsets of $X$ in 
$\bf\Gamma$ and $D_{2k+1}(U)\!\in\! D_{2k+1}({\bf\Gamma})$. We also set, for $\eta\! <\! 2$, 
$V_\eta\! :=\! O_{2k+1+\eta}$, which implies that $V$ is an increasing sequence of subsets of $X$ in 
$\bf\Gamma$ and $D_2(V)\!\in\! D_2({\bf\Gamma})$. Moreover, 
$X\!\setminus\! D_{2k+3}(O)\! =\!\big( X\!\setminus\! D_{2k+1}(U)\big)\cap\big( X\!\setminus\! D_2(V)\big)$ by the equivalences just above.\medskip

\noindent (d) Assume first that $\varepsilon\! =\! 1$. Let $U\! :=\! (U_\eta )_{\eta <2k+2}$ be an increasing sequence of subsets of $X$ in $\bf\Gamma$, $A\! :=\! D_{2k+2}(U)$, $V\! :=\! (V_\eta )_{\eta <2}$ be an increasing sequence of subsets of $X$ in $\bf\Gamma$, and ${B\! :=\! X\!\setminus\! D_2(V)}$. We then set 
$O_0\! :=\! U_0$, ${O_1\! :=\! (V_0\cap U_1)\cup U_0}$, 
${O_{2j}\! :=\! (V_1\cap U_{2j})\cup U_{2j-2}}$ and 
$$O_{2j+1}\! :=\! (V_0\cap U_{2j+1})\cup U_{2j-1}\cup (V_1\cap U_{2j})$$ 
whenever $0\! <\! j\!\leq\! k$, ${O_{2k+2}\! :=\! V_1\cup U_{2k}}$ and $O_{2k+3}\! :=\! V_1\cup U_{2k+1}$, so that 
$(O_\eta )_{\eta <2k+4}$ is an increasing sequence of subsets of $X$ in $\bf\Gamma$, and
$$\begin{array}{ll}
x\!\in\! A\!\cap\! B\!\!\!\!\!
& \Leftrightarrow\! (\exists j\!\leq\! k~~x\!\in\! U_{2j+1}\!\setminus\! U_{2j})\wedge 
x\!\in\!\big( V_0\cup (X\!\setminus\! V_1)\big)\cr
& \Leftrightarrow\!\exists j\!\leq\! k~~x\!\in\! (V_0\cap U_{2j+1})\!\setminus\! U_{2j}\cup 
U_{2j+1}\!\setminus\! (V_1\cup U_{2j})\cr
& \Leftrightarrow\! x\!\in\! (V_0\!\cap\! U_1)\!\setminus\! U_0\vee\exists j\!\in\! (0,k]~
x\!\in\!\big( (V_0\!\cap\! U_{2j+1})\!\setminus\! U_{2j})\big)\!\cup\!
\big( U_{2j-1}\!\setminus\! (V_1\!\cup\! U_{2j-2})\big)\cr
& \hfill{\vee x\!\in\! U_{2k+1}\!\setminus\! (V_1\cup U_{2k})}\cr
& \Leftrightarrow\! x\!\in\! (V_0\!\cap\! U_1)\!\setminus\! U_0\vee\exists j\!\in\! (0,k]~
x\!\in\!\big( (V_0\!\cap\! U_{2j+1})\cup U_{2j-1}\big)\!\setminus\!\big( (V_1\!\cap\! U_{2j})\!\cup\! U_{2j-2}\big)\cr
& \hfill{\vee x\!\in\! U_{2k+1}\!\setminus\! (V_1\cup U_{2k})}\cr
& \Leftrightarrow\! x\!\in\! D_{2k+4}(O)\mbox{,}
\end{array}$$
proving that $A\cap B\!\in\! D_{2k+4}({\bf\Gamma})$.\medskip

 Conversely, assume that $D_{2k+4}(O)\!\in\! D_{2k+4}({\bf\Gamma})$. We set $U_0\! :=\! O_0$ and, for $0\! <\!\eta\! <\! 2k\! +\! 2$, $U_\eta\! :=\! O_{\eta +2}$, which implies that $U$ is an increasing sequence of subsets of $X$ in $\bf\Gamma$ and $D_{2k+2}(U)$ is in $D_{2k+2}({\bf\Gamma})$. We also set, for $\eta\! <\! 2$, 
$V_\eta\! :=\! O_{\eta +1}$, which implies that $V$ is an increasing sequence of subsets of $X$ in $\bf\Gamma$ and $D_2(V)\!\in\! D_2({\bf\Gamma})$. Moreover, $D_{2k+4}(O)\! =\! D_{2k+2}(U)\cap\big( X\!\setminus\! D_2(V)\big)$ by the equivalences just above.\medskip

 Assume now that $\varepsilon\! =\! 0$. Let $U\! :=\! (U_\eta )_{\eta <2k+1}$ be an increasing sequence of subsets of $X$ in $\bf\Gamma$, $A\! :=\! D_{2k+1}(U)$, $V\! :=\! (V_\eta )_{\eta <2}$ be an increasing sequence of subsets of $X$ in $\bf\Gamma$, and 
${B\! :=\! X\!\setminus\! D_2(V)}$. We now define 
$O_0\! :=\! V_0\cap U_0$, $O_{2j+1}\! :=\! (V_1\cap U_{2j+1})\cup U_{2j-1}$ and 
${O_{2j+2}\! :=\! (V_0\cap U_{2j+2})\cup U_{2j}\cup (V_1\cap U_{2j+1})}$  
if $j\! <\! k$, $O_{2k+1}\! :=\! V_1\cup\bigcup_{\eta <2k}~U_\eta$ and finally 
${O_{2k+2}\! :=\! V_1\cup U_{2k}}$, so that $(O_\eta )_{\eta <2k+3}$ is an increasing sequence of subsets of $X$ in $\bf\Gamma$. Moreover, 
$$\begin{array}{ll}
x\!\in\! A\!\cap\! B\!\!\!\!\!
& \Leftrightarrow\!\big(\exists j\!\leq\! k~~x\!\in\! U_{2j}\!\setminus\! (\bigcup_{\theta <2j}~U_\theta)\big)\wedge 
x\!\in\!\big( V_0\cup (X\!\setminus\! V_1)\big)\cr
& \Leftrightarrow\!\exists j\!\leq\! k~~x\!\in\! (V_0\cap U_{2j})\!\setminus\! (\bigcup_{\theta <2j}~U_\theta )\cup 
U_{2j}\!\setminus\! (V_1\cup\bigcup_{\theta <2j}~U_\theta )\cr
& \Leftrightarrow\! x\!\in\! V_0\!\cap\! U_0\vee\exists j\!\in\! (0,k]~
x\!\in\!\big( (V_0\!\cap\! U_{2j})\!\setminus\! U_{2j-1}\big)\!\cup\!
\big( U_{2j-2}\!\setminus\! (V_1\!\cup\!\bigcup_{\theta <2j-2}~U_\theta )\big)\cr
& \hfill{\vee x\!\in\! U_{2k}\!\setminus\! (V_1\cup\bigcup_{\theta <2k}~U_\theta )}\cr
& \Leftrightarrow\! x\!\in\! V_0\!\cap\! U_0\vee\exists j\!\in\! (0,k]~
x\!\in\!\big( (V_0\!\cap\! U_{2j})\!\cup\! U_{2j-2}\big)\!\setminus\!\big( (V_1\!\cap\! U_{2j-1})\!\cup\!
\bigcup_{\theta <2j-2}~U_\theta\big)\cr
& \hfill{\vee x\!\in\! U_{2k}\!\setminus\! (V_1\cup\bigcup_{\eta <2k}~U_\eta )}\cr
& \Leftrightarrow\! x\!\in\! D_{2k+3}(O)\mbox{,}
\end{array}$$
proving that $A\cap B\!\in\! D_{2k+3}({\bf\Gamma})$.\medskip

 Conversely, assume that $D_{2k+3}(O)\!\in\! D_{2k+3}({\bf\Gamma})$. We set, for $\eta\! <\! 2k\! +\! 1$, 
$U_\eta\! :=\! O_{\eta +2}$, which implies that $U$ is an increasing sequence of subsets of $X$ in 
$\bf\Gamma$ and $D_{2k+1}(U)\!\in\! D_{2k+1}({\bf\Gamma})$. We also set, for $\eta\! <\! 2$, 
$V_\eta\! :=\! O_\eta$, which implies that $V$ is an increasing sequence of subsets of $X$ in 
$\bf\Gamma$ and $D_2(V)\!\in\! D_2({\bf\Gamma})$. Moreover, 
$D_{2k+3}(O)\! =\! D_{2k+1}(U)\cap\big( X\!\setminus\! D_2(V)\big)$ by the equivalences just above.\medskip

\noindent (e) By [FL24b, Lemma 3.1], we may assume that $\varepsilon\! =\! 1$. Assume that 
$C\!\in\!\borone (X)$, $A\!\in\! D_{2k+2}({\bf\Gamma})$, $B\!\in\!\checked{D}_{2k+2}({\bf\Gamma})$, and $D\!\in\!\checked{D}_2({\bf\Gamma})$. Then $\big( (A\cap C)\cup (B\!\setminus\! C)\big)\cap D\! =\! 
(A\cap D\cap C)\cup (B\cap D\!\setminus\! C)$. By (d) and (c), $A\cap D\!\in\! D_{2k+4}({\bf\Gamma})$ and $B\cap D\!\in\!\checked{D}_{2k+4}({\bf\Gamma})$, so that 
$\big( (A\cap C)\cup (B\!\setminus\! C)\big)\cap D$ is in 
$D_{2k+4}({\bf\Gamma})\oplus\checked{D}_{2k+4}({\bf\Gamma})$.\medskip

 Conversely, assume that 
$A'\!\in\!D_{2k+4}({\bf\Gamma})$ and $B'\!\in\!\checked{D}_{2k+4}({\bf\Gamma})$. By (c) and (d), we can find $A$ in $D_{2k+2}({\bf\Gamma})$, $D_A\!\in\!\checked{D}_2({\bf\Gamma})$, 
$B\!\in\!\checked{D}_{2k+2}({\bf\Gamma})$ and $D_B\!\in\!\checked{D}_2({\bf\Gamma})$ such that 
$A'\! =\! A\cap D_A$ and $B'\! =\! B\cap D_B$. Thus 
$$(A'\cap C)\cup (B'\!\setminus\! C)\! =\! (A\cap D_A\cap C)\cup (B\cap D_B\!\setminus\! C)\! =\! 
\big( (A\cap C)\cup (B\!\setminus\! C)\big)\cap\big( (D_A\cap C)\cup (D_B\!\setminus\! C)\big) .$$ 
Now $(A\cap C)\cup (B\!\setminus\! C)\!\in\! 
D_{2k+2}({\bf\Gamma})\oplus\checked{D}_{2k+2}({\bf\Gamma})$, while 
$$(D_A\cap C)\cup (D_B\!\setminus\! C)\!\in\!\checked{D}_2({\bf\Gamma})\oplus
\checked{D}_2({\bf\Gamma})\! =\!\checked{D}_2({\bf\Gamma})$$ 
(since $\big( (V_0\cup\neg V_1)\cap C\big)\cup\big( (W_0\cup\neg W_1)\!\setminus\! C\big)\! =\!
\big( (V_0\cap C)\cup (W_0\!\setminus\! C)\big)\cup
\big( (C\!\setminus\! V_1)\cup\neg (W_1\cup C)\big)$ and 
$\mathbf{\Gamma} = \mathbf{\Delta}^{0}_{1}\mbox{-PU}(\mathbf{\Gamma})$ for any non self-dual Wadge class of Borel sets, by \cite{LouveauSaintRaymond88}), so we are done. This finishes the proof.\boxlem{set}

\noindent\bf Notation.\rm\ We set, for 
$\alpha\!\in\! 2^\omega$ and $\varepsilon\!\in\! 2$, 
$(\alpha )_\varepsilon\! :=\!(\alpha (\varepsilon ),\alpha (\varepsilon\! +\! 2),\alpha (\varepsilon\! +\! 4),\cdots )$. Similarly, if $w\!\in\! 2^{<\omega}$ has even length $2l$ and $\varepsilon\!\in\! 2$, then we set $(w)_\varepsilon\! :=\!\big( w(\varepsilon ),w(\varepsilon\! +\! 2),w(\varepsilon\! +\! 4),\cdots ,w(\varepsilon\! +\! 2l\! -\! 2)\big)$. We set, for 
$A\!\subseteq\! 2^{<\omega}$, 
$A^*\! :=\!\{ w_1\!\cdots\! w_l\mid l\!\in\!\omega\wedge\forall i\! <\! l~~w_{i+1}\!\in\! A\}$, and\medskip

 By [FL24b, Lemma 3.5], if $A\!\subseteq\! 2^{<\omega}$ is a regular language, then too is the following language:
$$L_2\! :=\!\big\{ w\!\in\! 2^{<\omega}\mid\vert w\vert\mbox{ is even }\wedge (w)_0\!\in\! A^*
\wedge\big( (w)_1\!\subseteq\!\mathbbold{0}^\infty\vee\exists p,q\!\in\!\omega ~~
\mathbbold{0}^p\mathbbold{1}\mathbbold{0}^q\mathbbold{1}\!\subseteq\! (w)_1\big)\big\} .$$
The next lemma shows that the completeness can be propagated in the difference hierarchy.

\begin{lemma} \label{propagation} Let $k\in\omega$. If there is $A\!\subseteq\! 2^{<\omega}$ such that $A^\infty$ is $D_{2k+2}(\boraone )\oplus\checked{D}_{2k+2}(\boraone )$-complete, then 
$L_2^\infty$ is $D_{2k+4}(\boraone )\oplus\checked{D}_{2k+4}(\boraone )$-complete.\end{lemma}

\prooflem{propagation} Let us check that 
$$L_2^\infty\! =\!\big\{\alpha\!\in\! 2^\omega\mid 
(\alpha )_0\!\in\! A^\infty\wedge\big( (\alpha )_1\! =\!\mathbbold{0}^\infty\vee\exists p\!\not=\! q~~
(\alpha )_1(p)\! =\! (\alpha )_1(q)\! =\!\mathbbold{1}\big)\big\} .$$ 
Assume that ${\alpha\!\in\! L_2^\infty}$, which gives a sequence $(w_i)_{i\in\omega}$ of nonempty words in $L_2$ with the property that ${\alpha\! =\! w_0w_1\cdots}$. Note that 
$(\alpha )_\varepsilon\! =\! (w_0)_\varepsilon (w_1)_\varepsilon\cdots$ if $\varepsilon\!\in\! 2$ since the $w_i$'s are even. This implies that $(\alpha )_0\!\in\! A^\infty$, and 
$\big( (\alpha )_1\! =\!\mathbbold{0}^\infty\mbox{ or }
\exists p\!\not=\! q~~(\alpha )_1(p)\! =\! (\alpha )_1(q)\! =\!\mathbbold{1}\big)$.\medskip

 Conversely, assume that these two properties hold. If $(\alpha )_1$ has finitely many $\mathbbold{1}$'s, then we choose $l$ in such a way that $(\alpha )_1\vert l$ contains all the $\mathbbold{1}$'s in $(\alpha )_1$, and either 
$(w_0)_1\!\subseteq\!\mathbbold{0}^\infty$, or ${\mathbbold{0}^p\mathbbold{1}\mathbbold{0}^q\mathbbold{1}\!\subseteq\! (w_0)_1}$ for some $p,q\!\in\!\omega$. We then continue the decomposition of $(\alpha )_0$ into words of $A$, which gives $(w_{i+1})_0$. Setting 
$(w_{i+1})_1\! :=\!\mathbbold{0}^{\vert (w_{i+1})_0\vert}$, we get $w_{i+1}\!\in\! L_2$ and 
$\alpha\! =\! w_0w_1\cdots$, proving that $\alpha\!\in\! L_2^\infty$. If $(\alpha )_1$ has infinitely many $\mathbbold{1}$'s, then we choose $(w_{i+1})_1$ of a correct (with respect to $A^*$) length, big enough to ensure the existence of at least two ones in $(w_{i+1})_1$. Again, $\alpha\!\in\! L_2^\infty$. Thus 
$L_2^\infty\!\in\!\big( D_{2k+2}(\boraone )\oplus\checked{D}_{2k+2}(\boraone )\big)\sqcap
\checked{D}_2(\boraone )$, which is $D_{2k+4}(\boraone )\oplus\checked{D}_{2k+4}(\boraone )$ by Lemma \ref{set}.(e). We set 
$O_0\! :=\!\{\alpha\!\in\! 2^\omega\mid\exists p\!\not=\! q~~\alpha (p)\! =\!\alpha (q)\! =\!\mathbbold{1}\}$ and $O_1\! :=\!\{\alpha\!\in\! 2^\omega\mid\alpha\!\not=\!\mathbbold{0}^\infty\}$, so that 
$2^\omega\!\setminus\! D_2(O)$ is $\checked{D}_2(\boraone )$-complete by [FL24b, Example 3.3.3]. 
Assume now that $X$ is a zero-dimensional Polish space, and $E$ is a 
$D_{2k+4}(\boraone )\oplus\checked{D}_{2k+4}(\boraone )$ subset of $X$. Lemma \ref{set}.(e) provides a 
$D_{2k+2}(\boraone )\oplus\checked{D}_{2k+2}(\boraone )$ subset $S$ of $X$ and a $D_2(\boraone )$ subset $D$ of $X$ such that $E\! =\! S\!\setminus\! D$. Let 
$f_0\! :\! X\!\rightarrow\! 2^\omega$ be a continuous map with $S\! =\! f_0^{-1}(A^\infty )$, and 
$f_1\! :\! X\!\rightarrow\! 2^\omega$ be a continuous map with 
$X\!\setminus\! D\! =\! f_1^{-1}\big( 2^\omega\!\setminus\! D_2(O)\big)$. Then the map 
$f\! :\! X\!\rightarrow\! 2^\omega$ defined by $\big( f(x)\big)_\varepsilon\! :=\! f_\varepsilon (x)$ is continuous and satisfies $E\! =\! f^{-1}(L_2^\infty )$, showing the completeness of $L_2^\infty$ and finishing the proof.\boxlem{propagation}

\proofthm{condelta} We argue by induction on $n$. Our assumption provides a regular language 
$A\!\subseteq\! 2^{<\omega}$ with the property that $A^\infty$ is complete for the class $D_2(\boraone )\oplus\checked{D}_2(\boraone )$. Inductively, Lemma \ref{propagation} solves our problem.\boxthm{condelta}

\section{Concluding remarks}\indent

 We have got new Wadge degrees  of  $\omega$-powers of one-counter languages. Notice that they are also new Wadge degrees  of  $\omega$-powers, even in the general case of $\omega$-powers of finitary languages.  On the other hand, recall that the results of 
\cite{Fink-Lec5,DBLP:journals/tcs/FinkelGD26} implied that the length of the Wadge hierarchy of 
$\omega$-powers of rational languages is greater than or equal to $\omega + \omega$.  Moreover,  with  the results of \cite{Fink-Lec4}, we obtain that the length of the Wadge hierarchy of $\omega$-powers of  one-counter  languages is greater than or equal to 
$\omega + \omega + \omega =\omega\cdot 3$.  Then the main result we get in this paper leads to the following  improvement: the length of the Wadge hierarchy of $\omega$-powers of  one-counter  languages is greater than or equal to $\omega^2$.  This could still be extended to $\omega^2 + 1$ if we add the example of $\omega$-power of a one-counter  language which is ${\bf\Sigma}^1_1$-complete (hence in particular non-Borel), which was obtained in \cite{Fin03a}. Moreover, if we add the example of $\omega$-power of a context-free language which is Borel of infinite rank, given in \cite{Fin-Dup07}, then we obtain that the Wadge hierarchy of $\omega$-powers of context free languages is greater than or equal to $\omega^2 + 2$. 

\vfill\eject\medskip 

  Despite our efforts, the fact remains that the question of fully determining the topological complexity of  $\omega$-powers of one-counter languages  remains unresolved. Recall that, in a previous article, it has been demonstrated   that for each recursive ordinal $\xi <\omega_1^{\text{CK}}$, there exist some $\omega$-languages $P_\xi$ and $S_\xi$  accepted by B\"uchi one-counter automata such that $P_\xi$ is $\mathbf{\Pi}^{0}_{\xi}$-complete and $S_\xi$ is $\mathbf{\Sigma}^{0}_{\xi}$-complete \cite{Fin-mscs06}.\medskip
  
  Besides, each $\omega$-language $L\subseteq\Alphabet^{\omega}$ accepted by a B\"uchi one-counter automaton is of the form $L=\bigcup_{1\leq j \leq n} U_j\cdot V_j^\infty$, for some one-counter finitary languages $U_j$ and $V_j$, $1\leq j\leq n$. Therefore it seems plausible that there exist complete $\omega$-powers of a one-counter language, for each Borel class of recursive rank.   Actually the situation is even more complicated, since it has been proved in \cite{Fin-mscs06} that there exist some  B\"uchi one-counter automata accepting some Borel $\omega$-languages of non-recursive ranks strictly greater than $\omega_1^{\text{CK}}$.  So the question naturally arises whether there exist some one counter finitary language whose $\omega$-power is Borel of non-recursive rank.

\nocite{Fink-Lec3}

\bibliographystyle{alpha}
\addcontentsline{toc}{section}{References}
\begin{footnotesize}
\bibliography{Duparc-Finkel-Lecomte-one-counter-omega-powers_version_7_indices.bib}

\begin{thebibliography}{HMU01}

\bibitem[ABB97]{ABB96}
Jean{-}Michel Autebert, Jean Berstel, and Luc Boasson.
\newblock Context-free languages and pushdown automata.
\newblock In Grzegorz Rozenberg and Arto Salomaa, editors, {\em Handbook of
  Formal Languages, Volume 1: Word, Language, Grammar}, pages 111--174.
  Springer, 1997.

\bibitem[CG77a]{CohengGold77a}
Rina~S. Cohen and Arie~Y. Gold.
\newblock Theory of omega-languages. {I.} characterizations of
  omega-context-free languages.
\newblock {\em J. Comput. Syst. Sci.}, 15(2):169--184, 1977.

\bibitem[CG77b]{CohengGold77b}
Rina~S. Cohen and Arie~Y. Gold.
\newblock Theory of omega-languages. {II.} {A} study of various models of
  omega-type generation and recognition.
\newblock {\em J. Comput. Syst. Sci.}, 15(2):185--208, 1977.

\bibitem[DF07]{Fin-Dup07}
Jacques Duparc and Olivier Finkel.
\newblock An omega-power of a context-free language which is {B}orel above
  {$\Delta^{0}_{\omega}$}.
\newblock In Stephan Bold, Benedikt L\"{o}we, and Thoralf R\"{a}sch, editors,
  {\em Foundations of the Formal Sciences V - Infinite Games (FotFS V)},
  volume~11 of {\em Studies in Logic}, pages 109--122. College Publications,
  London, 2007.

\bibitem[Dup01]{Duparc01}
Jacques Duparc.
\newblock Wadge hierarchy and {V}eblen hierarchy part {I:} {B}orel sets of
  finite rank.
\newblock {\em Journal of Symbolic Logic}, 66(1):56--86, 2001.

\bibitem[FGL26]{DBLP:journals/tcs/FinkelGD26}
Olivier Finkel, Johan Girardot, and Dominique Lecomte.
\newblock Some regular {\(\omega\)}-powers in the {H}ausdorff-{K}uratowski
  hierarchy.
\newblock {\em Theor. Comput. Sci.}, 1072:115838, 2026.

\bibitem[Fin01a]{Fin01a}
Olivier Finkel.
\newblock Topological properties of omega context-free languages.
\newblock {\em Theor. Comput. Sci.}, 262(1):669--697, 2001.

\bibitem[Fin01b]{Fin01b}
Olivier Finkel.
\newblock Wadge hierarchy of omega context free languages.
\newblock {\em Theor. Comput. Sci.}, 269(1--2):283--315, 2001.

\bibitem[Fin03]{Fin03a}
Olivier Finkel.
\newblock Borel hierarchy and omega context-free languages.
\newblock {\em Theor. Comput. Sci.}, 290(3):1385--1405, 2003.

\bibitem[Fin04]{Fin04-FI}
Olivier Finkel.
\newblock An omega-power of a finitary language which is a {B}orel set of
  infinite rank.
\newblock {\em Fundam. Informaticae}, 62(3-4):333--342, 2004.

\bibitem[Fin05]{Fin-mscs06}
Olivier Finkel.
\newblock Borel ranks and {W}adge degrees of context free omega-languages.
\newblock In S.~Barry Cooper, Benedikt L{\"{o}}we, and Leen Torenvliet,
  editors, {\em New Computational Paradigms, First Conference on Computability
  in Europe, CiE 2005, Amsterdam, The Netherlands, June 8-12, 2005,
  Proceedings}, volume 3526 of {\em Lecture Notes in Computer Science}, pages
  129--138. Springer, 2005.

\bibitem[FL07]{Fin-Lec1}
Olivier Finkel and Dominique Lecomte.
\newblock There exist some \emph{omega} -powers of any {B}orel rank.
\newblock In Jacques Duparc and Thomas~A. Henzinger, editors, {\em Computer
  Science Logic, 21st International Workshop, {CSL} 2007, 16th Annual
  Conference of the EACSL, Lausanne, Switzerland, September 11-15, 2007,
  Proceedings}, volume 4646 of {\em Lecture Notes in Computer Science}, pages
  115--129. Springer, 2007.

\bibitem[FL09]{Fink-Lec2}
Olivier Finkel and Dominique Lecomte.
\newblock Classical and effective descriptive complexities of omega-powers.
\newblock {\em Ann. Pure Appl. Log.}, 160(2):163--191, 2009.

\bibitem[FL20]{Fink-Lec3}
Olivier Finkel and Dominique Lecomte.
\newblock Descriptive set theory and {\(\omega\)}-powers of finitary languages.
\newblock In Adrian Rezus, editor, {\em Landscapes in Logic}, volume~1 of {\em
  Studies in Logic}, pages 518--541. College Publications, London, 2020.

\bibitem[FL21]{Fink-Lec4}
Olivier Finkel and Dominique Lecomte.
\newblock Some complete $\omega$-powers of a one-counter language, for any
  {B}orel class of finite rank.
\newblock {\em Arch. Math. Log.}, 60(1-2):161--187, 2021.

\bibitem[FL24]{Fink-Lec5}
Olivier Finkel and Dominique Lecomte.
\newblock Wadge degrees of $\mathbf{\Delta}^0_2$ omega-powers.
\newblock {\em Mathematical Logic Quarterly}, 70(3):286--293, 2024.

\bibitem[HMU01]{HopcroftMotwaniUllman2001}
John~E. Hopcroft, Rajeev Motwani, and Jeffrey~D. Ullman.
\newblock {\em Introduction to automata theory, languages, and computation, 2nd
  Edition}.
\newblock Addison-Wesley series in computer science. Addison-Wesley-Longman,
  2001.

\bibitem[Kec95]{kechris1995classical}
A.S. Kechris.
\newblock {\em Classical descriptive set theory}, volume 156.
\newblock Springer, 1995.

\bibitem[Lec05]{Lecomte05}
Dominique Lecomte.
\newblock {\(\omega\)}-powers and descriptive set theory.
\newblock {\em Journal of Symbolic Logic}, 70(4):1210--1232, 2005.

\bibitem[LR88]{LouveauSaintRaymond88}
Alain Louveau and Jean~Saint Raymond.
\newblock Les propri\'et\'es de r\'eduction et de norme pour les classes de
  bor\'eliens.
\newblock {\em Fundamenta Mathematicae}, 131(3):223--243, 1988.

\bibitem[Mos09]{moschovakis2009descriptive}
Yiannis~Nicolas Moschovakis.
\newblock {\em Descriptive set theory}, volume 155.
\newblock American Mathematical Society, 2009.

\bibitem[Niw90]{Niwinski90}
Damian Niwinski.
\newblock Problem on $\omega$-powers.
\newblock Proceedings of the \emph{Logics and Recognizable Sets} Workshop at
  Kiel University, 1990.

\bibitem[PP04]{perrin2004infinite}
Dominique Perrin and Jean-{\'E}ric Pin.
\newblock {\em Infinite words: automata, semigroups, logic and games}, volume
  141.
\newblock Academic Press, 2004.

\bibitem[Sel85]{Selivanov85}
Victor~L. Selivanov.
\newblock Ershov hierarchy.
\newblock {\em Siberian Math. Journal}, 26(1):134--149, 1985.

\bibitem[Sem09]{Semmes09}
Brian~Thomas Semmes.
\newblock {\em A Game for the Borel Functions}.
\newblock Phd thesis, Institute for Logic, Language and Computation,
  Universiteit van Amsterdam, Agnietenkapel, May 2009.
\newblock Available at
  \url{https://eprints.illc.uva.nl/id/eprint/199/1/PP-2006-24.text.pdf}.

\bibitem[Sim92]{Simonnet92}
Pierre Simonnet.
\newblock {\em Automates et th\'eorie descriptive}.
\newblock Phd thesis, Universit\'e Paris VII, Jussieu, 1992.
\newblock Available at \url{https://theses.fr/1992PA077187}.

\bibitem[Sta86]{Staiger86a}
Ludwig Staiger.
\newblock Hierarchies of recursive omega-languages.
\newblock {\em J. Inf. Process. Cybern.}, 22(5/6):219--241, 1986.

\bibitem[Sta97]{Staiger97}
Ludwig Staiger.
\newblock \emph{{\(\omega\)}}-languages.
\newblock In Grzegorz Rozenberg and Arto Salomaa, editors, {\em Handbook of
  Formal Languages, Volume 3: Beyond Words}, pages 339--387. Springer, 1997.

\bibitem[Tho90]{Thomas90}
Wolfgang Thomas.
\newblock Automata on infinite objects.
\newblock In Jan van Leeuwen, editor, {\em Handbook of Theoretical Computer
  Science, Volume {B:} Formal Models and Semantics}, pages 133--191. Elsevier
  and {MIT} Press, 1990.

\bibitem[Wad83]{Wadge83}
William~W. Wadge.
\newblock {\em Reducibility and determinateness on the Baire space}.
\newblock Phd thesis, University of California, Berkeley, 1983.
\newblock Available at
  \url{https://billwadge.com/wp-content/uploads/2021/02/wadgephd.pdf}.

\end{thebibliography}
%\bibliography{abbrevs,Duparc-Finkel-Lecomte-one-counter-omega-powers_version_5_indices.bib}
\end{footnotesize}

\end{document}